\documentclass[vecphys]{svmult}

\usepackage{makeidx}         
\usepackage{graphicx}        
\usepackage{multicol}        
\usepackage[bottom]{footmisc}
\usepackage{epsfig}
\usepackage{array}
\usepackage{longtable}
\usepackage{rotating}
\usepackage{subfigure}
\usepackage{amssymb}
\usepackage{mathrsfs}
\usepackage{calrsfs}
\usepackage{graphics}
\usepackage{txfonts}
\usepackage{pslatex}

\newcommand{\sqrtsnn}{\sqrt{s_{\mbox{\tiny{\it{NN}}}}}}
\newcommand{\sigmaNN}{\sigma_{\mbox{\tiny{\it{NN}}}}}
\newcommand{\AaAa}{$AA$}
\newcommand{\AuAu}{$AuAu$}
\newcommand{\CuCu}{$CuCu$}
\newcommand{\PbPb}{$PbPb$}

\newcommand{\dAu}{$dAu$}
\newcommand{\pp}{$pp$}
\newcommand{\ppbar}{$p\bar{p}$}
\newcommand{\sqrts}{\sqrt{s}}
\newcommand\mqhat{{\mean{\hat{q}}}}
\newcommand\qhat{{\hat{q}}}

\newcommand{\dETdeta}{dE_{T}/d\eta|_{\eta=0}}
\newcommand{\ecrit}{\varepsilon_{\mbox{\tiny{\it crit}}}}

\newcommand{\Tcrit}{T_{\mbox{\tiny{\it crit}}}}

\newcommand{\Qeff}{Q_{\mbox{\tiny{\it eff}}}}

\newcommand{\pythia}{{\sc pythia}}
\newcommand{\herwig}{{\sc herwig}}
\newcommand{\qpythia}{{\sc q-pythia}}
\newcommand{\qherwig}{{\sc q-herwig}}
\newcommand{\hydjet}{{\sc hydjet}}
\newcommand{\hijing}{{\sc hijing}}
\newcommand{\jewel}{{\sc jewel}}
\newcommand{\fastjet}{{\sc FastJet}}

\def\ttt#1{\texttt{\small #1}}
\def\mean#1{\ensuremath{\left<#1\right>}}

\def\gampi{\gamma$--$h\ } 
\def\z{z_{_{\gamma h}}}

\newcommand{\IAA}{I_{AA}}
\newcommand{\IAAa}{I^{away}_{AA}}

\newcommand{\etg}{E_T^{\gamma}}
\newcommand{\etj}{E_T^{jet}}

\newcommand {\llb} { \left[ \frac{\mbox{}}{\mbox{}} \right.}
\newcommand {\lrb} { \left. \frac{\mbox{}}{\mbox{}} \right] }

\makeindex             

\begin{document}

\title*{Jet quenching} 

\author{David d'Enterria}
\institute{CERN, PH-EP, CH-1211 Geneva 23, Switzerland\\
LNS, MIT, Cambridge, MA 02139-4307, USA}

\maketitle

\begin{abstract}
We present a comprehensive review of the physics of hadron and jet production at large 
transverse momentum in high-energy nucleus-nucleus collisions. Emphasis is put on experimental 
and theoretical ``jet quenching'' observables that provide direct information on the (thermo)dynamical 
properties of hot and dense QCD matter.
\end{abstract}


\section{Introduction}
\label{sec:intro}


The research programme of high-energy nucleus-nucleus physics is focused on the study 
of the fundamental theory of the strong interaction -- Quantum Chromo Dynamics (QCD) --
in extreme conditions of temperature, density and small parton momentum fraction (low-$x$) 
-- see e.g.~\cite{d'Enterria:2006su} for a recent review.
By colliding two heavy nuclei at ultrarelativistic energies one expects to form a hot and dense 
deconfined medium whose collective (colour) dynamics can be studied experimentally.
Lattice QCD calculations~\cite{latt} predict a new form of matter at energy densities (well)
above $\ecrit \approx$~1~GeV/fm$^3$ consisting of an extended volume of deconfined and
chirally-symmetric (bare-mass) quarks and gluons: the Quark Gluon Plasma (QGP)~\cite{shuryak77}.\\

Direct information on the thermodynamical properties (like temperature, energy or particle densities, ...) 
and transport properties (such as viscosity, diffusivity and conductivity coefficients) of the QGP can be 
obtained by comparing the results for a given observable $\Phi_{AA}$ measured in nucleus-nucleus (\AaAa, ``QCD medium'')
to those measured in proton-proton ($pp$, ``QCD vacuum'') 
collisions as a function of centre-of-mass (c.m.) energy $\sqrtsnn$, transverse momentum $p_T$, rapidity $y$, reaction centrality
(impact parameter $b$), and particle type (mass $m$). Schematically:
\begin{eqnarray}
\!\!\!\!\!\!\!\!\!\!\!\!\!\!\!\!\!\!R_{AA}(\sqrtsnn,p_T,y,m;b) &=&\frac{\mbox{\small{``hot/dense QCD medium''}}}{\mbox{\small{``QCD vacuum''}}}
\, \propto \,\frac{\Phi_{AA}(\sqrtsnn,p_T,y,m;b)}{\Phi_{pp}(\sqrts,p_T,y,m)}
\label{eq:Ratio_AA}
\end{eqnarray}
Any observed {\it enhancement} ($R_{AA}>1$) and/or {\it suppression} ($R_{AA}<1$) in this ratio 
can then be linked to the properties of strongly interacting matter after accounting 
for a realistic (hydrodynamical) modeling of the space-time evolution of the expanding 
system of quarks and gluons 
produced in the collision.\\

This article presents an up-to-date review of experimental and theoretical studies of 
quantities such as $R_{AA}(\sqrtsnn,p_T,y,m;b)$ for high-$p_T$ hadrons 
and jets produced at RHIC collider energies and prospects for the LHC.


\section{Jet quenching and parton energy loss in QCD matter}

In this first Section, we introduce the general concepts, basic variables and formulas of energy loss 
of a fast charged particle in a dense thermalised plasma (starting with the somewhat simpler QED case),
and we enumerate the expected phenomenological consequences of QCD energy loss for gluons and 
light- and heavy-quarks traversing a hot and dense QGP.

\subsection{Hard probes of hot and dense QCD matter}

\begin{figure}[htbp]
\centering
\includegraphics[width=8cm,clip]{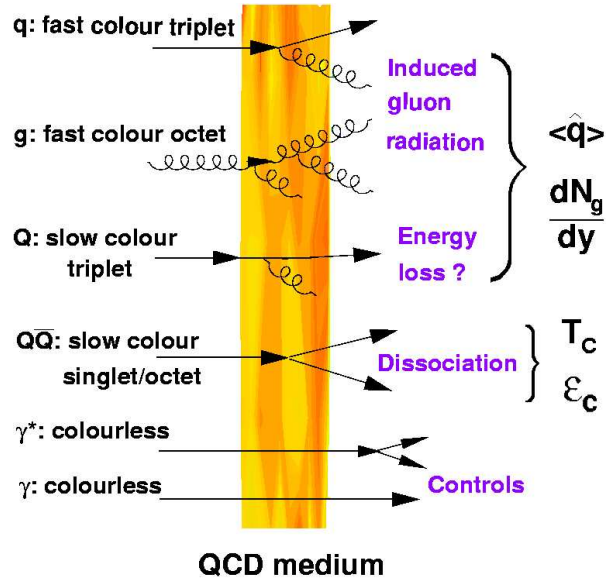}
\caption{Examples of hard probes whose modifications in high-energy \AaAa\ collisions provide 
direct information on properties of QCD matter such as the transport coefficient $\qhat$, the
initial gluon rapidity density $dN^g/dy$, and the critical temperature $\Tcrit$ and energy 
density $\ecrit$~\cite{d'Enterria:2006su}.}
\label{fig:qgp_probes}
\end{figure}

Among all available observables 
in high-energy nuclear collisions, particles with large transverse momentum and/or mass, 
$p_T,m\gtrsim Q_{_{0}} \gg \Lambda_{_{QCD}}$, where~$Q_{_{0}}$~=~$\mathscr{O}(1$~GeV$)$
and $\Lambda_{_{QCD}}\approx$~0.2~GeV is the QCD scale, constitute valuable tools 
to study ``tomographically'' the hottest and densest phases 
of the reaction (Fig.~\ref{fig:qgp_probes}). Indeed, such ``hard probes''
(i) originate from partonic scatterings with large momentum transfer $Q^2$ and thus are directly 
coupled to the fundamental QCD degrees of freedom, (ii) 
are produced in very short time-scales, $\tau\sim 1/p_T \ll 1/Q_{_{0}} \sim$~0.1 fm/c, allowing them to 
propagate through (and be potentially affected by) the medium, and (iii) their cross sections can be 
theoretically predicted using the perturbative QCD (pQCD) framework~\cite{yr_hardprobes_lhc}.

Jet production in hadronic collisions is a paradigmatic hard QCD process. An elastic ($2 \to 2$) or inelastic 
($2 \to 2 + X$) scattering of two partons from each one of the colliding hadrons (or nuclei) results in the 
production of two or more partons in the final-state. At high $p_T$, the outgoing partons have a large virtuality 
$Q$ which they reduce by subsequently radiating gluons and/or splitting into quark-antiquark pairs. 
Such a parton branching evolution is governed by the QCD ``radiation probabilities'' given by the 
Dokshitzer-Gribov-Lipatov-Altarelli-Parisi (DGLAP) equations~\cite{dglap}
down to virtualities $\mathscr{O}$(1~GeV$^2$). At this point,
the produced partons fragment non-perturbatively into a set of final-state hadrons. The characteristic collimated 
spray of hadrons resulting from the fragmentation of an outgoing parton is called a ``jet''.
\begin{figure}[htbp]
\centering
\includegraphics[height=6.5cm,width=6.7cm,clip]{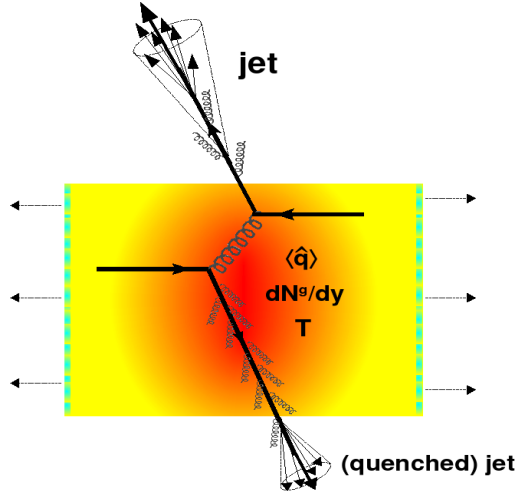}
\caption{``Jet quenching'' in a head-on nucleus-nucleus collision. Two quarks suffer a hard scattering: one goes out 
directly to the vacuum, radiates a few gluons and hadronises, the other goes through the dense plasma created 
(characterised by transport coefficient $\qhat$, gluon density $dN^g/dy$ and temperature $T$), suffers 
energy loss due to medium-induced gluonstrahlung and finally fragments outside into a (quenched) jet.}
\label{fig:jet_quench}
\end{figure}

One of the first proposed ``smoking guns'' of QGP formation was ``jet quenching''~\cite{Bjorken:1982tu} 
i.e. the attenuation or disappearance of the spray of hadrons resulting from the fragmentation of a 
parton having suffered energy loss in the dense plasma produced in the reaction (Fig.~\ref{fig:jet_quench}).
The energy lost by a particle in a medium, $\Delta E$, provides fundamental information on its properties.
In a general way, $\Delta E$ depends both on the characteristics of the particle traversing it (energy $E$, mass $m$, 
and charge) and on the plasma properties (temperature $T$,  particle-medium interaction coupling\footnote{The QED 
and QCD coupling ``constants'' are $\alpha_{em} = e^2/(4\pi)$ and $\alpha_{s} = g^2/(4\pi)$ respectively.}
$\alpha$, and thickness $L$), i.e. $\Delta E(E, m, T, \alpha, L)$. The following (closely related) variables are 
extremely useful to characterise the interactions of a particle inside a medium: 
\begin{itemize}
\item the {\it mean free path} $\lambda=1/(\rho\sigma)$, where $\rho$ is the medium density 
($\rho\propto T^3$ for an ideal gas) and $\sigma$ the integrated cross section of the particle-medium
interaction\footnote{One has $\lambda\sim(\alpha T)^{-1}$ since the QED,QCD {\it screened} Coulomb scatterings are $\sigma_{el}\propto \alpha/T^2$.},
\item the {\it opacity} $N = L/\lambda$ or number of scatterings experienced by the particle in a medium of thickness~$L$,
\item the {\it Debye mass} 
$m_{D}(T) \sim g\,T$ (where $g$ is the coupling parameter) 
is the inverse of the screening length of the (chromo)electric fields in the plasma. $m_D$ characterises the typical
momentum exchanges with the medium and also gives the order of the ``thermal masses'' of the plasma constituents,
\item the {\it transport coefficient} $\qhat \equiv m_D^2/\lambda$ encodes the ``scattering power'' of the medium
through the average transverse momentum squared transferred to the traversing particle per unit path-length. 
$\qhat$ combines both thermodynamical ($m_D,\rho$) and dynamical ($\sigma$) properties of the 
medium~\cite{Arnold:2008iy,Arnold:2008vd,Baier:2008js}:
\begin{equation}
\qhat \; \equiv m_D^2/\lambda \; = \; m_D^2\; \rho \; \sigma \; .
\label{eq:qhat}
\end{equation}
As a numerical QCD example\footnote{
For unit conversion, multiply by powers of $\hslash\,c\simeq$~0.2\,GeV\,fm
(other useful equalities: 10~mb~=~1~fm$^{2}$, and 1~GeV$^{-2}$~=~0.389~mb).}, 
let us consider an equilibrated {\it gluon} plasma 
at $T=0.4$~GeV and a strong coupling $\alpha_s\approx$~0.5~\cite{Baier:2006fr}. At this temperature, 
the particle (energy) density is $\rho_g = 16/\pi^2 \;\zeta (3)\cdot T^3 \approx$~15~fm$^{-3}$ 
($\epsilon_g = 8 \pi^2/15\cdot T^4$ $\approx$~17~GeV/fm$^3$), i.e. 100 times denser than
normal nuclear matter ($\rho$~=~0.15~fm$^{-3}$). 
At leading order (LO), the Debye mass is $m_D = (4 \pi \alpha_s)^{1/2} T\approx$~1~GeV. 
The LO gluon-gluon cross section is 
$\sigma_{gg} \simeq 9 \pi \alpha^2_s/(2 m_D^2) \approx$~1.5~mb. 
The gluon mean free path in such a medium is  $\lambda_g = 1/(\rho_g \sigma_{gg}) \simeq $~0.45~fm
(the quark mean-free-path is $\lambda_q = C_A/C_F \,\lambda_g\approx$~1~fm,
where $C_A/C_F = 9/4$ is the ratio of gluon-to-quark colour factors).
The transport coefficient is therefore $\hat q \simeq m_D^2 / \lambda_g \simeq 2.2$~GeV$^2$/fm. 
Note that such a numerical value has been obtained with a LO expression in $\alpha_s$ 
for the parton-medium cross section. Higher-order scatterings (often encoded in a ``$K$-factor''~$\approx$~2~--~4) 
could well result in much larger values of $\qhat$.
\item the {\it diffusion constant} $D$, characterising the dynamics of {\it heavy} non-relativistic particles 
(mass $M$ and speed $v$) traversing the plasma, is connected, via the Einstein relations
\begin{equation}
D = \; 2T^2/\kappa\; = T/(M\;\eta_D) \;
\label{eq:diff_eq}
\end{equation}
to the {\it momentum diffusion coefficient} $\kappa$ -- the average momentum squared gained by 
the particle per unit-time (related to the transport coefficient as $\kappa\approx \qhat\,v$) -- 
and the {\it momentum drag coefficient} $\eta_D$.
\end{itemize}


\subsection{Mechanisms of in-medium energy loss}
\label{sec:intro_theory}

In a general way, the total energy loss of a particle traversing a medium is the sum of 
collisional and radiative terms\footnote{In addition, synchrotron-, \v{C}erenkov- and transition-radiation energy losses can 
take place respectively if the particle interacts with the medium magnetic field, if its velocity is greater 
than the local phase velocity of light, or if it crosses suddenly from one medium to another. Also, plasma instabilities
may lead to energy losses. Yet, those effects -- studied e.g. in~\cite{Zakharov:2008uk,Dremin:2009kf,Djordjevic:2005nh,Mannarelli:2008ib} 
for QCD plasmas -- are generally less important in terms of the amount of $E_{loss}$.}: $\Delta E = \Delta E_{coll} +\Delta E_{rad}$. 
Depending on the kinematic region, a (colour) charge can lose energy\footnote{Note that 
if the energy of the particle is similar to the plasma temperature, $E\sim \mathscr{O}(T)$,
the particle can also {\it gain} energy while traversing it.} in a plasma 
with temperature $T$ mainly by two mechanisms\footnote{Note 
that the separation is not so clear-cut since the diagrams assume well-defined asymptotic out states, but the 
outgoing particles may still be in the medium and further rescatter.}.
\begin{figure}[htbp]
\centering
\includegraphics[height=3.4cm,clip]{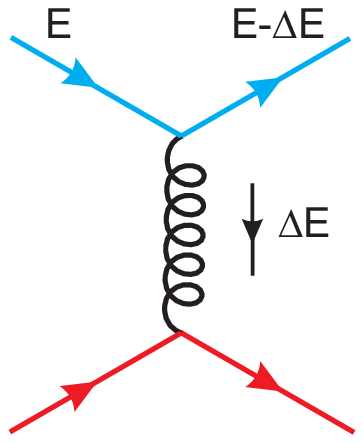}\hspace*{2cm}
\includegraphics[height=3.4cm,clip]{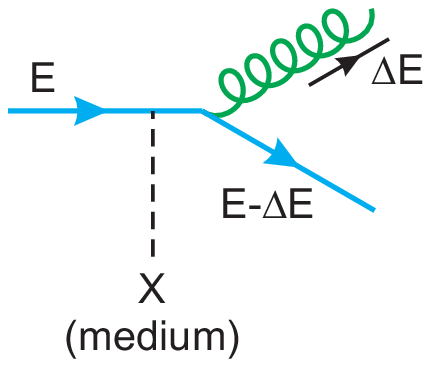}
\caption{Diagrams for collisional  (left) and radiative (right) energy losses of a quark of energy $E$
traversing a quark-gluon medium.}
\label{fig:FeynmanDiagrams}
\end{figure}
\begin{itemize}
\item  {\bf Collisional energy loss} through {\it elastic} scatterings with the medium constituents 
(Fig.~\ref{fig:FeynmanDiagrams}, left) 
dominates at low particle momentum.
The {\it average} energy loss in one scattering (with cross section $d\sigma/dt$, where $t=Q^2$ is the momentum 
transfer squared) in a medium of temperature $T$, is:
\begin{equation}
\mean{\Delta E_{coll}^{1 scat}}\approx\frac{1}{\sigma\,T}\;\int_{m_D^2}^{t_{max}} t \,\frac{d\sigma}{dt}\; dt \;\;.
\label{eq:DE_dt}
\end{equation}
\item {\bf Radiative energy loss} through {\it inelastic} scatterings 
within the medium (Fig.~\ref{fig:FeynmanDiagrams}, right), 
dominates at higher momenta. This loss can be determined from the corresponding single- or double-differential 
photon or gluon {\it Bremsstrahlung spectrum} ($\omega \; dI_{rad}/d\omega$ or $\omega \; d^2I_{rad}/d\omega \,dk_\perp^2$, 
where $\omega$, $k_\perp$ are respectively the energy and transverse momentum of the radiated photon or gluon):
\begin{eqnarray}
\Delta E_{rad}^{1 scat}=\int^E \omega \,\frac{dI_{rad}}{d\omega} \;d\omega\; , \;\; \mbox{ or }\;\;
\Delta E_{rad}^{1 scat}=\int^E \int^{k_{T,max}} \omega \,\frac{d^2I_{rad}}{d\omega\,dk_\perp^2} \;d\omega\,dk_\perp^2\;.
\label{eq:DE_domega}
\end{eqnarray}
\end{itemize}
For incoherent scatterings one has simply: $\Delta E^{tot} =N \cdot \Delta E^{1 scat}$, where $N=L/\lambda$ is the medium opacity.
The energy loss per unit length or {\it stopping power}\footnote{By `stopping power', one means 
a property of the matter, while `energy loss per unit length' describes what happens to the particle. 
For a given particle, the numerical value and units are identical (and both are usually written with 
a minus sign in front).} is: 
\begin{equation}
-\frac{dE}{dl} = \frac{\mean{\Delta E^{tot}}}{L}~,
\label{eq:dEdl}
\end{equation}
which for {\it incoherent} scatterings reduces to: $- dE/dl = \mean{\Delta E^{1 scat}}/{\lambda}$.


\subsubsection{Energy losses in QED}

\begin{figure}[htbp]
\centering
\includegraphics[height=5.5cm,width=10.cm,clip]{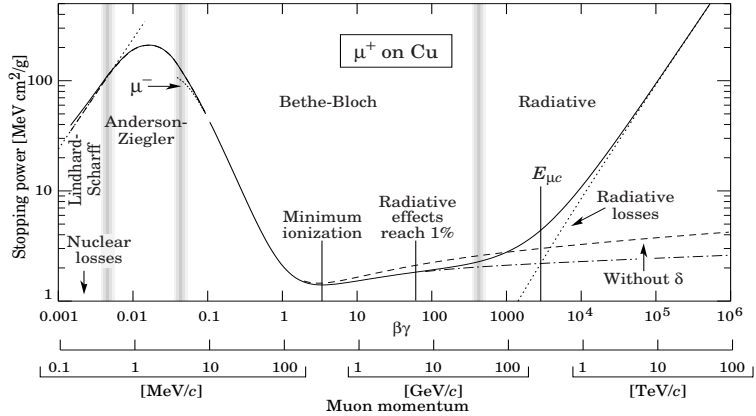}
\caption{Stopping power, $-dE/dl$, 
for positive muons in copper as a function of  $\beta\gamma = p/Mc$ (or momentum $p$). 
The solid curve indicates the total stopping power~\protect{\cite{Yao:2006px}}.}
\label{fig:BetheBloch}
\end{figure}

As an illustrative example, we show in Fig.~\ref{fig:BetheBloch} the stopping power of muons in copper. 
At low and high energies, the collisional (aka ``Bethe-Bloch'') and the radiative energy losses dominate respectively.

Yet, the hot and dense plasma environment that one encounters in ``jet quenching'' scenarios
is not directly comparable to the QED energy loss in {\it cold} matter represented in Fig.~\ref{fig:BetheBloch}.
A recent review by Peign\'e and Smilga~\cite{Peigne:2008wu} presents the parametric dependences of the 
energy loss of a lepton traversing a {\it hot} QED plasma with temperature $T$ and Debye-mass $m_D$.
In a simplified manner, inserting the Coulomb (lepton-lepton) and Compton (lepton-photon) 
scattering cross sections in Eq.~(\ref{eq:DE_dt}) and using Eq.~(\ref{eq:dEdl}), one obtains the 
{\it collisional} energy losses per unit-length:
\begin{itemize}
\item Light lepton $(M^2\ll ET)$: $- \frac{d E_{coll}}{dl} \simeq \frac{\pi}{3}\alpha^2 T^2 \; \ln \left(\frac{E\,T}{m_{D}^2}\right)
\sim \alpha\;m_D^2 \; \ln \left(\frac{E\,T}{m_{D}^2}\right)$
\item Heavy lepton $(M^2\gg ET)$: $- \frac{d E_{coll}}{dl} \simeq \frac{2\pi}{3}\alpha^2 T^2 \; \ln \left(\frac{E\,T}{m_{D}\,M} \right)
\sim \alpha\; m_D^2 \; \ln \left(\frac{E\,T}{m_{D}\,M} \right)$
\end{itemize}
For {\it radiative} losses, the amount of photon emission, depends chiefly on the thickness of the plasma\footnote{We 
consider here the formulas where the charged particle is produced {\it inside} the plasma, as this is the typical situation 
encountered in a QGP.}. For thin media ($L\ll\lambda$), the traversing particle suffers at most one single scattering 
and the QED radiation spectrum is just given by the Bethe-Heitler (BH) bremsstrahlung expression~\cite{Bethe:1934za}. 
On the contrary, for thick media ($L\gg\lambda$) there are $N$(=opacity) scatterings and the Landau-Pomeranchuk-Migdal 
(LPM)~\cite{Migdal:1956tc} coherence effect\footnote{The LPM effect describes the fact that, since it takes a finite time 
to emit a photon, neighbouring medium particles interfere coherently (destructively) and act as one effective scattering centre, 
inducing {\it single} photon radiation.} reduces the amount of radiation compared to $N$ times the BH spectrum. 
Making use of Eq.~(\ref{eq:DE_domega}) one obtains, from each corresponding radiation spectra, the following parametric
expressions~\cite{Peigne:2008wu}:
\begin{itemize}
\item ``Bethe-Heitler''\footnote{Strictly speaking, as discussed in~\cite{Peigne:2008wu}, for light particles
produced {\it inside} a plasma, there is no BH regime whatsoever: the $L^2$-dependence appearing in 
$\Delta E_{rad}^{BH}$ is due to in-medium formation-time constraints absent in the truly (asymptotic) BH regime.} $(L\ll \lambda)$:
$\;\;\;\omega \frac{dI_{rad}}{d\omega} \sim \alpha\; (L^2 m_D^2/\lambda) \cdot \omega/E^2 \sim \alpha\; \qhat\,L^2\cdot \omega/E^2$
\begin{eqnarray}
\hspace*{-1.cm} 
\Delta E_{rad}^{BH} \sim \alpha\; \qhat\;L^2 \sim \alpha^3\, T^3\,L^2 \; \; \Longrightarrow \; \;
- \frac{d E_{rad}}{dl} = \frac{\Delta E_{rad}}{L} \sim \alpha^3\,T^3\,L   \; \; 
\label{eq:BH_QED}
\end{eqnarray}
\item Landau-Pomeranchuk-Migdal $(L\gg \lambda)$: $\;\;\; \omega \frac{dI_{rad}}{d\omega} \approx \alpha^2\, L \sqrt{T^3\,\frac{\omega}{E^2}\,\ln\left(E^2/(\omega T)\right)}$
\begin{eqnarray}
\hspace*{-0.8cm} \Delta E_{rad}^{LPM} \sim \alpha^2\, L\,\sqrt{ET^3\,\ln(E/T)}\; \; \Longrightarrow \; \; - \frac{d E_{rad}}{dl} 
\sim \alpha^2\,\sqrt{ET^3\,\ln(E/T)} 
\label{eq:LPM_QED}
\end{eqnarray}
\end{itemize}
We see that, in general, the radiative energy losses of an energetic lepton crossing a hot QED plasma are much larger
than their collisional losses.
Yet, if the particle is heavy, the amount of radiation at angles within a cone $\theta<M/E$ is suppressed 
by a factor $m_D^2/M^2$ (``dead cone'' effect, see later) resulting in a reduction of the bremsstrahlung emission
by a factor $m_D^2/M^2\sim\alpha\,T^2/M^2$.

\subsubsection{Energy losses in QCD}

The main differences between QED and QCD energy losses result from the {\it non-Abelian} nature of QCD: 
i.e. the fact that gluons can also interact with themselves (at variance with photons in QED)
introduces several important changes. First, 
the QCD coupling $\alpha_s$ runs more rapidly than $\alpha_{em}$ (at least for not 
asymptotically-high temperatures), and the scale $Q$ at which $\alpha_s(Q)$ is evaluated needs 
to be explicitly considered in all calculations of {\it collisional} energy losses~\cite{Peshier:2006hi,Peigne:2008nd}. 
Second, 
it is crucial to take  into account the different coupling of quarks and gluons
with the medium. The relative strengths of the three distinct QCD vertices: $\alpha_s C_F$ for $q \leftrightarrow qg$,  
$\alpha_s C_A$ for $g \leftrightarrow gg$, and $\alpha_s T_F$ for $g \leftrightarrow q{\bar q}$, are determined 
by the structure (Casimir factors $C_R$) of the gauge group describing the strong force\footnote{For $SU(N_c)$ 
with $N_c$ the number of colours: $C_A = N_c$, $C_F = (N_c^2-1)/2N_c$ and $T_F$ = 1/2~\cite{ellis96}.}. 
The probability for a gluon (quark) to radiate a gluon is proportional to the colour factor $C_A$ = 3 ($C_F$ = 4/3). 
In the asymptotic limit, in which the radiated gluons carry a small fraction of the original parton momentum
(and neglecting the gluon splitting into quark-antiquark pairs, proportional 
to the smaller factor $T_R$ = 1/2), the average number of gluons radiated by 
a gluon is $C_A/C_F$ = 9/4 times higher than that radiated by a quark (that is the 
reason why gluon jets have a larger (and softer) hadron multiplicity than quark jets).


\paragraph{QCD collisional energy loss}

The collisional energy loss due to elastic scattering 
of a parton of energy $E$ inside a QGP of temperature $T$ 
was originally estimated by Bjorken~\cite{Bjorken:1982tu} and Braaten-Thoma~\cite{Braaten:1991we} 
and later improved (including running coupling, finite energy kinematics, and quark-mass effects)
by various authors~\cite{Peshier:2006hi,Peigne:2008nd,Zakharov:2007pj}. 
Using Eq.~(\ref{eq:DE_dt}) with the 
momentum-transfer integral limits given by (i) the QGP Debye-mass squared 
$t_{min}=m_{D}^2 (T) \sim 4\pi \alpha _s T^2(1+N_f/6)$ and (ii) $t_{max}= s \sim ET$, and 
taking the dominant contribution to the parton-parton $t$-differential elastic cross section
\begin{equation} 
\frac{d\sigma }{dt} \approx C_i \; \frac{4\pi\,\alpha_s^2(t)}{t^2}~,~\mbox{ with } \alpha_s(t) = \frac{12\pi}{(33-2n_f)\ln{(t/\Lambda_{QCD}^2)}} \>
\label{eq:dsigdt} 
\end{equation} 
where $C_i = 9/4, 1, 4/9$ are the colour factors for $gg$, $gq$ and $qq$ scatterings respectively,
one finally obtains (for $E \gg M^2/T$)~\cite{Peigne:2008nd}:
\begin{itemize}
\item Light-quark, gluon: $-\frac{dE_{coll}}{dl}\big\vert_{q,g} = \frac{1}{4}\,C_R \,\alpha_s(ET)\;m_D^2\,\ln\left(\frac{ET}{m_D^2}\right)\;,$
\item Heavy-quark: $- \frac{d E_{coll}}{dl}\big|_{Q} = -\frac{d E_{coll}}{dl}\big|_{q} - \frac{2}{9}\,C_R\,\pi\,T^2\,\left[\alpha_s(M^2)\alpha_s(ET)\,\ln\left(\frac{ET}{M^2}\right) \right]\;,$
\end{itemize}
with $C_R$~=~4/3 (3) being the quark (gluon) colour charge. The amount of $\Delta E_{coll}$ is 
linear with the medium thickness, and it depends only logarithmically on the initial parton energy.
As a numerical example, taking $E=20$~GeV, $M=1.3$~GeV/c$^2$ (charm quark) and a medium with $T=0.4$~GeV 
and $m_D=1$~GeV/c$^2$, the elastic energy losses per unit-length are $-dE_{coll}/dl\big\vert_{q} = 2.3$~GeV/fm 
and $-dE_{coll}/dl\big|_{Q} = 2.6$~GeV/fm.


\paragraph{QCD radiative energy loss}

The dominant mechanism of energy loss of a fast parton in a QCD environment is of radiative 
nature (``gluonstrahlung'')~\cite{gyulassy90,bdmps,glv,Wiedemann:2000za}: a parton traversing a QGP 
loses energy mainly by medium-induced multiple gluon emission. 
The starting point to determine the radiation probabilities in QCD are the DGLAP splitting 
functions 
in the vacuum~\cite{dglap}
\begin{equation}
P_{{\rm q}\to {\rm qg}}(z) = C_F \, \frac{[1+(1-z)^2]}{z} \,,\;\mbox{ and }\;\;  P_{{\rm g}\to {\rm gg}}(z) = C_A \, \frac{[1 + z^4 + (1-z)^4]}{z(1-z)} \,,
\label{eq:splitt}
\end{equation}
(where $z=\omega/E$ is the fraction of energy of the parent parton taken by the radiated gluon),
modified to take into account the enhanced medium-induced radiation. 
The resulting radiated gluon spectrum, $\omega \;dI_{rad}/d\omega \propto P_{q,g\to g}(\omega/E)$,
has been computed by various groups under various approximations 
(see Section~\ref{sec:Eloss_models}). All medium modifications are often 
encoded into the ``transport coefficient'' parameter, $\qhat$, introduced previously, see Eq.~(\ref{eq:qhat}). 
For thin (thick) media, i.e. for $L \ll \lambda$ ($L \gg \lambda$), one deals with the Bethe-Heitler 
(Landau-Pomeranchuk-Migdal) gluonsstrahlung spectrum. In the LPM case, one further differentiates between the soft 
or hard gluon emission cases with respect to the characteristic gluonstrahlung energy\footnote{Up 
to prefactors, $\omega_c$  is the average energy lost in the medium (for $\omega_c < E$): 
$\omega_c \simeq  2 \mean{\Delta E_{rad}} / (\alpha_s C_R)$.}
$\omega_c = \frac{1}{2}\qhat\;L^2$.
Making use of Eq.~(\ref{eq:DE_domega}), 
the basic QCD radiative energy loss formulas read~\cite{Peigne:2008wu}:
\begin{itemize}
\item ``Bethe-Heitler''\footnote{In reality, BH usually refers to the single scattering spectrum of an {\it asymptotic} particle produced {\it outside}
the medium (see footnote for the analogous Bethe-Heitler QED formula).} (BH) regime $(L\ll \lambda)$: 
\begin{eqnarray}
\hspace*{-1.9cm} 
\omega \frac{dI_{rad}}{d\omega} \approx \alpha_s\; \qhat\,L^2/\omega \; \; \Longrightarrow \; \;
\Delta E_{rad}^{BH} 
\approx \alpha_s\; \qhat\, L^2 \ln(E/(m_D^2 \,L))\; \; 
\label{eq:BH_QCD}
\end{eqnarray}
\item Landau-Pomeranchuk-Migdal (LPM) regime $(L\gg \lambda)$: 
\begin{eqnarray}
\hspace*{-1.cm} 
\omega \frac{dI_{rad}}{d\omega} \approx \alpha_s\, 
\left\{ \begin{array} 
  {l}                 
  \hspace*{-0.1cm}\sqrt{\qhat\,L^2/\omega}    \\ 
   \qhat\,L^2/\omega 
   \end{array} \right.
\Longrightarrow 
\Delta E_{rad}^{LPM} & \approx & \alpha_s\,
\left\{ \begin{array} 
  {l@{\;\;}l}                 
  \qhat\,L^2 & (\omega < \omega_c) \\ 
  \qhat\,L^2\,\ln(E/(\qhat\,L^2)) & (\omega > \omega_c) 
\end{array} \right.
\label{eq:LPM_QCD}
\end{eqnarray}
\end{itemize}
We note two things. First, because of the destructive interference, the LPM spectrum, $\omega \,dI_{rad}/d\omega\propto\omega^{-1/2}$, 
is suppressed in the infrared (i.e. for small $\omega$'s) compared to the independent 
Bethe-Heitler gluon spectrum, $\omega \,dI_{rad}/d\omega\propto\omega^{-1}$. Note also that, 
due to the steeply falling spectrum of the radiated gluons, the integrated LPM energy loss is dominated 
by the region $\omega\lesssim\omega_c$. Second, the QCD energy loss shows a characteristic $L^2$ 
dependence on the plasma thickness which however, as noted in~\cite{Peigne:2008wu}, is also present 
in the case of Abelian (QED) plasmas, see Eq.~(\ref{eq:BH_QED}), and is a general feature of the 
medium-induced energy loss of any {\it in-medium newborn} particle. 
The main distinctions of the energy loss in a QCD compared to a QED plasma are the presence of  
different colour factors for the $q$ and $g$ charges ($\Delta E_{rad}\propto\qhat\propto \sigma_{particle-medium}$ 
which in the QCD case is proportional to $C_R$) and the extra logarithmic dependence of 
$\Delta E_{rad}$ on the energy $E$ of the traversing particle.

\begin{figure}[htpb]
\includegraphics[height=4.5cm]{figs/denterria_Eloss_coll_rad_3D_hydro_qin08.eps}\hspace{0.2cm}
\includegraphics[height=4.4cm,width=5.2cm]{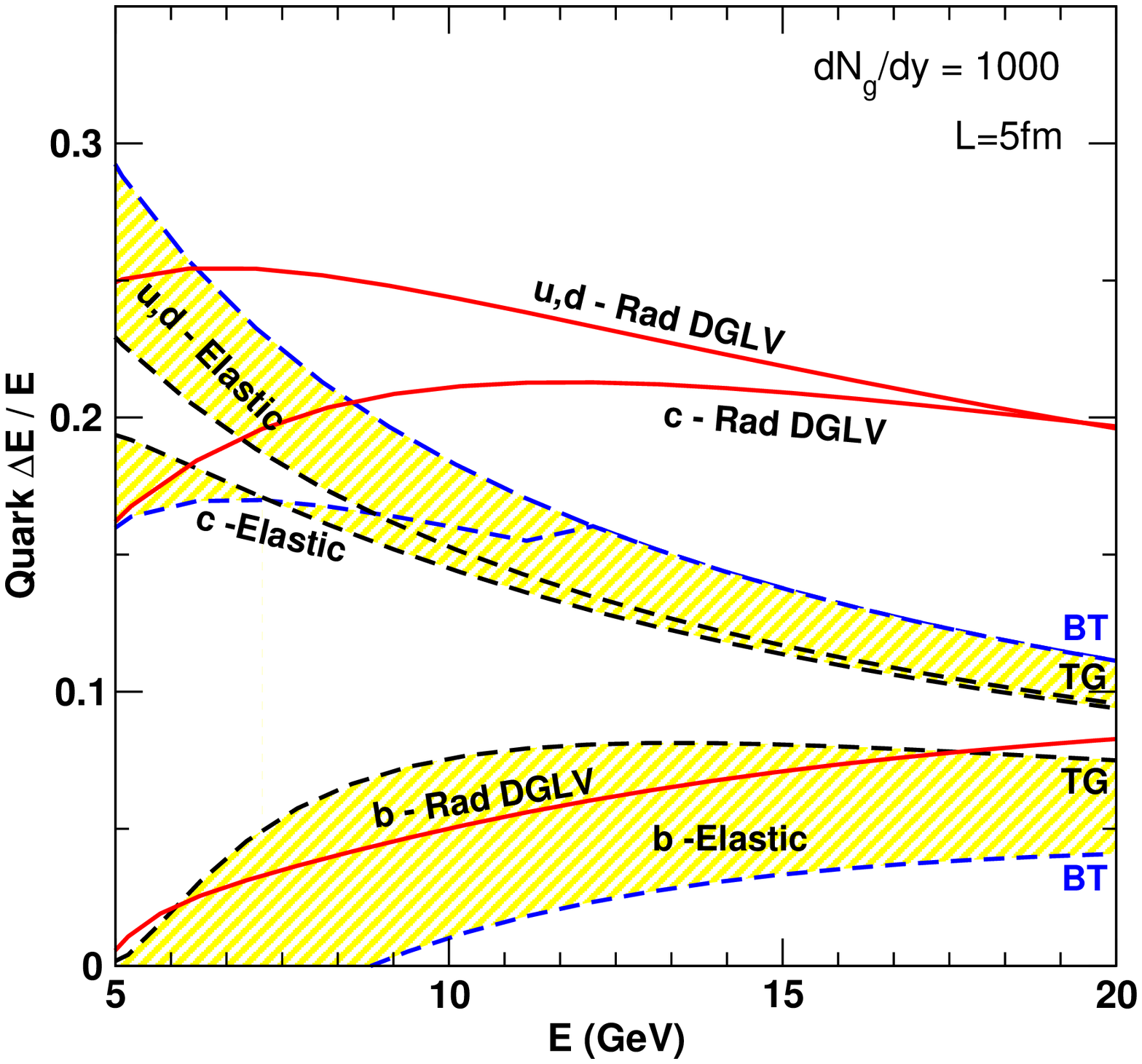}
\caption{Comparison of the average radiative and elastic energy losses of light-quarks (left) 
and light- and heavy-quarks (right) passing through the medium produced in central \AuAu\ collisions at RHIC energies
as obtained by the AMY~\cite{Qin:2007rn} and DGLV~\cite{Wicks:2005gt} models (see later).}
\label{fig:Coll_vs_Rad}
\end{figure}

For a gluon with $E$~=~20~GeV in a medium with $\qhat$~=~2~GeV$^2$/fm and $L = 6$~fm,
$d E_{rad}/dl$ is $\mathscr{O}(10$~GeV/fm$)$ (to be compared with the elastic losses of 
$\mathscr{O}$(2~GeV/fm) estimated before).
As seen in Fig.~\ref{fig:Coll_vs_Rad} for a more realistic phenomenological case, $\Delta E_{coll}$ 
is in general a small correction compared to $\Delta E_{rad}$ 
for light quarks and gluons but it can be an important contribution 
for slower heavy-quarks (see next). 




\paragraph{Heavy-quark radiative energy loss (``dead cone'' effect)}

Gluon bremsstrahlung off a heavy quark differs from that of a massless parton, already 
in the vacuum. Due to kinematics constraints, the radiation is suppressed at angles smaller 
than the ratio of the quark mass $M$ to its energy $E$. 
The double-differential distribution of gluons of transverse momentum $k_\perp$ and 
energy $\omega$ radiated by a heavy quark at small-angles ($k_\perp\approx \omega\,\theta$), 
differs from the standard bremsstrahlung spectrum by the factor
\begin{equation}
\omega \frac{dI_{rad,Q}}{d\omega\,dk_\perp^2} = {\alpha_s\ C_F \over \pi}\ {k_{\perp}^2 \,\over (k_{\perp}^2 + \omega^2 \theta_0^2)^2} \;\approx \;\omega \frac{dI_{rad}}{d\omega\,dk_\perp^2} \cdot \left( 1+\frac{\theta_0^2}{\theta^2}\right)^{-2}\,,\;\;  \theta_0\equiv\frac{M}{E}=\frac{1}{\gamma}\,.
\label{eq:dI_hQ}
\end{equation}
This effect, known as the ``dead cone''~\cite{deadcone}, results in a reduction
of the total gluon radiation emitted off heavy-quarks. In the medium, the total amount of reduction depends
on a non-trivial way on the various scales ($E, M, L$) of the problem~\cite{Peigne:2008wu}.
In a simplified way, the effect is $\mathscr{O}(m_D/M)$ -- compared to $\mathscr{O}(m_D^2/M^2)$ in the QED case --
i.e. for a plasma with Debye-mass $m_D$~=~1 GeV/c$^2$, the reduction of radiative energy loss
for a charm (bottom) quark of mass 1.3 (4.2)~GeV/c$^2$ is of order $\sim$25\% (75\%).


\subsection{Phenomenological consequences of parton energy loss}
\label{sec:observables}

Medium-induced parton energy loss in \AaAa\ reactions results in various observable 
consequences compared to the same ``free space'' measurements in proton-proton (\pp) collisions. 
The presence of jet quenching manifests itself via :
\begin{description}
\item (i) a suppression of the spectrum  ($dN_{AA}/dp_T$) of high-$p_T$ 
hadrons~\cite{gyulassy90},  
\item(ii) unbalanced back-to-back high-$p_T$ dihadron azimuthal correlations
($dN_{pair}/d\phi$)~\cite{appel86_blaizot_mclerran86}, 
\item(iii) modified energy-particle flow (softer hadron spectra, larger multiplicity, increased 
angular broadening, ...) within the final jets~\cite{Salgado:2003rv,Borghini:2005em,Armesto:2008qh,Armesto:2008qe,Ramos:2008qb,Ramos:2008as}.
\end{description}
In addition, due to the aforementioned
hierarchy of flavour-dependent radiative energy losses:
$\Delta E_{rad}(g) > \Delta E_{rad}(q) > \Delta E_{rad}(c) > \Delta E_{rad}(b)$,
all these medium effects are expected to be larger for gluons and $u$, $d$, $s$ quarks than for $c$ or $b$ 
quarks\footnote{The heaviest top-quark decays into $W\,b$ immediately ($\tau<0.1$~fm/c) after production.}. 


\paragraph{(i) High-$p_T$ (leading) hadron spectra:}

\noindent 
The {\it leading hadron} of a jet is the hadron that carries the largest fraction of the energy
of the fragmenting parton\footnote{High-$p_T$ inclusive hadron spectra are dominated
by particles with $\mean{z}\approx$ 0.5~--~0.7~\cite{Kretzer:2004ie}.}.
In a heavy-ion collision, if the parent parton suffers energy loss, 
the energy available for such hadrons is reduced and consequently their spectrum is depleted
compared to \pp. From the measured suppression factor one can determine 
$\Delta E_{loss}$ (see Eq.~(\ref{eq:eloss}) later) 
and estimate various properties of the produced plasma  such as:
\begin{itemize}
\item the average {\bf transport coefficient} $\mqhat$, from 
Eqs.~(\ref{eq:BH_QCD}),~(\ref{eq:LPM_QCD}) via $\langle\Delta E\rangle \propto \alpha_s\,\langle\hat{q}\rangle\,L^2$,
\item the initial {\bf gluon density} $dN^g/dy$ of the expanding plasma (with {\it original} transverse area 
$A_\perp=\pi\,R_A^2\approx$~150~fm$^2$ and thickness $L$), 
from~\cite{glv}: 
\begin{equation}
\Delta E \propto \alpha_s^3\,C_R\,\frac{1}{A_\perp}\frac{dN^g}{dy}\,L\mbox{ .}
\label{eq:glv}
\end{equation}
\end{itemize}
High-$p_T$ spectra in \pp\ and \AaAa\ collisions are discussed in detail in Chapter~\ref{sec:highpT_hadrons}.


\paragraph{(ii) High-$p_T$ di-hadron correlations:}

Parton-parton $2 \to 2$ scatterings are balanced in $p_T$ i.e. they are 
back-to-back in azimuthal angle ($\Delta \phi\approx \pi$). Such azimuthal correlation is
smeared out if one or both partons suffer rescatterings (``$k_T$ broadening'') in a dense plasma. 
The dijet-acoplanarity arising from the interactions of a parton in an expanding QGP 
is $ \mean{k_{T}^2}_{med} \simeq (m_D^2/\lambda) L\,\ln (L/\tau_0)\propto \qhat L$~\cite{Qiu:2003pm}
and, thus, the final azimuthal correlations between the hadrons issuing from quenched partons will 
show a dependence on the {\bf transport coefficient} and thickness of the medium: 
$d^2N_{pair}/d\Delta\phi = f(\qhat,L)$.\\

In addition, it has been proposed that a parton propagating through a QGP with supersonic ($\beta>c_s$) 
or ``superluminal'' ($\beta>1/n$) velocities 
can generate a wake of lower energy gluons with either Mach-~\cite{mach1,mach2,rupp05} or 
\v{C}erenkov-like~\cite{rupp05,cerenkov1,cerenkov2} conical angular patterns. Such a conical 
emission can propagate, after hadronisation, into the final correlations of the measured 
hadrons with respect to the jet axis:
\begin{itemize}
\item In the first case, the {\bf speed of sound} of the traversed matter\footnote{
The speed-of-sound -- namely the speed of a small disturbance through the medium -- 
for an {\it ideal} QGP (with $\varepsilon= 3 P$, where $P$ is the pressure) is simply $c_s=1/\sqrt{3}$.}, 
$c_s^2 = \partial P/\partial\varepsilon$, can be determined from the characteristic 
Mach angle $\theta_{M}$ of the secondary hadrons:
\begin{equation}
\cos(\theta_{M}) = \frac{c_s}{\beta} \;, 
\label{eq:mach}
\end{equation}
\item In the second scenario, the 
{\bf refractive index} of the medium, $n \approx \sqrt{\varepsilon_r}$ where $\varepsilon_r$ is the 
gluon dielectric constant, can be estimated 
from the \v{C}erenkov angle of emission $\theta_{c}$ of the hadrons: 
\begin{equation}
\cos(\theta_{c}) = \frac{1}{n\,\beta} = \frac{1}{\sqrt{\epsilon_r}\,\beta}. 
\label{eq:cerenkov}
\end{equation}
\end{itemize}
High-$p_T$ hadron correlations in nucleus-nucleus collisions are covered in Chapter~\ref{sec:highpTcorrs}.


\paragraph{(iii) Jet spectra, jet shapes, and fragmentation functions:}

\noindent 
The measurements of fully reconstructed (di)jets or of jets tagged by an away-side photon~\cite{Arleo:2004xj} 
or $Z$-boson~\cite{Lokhtin:2004zb} in heavy-ion collisions allow one to 
investigate -- in much more detail than using single- or double- hadron observables -- the mechanisms 
of in-medium parton radiation as well as to obtain the {\bf transport coefficient} $\qhat$ via: 
\begin{itemize}
\item Medium-modified jet profiles~\cite{Salgado:2003rv,Vitev:2008rz}, through the 
differential $\rho^{med}(r;\qhat)$ and integrated $\Psi^{med}(r;\qhat)$ jet-shapes 
($r$ is the distance to the jet axis) and the thrust $\mathcal{T}(\qhat)$ variables, 
\item Medium-modified fragmentation functions~\cite{Arleo:2008dn}, $\mathcal{D}^{med}_{parton\to hadron}(z,Q^2)$,
where $z$~=~$p_{hadron}/p_{parton}$ is the fraction of the jet energy carried by a hadron, 
which very schematically can be written as
\begin{equation}
\mathcal{D}_{i\to h}^{med}(z;\hat{q},L) \simeq P(z;\hat{q},L) \otimes \mathcal{D}_{i\to h}^{vac}(z) \;,
\label{eq:mFF}
\end{equation}
where the correction $P(z,\hat{q},L)$ can be connected to the QCD splitting functions, Eq.~(\ref{eq:splitt}),
modified according to various possible prescriptions (see Section~\ref{sec:MCs}) to take into account 
medium-induced gluon radiation. 
\end{itemize}
Jet physics studies in heavy-ion collisions are covered in Chapter~\ref{sec:jets}.

%
\section{Parton energy loss phenomenology}
\label{sec:phenomenology}

The use of fast partons as calibrated tomographic probes of hot and dense QCD matter 
in heavy-ion collisions relies on the possibility to compute theoretically 
(i) their perturbative production cross sections, 
and (ii) their modifications suffered while propagating through a strongly-interacting medium. 
We discuss here the basic pQCD principles used to compute high-$p_T$ hadron (and jet) cross sections, 
and then we outline the 
various existing parton energy loss schemes.


\subsection{QCD factorisation in high-$p_T$ hadron and jet production in \AaAa\ collisions}
\label{sec:factorisation}

\begin{figure}[htbp]
\centering
\vskip -0.3cm
\includegraphics[width=8.7cm,height=6.2cm]{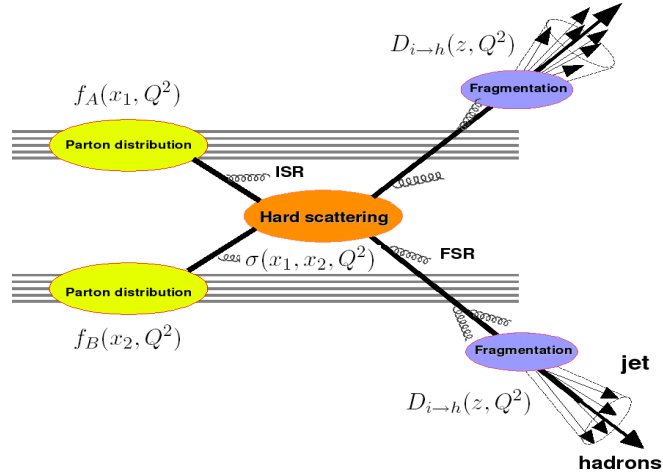}
\caption{Sketch of dijet production and pQCD collinear factorisation in hadronic collisions: $f_{a/A}(x)$
are the PDFs, $\mathcal{D}_{i\rightarrow h}(z)$ the FFs, and ISR (FSR) represents initial (final)-state 
radiation.}
\label{fig:pQCD_factorization}
\end{figure}

Because of asymptotic freedom, the QCD coupling $\alpha_s$ is small for high-energy 
(short distance) parton interactions: $\alpha_s(Q^2\to\infty)\rightarrow 0$. The single inclusive\footnote{{\it Inclusive} 
refers to the consideration of {\it all} possible channels that result in the production of a given particle 
$c$, without any particular selection of the final-state $X$.} production of a high-$p_T$ parton $c$ 
in a parton-parton collision, $ab\rightarrow c + X$, can thus be computed using perturbation
theory techniques. Over short distances, the infinite number of Feynman diagrams that would theoretically
result in the production of the outgoing parton $c$, can be approximated accurately by a much more 
manageable number of terms. In high-energy 
{\it hadron-hadron} collisions, the production of high-$p_T$ particles can be computed 
from the underlying {\it parton-parton} processes using the QCD ``factorisation theorem'' (see sketch in 
Fig.~\ref{fig:pQCD_factorization})~\cite{factor}. The production cross section of a high-$p_{T}$ hadron $h$ 
can be written 
as the product 
\begin{equation}
d\sigma^{hard}_{AB\rightarrow h}= f_{a/A}(x_1,Q^2)\otimes f_{b/B}(x_2,Q^2)\otimes
d\sigma_{ab\rightarrow c}^{hard}(x_1,x_2,Q^2) \otimes \mathcal{D}_{c\rightarrow h}(z,Q^2) \;, 
\label{eq:factorisation}
\end{equation}
where $\sigma_{ab\rightarrow cX}(x_1,x_2,Q^2)$ is the 
perturbative partonic cross section computable up to a given order in $\alpha_s$, 
and where the two non--perturbative terms:
\begin{itemize}
\item $f_{a/A}(x,Q^2)$: parton distribution functions (PDF), encoding the probability of finding a parton of 
flavour $a$ and momentum fraction $x=p_{parton}/p_{nucleus}$ inside the nucleus $A$,
\item $\mathcal{D}_{c\rightarrow h}(z,Q^2)$: fragmentation function (FF), describing the ``probability'' that the outgoing 
parton $c$ fragments into a final hadron $h$ with fractional momentum $z=p_{hadron}/p_{parton}$,
\end{itemize}
are universal (i.e. process-independent) objects that can be determined experimentally e.g. in deep-inelastic $e^\pm$-nucleus 
and $e^+e^-$ collisions, respectively. For the case of  total parton (i.e. jet) cross section, 
one simply sets $\mathcal{D}_{c\rightarrow h}=\delta(1-z)$ in Eq.~(\ref{eq:factorisation}).

The basic assumption underlying the factorised form of Eq.~(\ref{eq:factorisation}) is that the characteristic time 
of the parton-parton interaction is much shorter than any long-distance interaction occurring before (among partons 
belonging to the same PDF) or after (during the evolution of the struck partons into their hadronic final-state) 
the hard collision itself. The validity of Eq.~(\ref{eq:factorisation}) holds thus on the possibility to separate 
long- and short-distance effects with independent QCD time- (length-) scales, 
as well as on the ``leading-twist''\footnote{Processes in which more than one parton from the same 
hadron/nucleus interact coherently, are called ``higher-twist'' processes.} assumption of {\it incoherent} 
parton-parton scatterings. Since partons are effectively ``frozen'' during the hard scattering,
one can treat each nucleus as a collection of free partons. Thus, {\it with regard to high $p_T$ production}, 
the density of partons in a nucleus with mass number $A$ is expected
to be simply equivalent to that of a superposition of $A$ independent nucleons\footnote{In reality, nuclear PDFs
are modified compared to proton PDFs by initial-state ``(anti)shadowing'' effects (see~\cite{Armesto:2006ph} for a recent review).}: 
$f_{a/A}(x,Q^2) \approx A\cdot f_{a/N}(x,Q^2)$. Thus,
\begin{equation}
d\sigma^{hard}_{AB\rightarrow h} \approx A\cdot B\;\cdot f_{a/p}(x,Q^2)\otimes \;f_{b/p}(x,Q^2)\otimes
d\sigma_{ab\rightarrow c}^{hard}\otimes \mathcal{D}_{c\rightarrow h}(z,Q^2)\,.
\label{eq:factorisation2}
\end{equation}
From (\ref{eq:factorisation}), it is clear that QCD factorisation
implies that hard inclusive cross sections in a $A\;B$ reaction 
scale simply as $A\cdot B$ times the corresponding \pp\ cross sections:
\begin{equation}
d\sigma^{hard}_{AB} = A\cdot B\cdot d\sigma_{pp}^{hard}.
\label{eq:factorisation3}
\end{equation}
Since nucleus-nucleus experiments usually measure invariant {\it yields} 
for a given centrality bin (or impact parameter $b$), one writes instead: 
\begin{equation}
dN^{hard}_{AB}(b) = \langle T_{AB}(b)\rangle \cdot d\sigma_{pp}^{hard},
\label{eq:TAB_scaling}
\end{equation}
where the nuclear overlap function at $b$, $T_{AB}(b)$, is determined within a geometric 
Glauber eikonal\footnote{The `eikonal' approximation assumes that the particle trajectories are simple straight lines.} 
model from the measured Woods-Saxon distribution for the interacting 
nuclei~\cite{dde_glauber}. 
Intuitively, one can think of the nuclear-overlap $T_{AA}(b)$ as
a function that characterises the surface profile of two ``beams'' of nucleons
colliding at a distance $b$. The units of [area]$^{-1}$ of $T_{AA}$ indicate 
that it represents somehow the effective ``parton (integrated) luminosity''
of the collision.
Since the number of inelastic nucleon-nucleon ($NN$) collisions 
at $b$, $N_{coll}(b)$, is proportional to $T_{AB}(b)$: 
$N_{coll}(b) = T_{AB}(b)\cdot \sigmaNN^{inel}$,
one also writes often Eq.~(\ref{eq:TAB_scaling}) as: 
\begin{equation}
dN^{hard}_{AB}(b) = \langle N_{coll}(b)\rangle \cdot dN_{pp}^{hard}.
\label{eq:TAB_scaling2}
\end{equation}
For minimum-bias\footnote{{\it Minimum-bias} collisions are those where there is no specific 
selection of the final-state (e.g. in particular  for heavy-ions, no centrality selection).} $A\,B$ collisions, the 
average nuclear-overlap and number of $NN$ collisions take a simple form\footnote{E.g. 
for \AuAu\  at $\sqrtsnn$~=~200~GeV 
($\sigmaNN^{inel} =$~41~mb, $\sigma_{AuAu}^{geo} =$~7000~mb):
$\langle T_{AuAu}\rangle$~=~5.5~(23.3) mb$^{-1}$ and $\langle N_{coll}\rangle$~=~230~(955) for
minimum-bias (10\% most central) collisions, resp.}: 
$\mean{T_{AB}}=A\,B\,/\sigma_{AB}^{geo}$ and $\mean{N_{coll}}=A\,B\cdot\sigmaNN/\sigma_{AB}^{geo}$.
The standard method to quantify the effects of the medium on the yield of a hard probe 
in a \AaAa\ reaction is thus given by the {\it nuclear modification factor}:
\begin{equation}
R_{AA}(p_{T},y;b)\,=\,\frac{d^2N_{AA}/dy dp_{T}}{\langle T_{AA}(b)\rangle\,\times\, d^2 \sigma_{pp}/dy dp_{T}},
\label{eq:R_AA}
\end{equation}
This factor -- a quantitative version of the ratio sketched in Eq.~(\ref{eq:Ratio_AA}) --
measures the deviation of \AaAa\ at 
$b$ from an incoherent superposition of $NN$ collisions ($R_{AA}$~=~1).
This normalisation is often known as ``binary collision scaling''.


\subsection{Jet quenching models}
\label{sec:Eloss_models}

\begin{figure}[htbp]
\centering
\includegraphics[width=7.5cm,height=1.5cm]{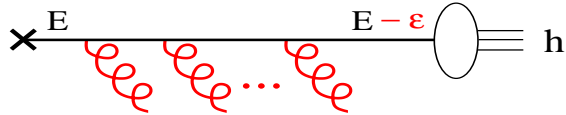}
\caption{Schematic representation of parton energy loss in a QGP, implemented via rescaling 
of the energy of the traversing parton at the point where it fragments into hadrons~\cite{Arleo:2008dn}.}
\label{fig:FF_erescaling}
\end{figure}

The energy-loss formulas presented in Section~\ref{sec:intro_theory} refer to an idealistic situation 
with an infinite-energy parton traversing a {\it static} and {\it uniform} QGP with an {\it ideal-gas} equation-of-state (EoS). 
Experimentally, the situation that one encounters with {\it realistic} plasmas in heavy-ion collisions is more complex:
\begin{itemize} 
\item first, there is no direct measurement of the traversing parton but (in the best case) only of the final-state 
{\it hadrons} issuing from its fragmentation,
\item the traversing partons can be produced at any initial point within the fireball and their energy spectrum is 
steeply (power-law) falling,
\item the temperature and density of the plasma, and correspondingly its Debye-mass and transport coefficient, 
are {\it position-dependent}: $m_D({\bf r}),\qhat({\bf r})$,
\item the produced plasma is expanding with large longitudinal (transversal) velocities, $\beta~\approx$~1~(0.7),
and so the medium properties are also {\it time-dependent}: $m_D({\tau}),\qhat({\tau})$,
\item the {\it finite-size} of the medium and associated energy loss {\it fluctuations}, have to be taken into account.
\end{itemize}
All those effects can result in potentially significant deviations from the analytical formulas of 
Section~\ref{sec:intro_theory} (e.g. in an expanding plasma the dependence of $\Delta E_{rad}$ 
on the medium thickness $L$, Eq.~(\ref{eq:LPM_QCD}), becomes effectively {\it linear}, Eq.~(\ref{eq:glv}), 
rather than quadratic). Four major phenomenological approaches (identified often with the initials of their authors) 
have been developed~\cite{Majumder:2007iu} to connect the QCD energy loss calculations with the experimental 
observables mentioned in Section~\ref{sec:observables}: 
\begin{itemize}
\item Path-integral approach to the opacity expansion (BDMPS-LCPI/ASW)~\cite{Zakharov:1996fv,Zakharov:1997uu,Zakharov:1998sv,Baier:1996kr,bdmps,Wiedemann:2000ez,Wiedemann:2000za,Salgado:2002cd,Salgado:2003gb}
\item Reaction Operator approach to the opacity expansion (DGLV)~\cite{Gyulassy:1999zd,glv,Gyulassy:2001nm,Djordjevic:2003zk,Wicks:2005gt}
\item Higher Twist (HT)~\cite{Guo:2000nz,Wang:2001ifa,Zhang:2003yn,Majumder:2004pt,Majumder:2007hx,Majumder:2007ne}
\item Finite temperature field theory approach (AMY)~\cite{Arnold:2001ba,Arnold:2000dr,Jeon:2003gi,Turbide:2005fk}
\end{itemize}
The models differ in their 
assumptions about the relationships between the relevant scales 
(parton energy $E$ and virtuality $Q^2$, and typical momentum $\mu\approx m_{D}$ 
and spatial extent $L$ of the medium), as well as by how they treat or approximate the space-time profile 
of the medium. In practical terms, all schemes are based on a pQCD factorised approach,
i.e. on Eq.~(\ref{eq:factorisation}), where the {\it entire} effect of energy loss is concentrated 
on the calculation of the medium-modified parton fragmentation functions into final hadrons: 
$\mathcal{D}^{vac}_{c\to h}(z) \to \mathcal{D}^{med}_{c\to h}(z',\qhat)$. 
The final hadronisation of the hard parton is always assumed to occur in the vacuum 
after the parton, with degraded energy ($z'<z$), has escaped from the system (Fig.~\ref{fig:FF_erescaling}). 

\paragraph{BDMPS-LCPI \& ASW}

\begin{figure}[htbp]
\centering
\includegraphics[angle=-90,width=6.0cm]{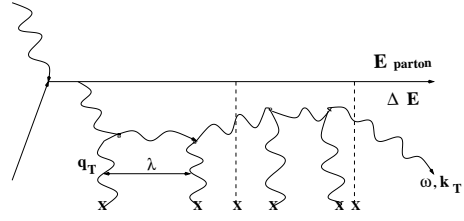}
\caption{Typical gluon radiation diagram in the BDMPS approach~\cite{bdmps}.}
\label{fig:bdmps_diag}
\end{figure}

The approaches of Baier, Dokshitzer, Mueller, Peign\'{e} and Schiff (BDMPS)~\cite{Baier:1996kr,bdmps,Baier:1998yf,Baier:1998kq} 
and the Light-Cone-Path-Integral (LCPI) by Zakharov~\cite{Zakharov:1996fv} 
compute energy loss in a coloured medium in a multiple soft-scatterings approximation. 
A hard parton traversing the medium interacts with various scattering centres and splits into an outgoing parton
as well as a radiated gluon (Fig.~\ref{fig:bdmps_diag}).
The propagation of the traversing parton and radiated gluons is expressed using Green's functions 
which are obtained by a path integral over the fields. The final outcome of the approach is a complex 
analytical expression for the radiated gluon energy distribution $\omega\,dI/d\omega$ as a function of 
the transport coefficient $\qhat$, Eq.~(\ref{eq:qhat}), 
defined perturbatively as~\cite{Baier:2002tc}: 
\begin{eqnarray}
\qhat\equiv \rho \int d^2k_\perp\;k_\perp^2\;\frac{d\sigma}{d^2k_\perp}\;,
\label{eq:qhat_pqcd}
\end{eqnarray}
where $\rho$ is the medium density of scattering centres (mainly gluons), 
$k_\perp$ is the transverse momentum of the radiated gluon, and $d\sigma$ is the 
differential parton-medium cross section. 
The medium-modified parton-to-hadron fragmentation functions are 
modelled as 
\begin{equation}
\mathcal{D}_{i\to h}^{med}(z',Q^2)=P_E(\epsilon;\qhat)\otimes \mathcal{D}_{i\to h}^{vac}(z,Q^2),
\label{eq:mff}
\end{equation}
where the {\it quenching weights} $P_E(\epsilon;\qhat)$ -- computed by Armesto, Salgado and 
Wiedemann (ASW)~\cite{Salgado:2003gb,Baier:2001yt,armesto05} --
encode the probability (assumed Poissonian) that the propagating parton loses a fraction of energy $\epsilon$~=~$\Delta E/E$
due to $n$ gluon emissions 
\begin{equation}
  P_E(\epsilon;\qhat) = \sum_{n=0}^\infty \frac{1}{n!}
  \left[ \prod_{i=1}^n \int d\omega_i \frac{dI^{med}(\qhat)}{d\omega}
    \right]
    \delta\left(\epsilon\; -\sum_{i=1}^n {\omega_i \over E} \right)
    \exp\left[ - \int d\omega \frac{dI^{med}}{d\omega}\right].
  \label{eq:qw}
\end{equation}
The quenching weights have been implemented in a Monte Carlo model, the 
Parton-Quenching-Model (PQM)~\cite{pqm,Loizides:2006cs} accounting for 
a realistic description of the parton production points in a {\it static} QGP.
The transport coefficient $\hat{q}$ is used as the fit parameter for the data.
The longitudinal expansion of the plasma 
can be taken into account by rescaling the transport coefficient according 
to the following law~\cite{Salgado:2002cd}:
\begin{equation}
  \mean{\hat{q}} = \frac{2}{L^2}\int_{\tau_0}^{\tau_0+L}
  d\tau\, \left(\tau - \tau_0\right) \,
  \hat{q}(\tau)
\label{eq:qhat_time}
\end{equation}
where $\qhat(\tau) = \qhat(\tau_0) \left(\tau_0/\tau\right)^\alpha$ and $\alpha$ 
characterises the time-dependence of the 
plasma density: 
$\rho(\tau)\propto \tau^{-\alpha}$.
A purely longitudinal (or Bjorken) expansion corresponds to $\alpha = 1$, and is 
often assumed in phenomenological applications.
When $\tau_0 \ll L$, Eq.~(\ref{eq:qhat_time}) reduces to 
$\langle\qhat\rangle\simeq \,\qhat(\tau_0)\, 2\tau_0/L$~\cite{Baier:2002tc}.


\paragraph{DGLV} 

\begin{figure}[htbp]
\centering
\includegraphics[height=4.6cm,width=6.cm]{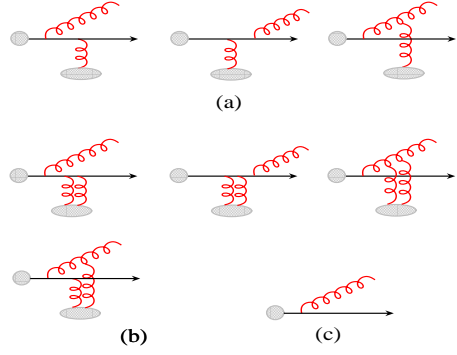}
\caption{Diagrams contributing to the lowest order in the opacity energy loss expansion~\cite{Vitev:2005ch}.}
\label{fig:ScatteringCenters}
\end{figure}

The Gyulassy--L\'evai--Vitev (GLV)~\cite{Gyulassy:1993hr,Gyulassy:1999zd,glv,Gyulassy:2001nm} 
(aka DGLV~\cite{Djordjevic:2003zk,Wicks:2005gt}) approach calculates the parton energy loss 
in a dense deconfined medium consisting, as in the BDMPS approach, of almost static (i.e. heavy) 
scattering centres (Fig.~\ref{fig:ScatteringCenters}) producing a screened Coulomb (Yukawa) potential. 
Although both approaches are equivalent~\cite{Wiedemann:2000za,Salgado:2003gb,Armesto:2003jh},
at variance with the BDMPS {\it multiple-soft} bremsstrahlung, GLV starts from the 
{\it single-hard} radiation spectrum which is then expanded to account for gluon emission 
from multiple scatterings via a recursive diagrammatic procedure~\cite{glv}. 
The traversing parton gains a transverse momentum ${\bf q}_\perp$ and radiates 
(before or after the scattering) a gluon with a certain momentum 
${\bf k}=(\omega,\frac{k^2_\perp}{\omega}{\bf k}_\perp)$. 
The gluon differential distribution at first-order in opacity
~\cite{Gyulassy:1999zd} is 
\begin{eqnarray}
\omega\frac{dI^{(1)}}{d\omega \,dk_\perp^2} &=& \omega \frac{dI^{(0)}}{d\omega \,dk_\perp^2} \frac{L}{\lambda_g} 
\int_0^{q_{max}^2} d^2 {q}_\perp \frac{m_{D}^2}{\pi ({q}_\perp^2 + m_D^2 )^2}
\frac{2 k_\perp \cdot {q}_\perp ( k- q_\perp)^2 L^2  }{ 16\omega^2 + (k - q)_\perp^4L^2},
\label{eq:GLV2}
\end{eqnarray}
where $\lambda_g$ is the mean free path of the radiated gluon. 
Applying the aforementioned recursive procedure, one obtains the gluon distribution to finite order 
($N\geq 1$) in opacity. Each emission at a given opacity is assumed independent and a 
probabilistic scheme is set up wherein, identically to Eq.~(\ref{eq:qw}), the parton loses an energy fraction 
$\epsilon$ in $n$ tries with a Poisson distribution~\cite{Gyulassy:2001nm}, 
\begin{equation}
P_n(\epsilon,E) = \frac{e^{-\mean{N^g} }}{n!} \Pi_{i=1}^n \llb \int d \omega_i \frac{dI}{d\omega_i} \lrb \delta(\epsilon\,E - \sum_{i=1}^{n} \omega_i  ) , 
\label{eq:GLV3}
\end{equation}
where, $\mean{N^g}$ is the mean number of gluons radiated per coherent interaction set.
Summing over $n$ gives the probability  $P(\epsilon)$ for  an incident parton to lose 
a momentum fraction $\epsilon$ due to its passage through the medium. This is then used to 
model a medium-modified FF, by shifting the energy fraction available to produce a hadron 
in a similar way as Eq.~(\ref{eq:mff}). 
The key medium property to be obtained from the fits to the experimental data,
is the initial gluon density $dN^g/dy$, after accounting for longitudinal expansion of the plasma.
Note that the density of colour charges of a cylinder of plasma with ``length'' $\tau$ and
surface $A_\perp$, is $\rho\,\approx \,dN^g/dy/(\tau \;A_\perp)$.


\paragraph{Higher Twist (HT)} 

\begin{figure}[htbp]
\centering
\includegraphics[height=3.0cm,width=\linewidth]{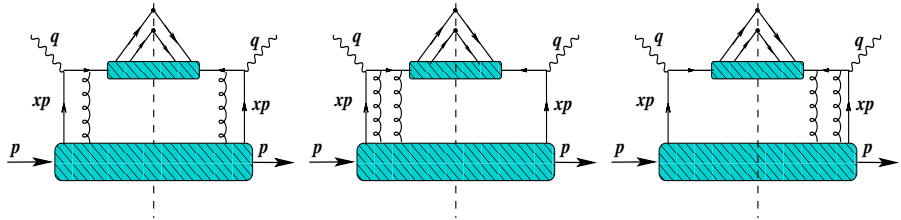}
\caption{Higher-twist contributions to quark scattering in a medium (hatched area)~\cite{Majumder:2008jy}.}
\label{fig:HT_Majumder}
\end{figure}

The higher-twist approximation~\cite{Qiu:1990xxa,Qiu:1990xy,Luo:1994np,Guo:2000nz,
Wang:2001ifa,Zhang:2003yn} describes the multiple scattering of a parton as
power corrections to the leading-twist cross section (Fig.~\ref{fig:HT_Majumder}). 
These corrections are enhanced by the medium-length $L$ and suppressed by the 
power of the hard scale $Q^2$. Originally, this approach was applied to calculate the medium 
corrections to the total cross section in nuclear deep-inelastic $e\,A$ scattering.

The scheme allows one to compute multiple Feynman diagrams such as those in Fig.~\ref{fig:HT_Majumder}
which are then combined coherently to calculate the modification of the fragmentation function directly
as a medium-dependent {\it additive} contribution, $\mathcal{D}_{i\to h}^{med} = \mathcal{D}_{i\to h}^{vac} + \Delta \mathcal{D}_{i\to h}^{med}$, with
\begin{equation}
\Delta \mathcal{D}_{i\to h}^{med}(z,Q^2) = \int_0^{Q^2} \frac{dk_{\perp}^2}{k_{\perp}^2} \frac{\alpha_s}{2\pi} \left[ \int_{z_h}^1 \frac{dx}{x} 
\sum_{j=q,g}\left\{ \Delta P_{i \to j}^{med} 
\; \mathcal{D}_{j\to h} \left(\frac{z_h}{x} \right)  \right\} \right].
 \label{eq:med_mod}
\end{equation}
Here, $\Delta P_{i\to j} \propto P_{i \to j} \; C_A \alpha_s  T^A_{qg}$
represents the medium-modified splitting function of parton $i$ into $j$
(a momentum fraction $x$ is left in parton $j$ and the radiated gluon or quark carries away 
a transverse momentum $k_\perp$). The entire medium effects are 
incorporated in the nuclear quark-gluon correlation $T^A_{qg}$ term. 
The normalisation $C$ of this correlator is set by fitting to one data point 
from which one can directly calculate the medium-modified FFs and then the final hadron spectrum.
The parameter $C$ can also be used to calculate the average energy loss suffered by the parton.


\paragraph{AMY}

\begin{figure}
\centering \includegraphics[height=2.35cm,width=0.8\linewidth]{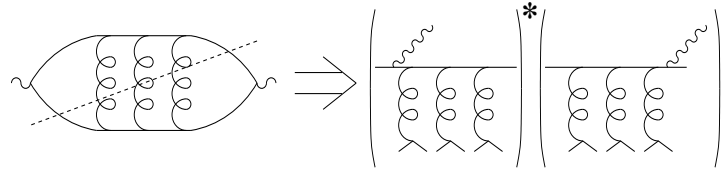}
\caption{A typical ladder diagram in a thermal medium in the AMY formalism~\cite{Arnold:2002ja}.}
\label{fig:amy}
\end{figure}

The Arnold--Moore--Yaffe (AMY)~\cite{Arnold:2001ba,Arnold:2001ms,Arnold:2002ja,Jeon:2003gi,
Turbide:2005fk} approach describes parton energy loss in a hot equilibrated QGP, where a
hierarchy $T\gg gT\gg g^2T$ is assumed. 
The hard parton scatters off other partons in the medium, leading to momentum transfers of 
$\mathscr{O}(gT)$ and inducing collinear radiation. 
Multiple scatterings of the incoming (outgoing) parton and the radiated gluon 
are combined to get the leading-order gluon radiation rate. 
One essentially calculates the imaginary parts of ladder diagrams such as those shown in Fig.~\ref{fig:amy}, 
by means of integral equations which yield the $1\to 2$ transition rates $\Gamma_{bg}^a$ of a hard parton $a$ 
into a radiated gluon $g$ and another parton $b$. 
These rates, with $T$-dependent Bose-Einstein (for gluons)
and Fermi-Dirac (for quarks) exponential factors for the medium partons, are then used to evolve the original distributions 
over the medium length by means of a Fokker-Planck like equation~\cite{Jeon:2003gi}
\begin{equation}
\frac{d P_a(p)}{d t} = \int dk \sum_{b,c} \left[ P_b(p+k) \frac{d\Gamma^b_{ac}(p+k,p)}{dk dt} 
- P_a(p) \frac{d \Gamma^a_{bc}(p,k)}{dk dt}  \right] .
\label{eq:AMY1}
\end{equation}
The medium modified FF are obtained from the convolution of the vacuum FF with the hard parton 
distributions when exiting the plasma~\cite{Turbide:2005fk}:
\begin{equation}
\mathcal{D}_{a\to h}^{med}(z) = \int dp_f \frac{z^{\prime}}{z} \sum_a P_{a} (p_f;p_i) \,\mathcal{D}^{vac}_{a\to h}(z^{\prime})\,,
\label{eq:AMY2}
\end{equation}
where $z=p_h/p_i$ and $z^{\prime}=p_h/p_f$, with $p_i$ and $p_f$ the momenta of the hard partons 
immediately after the hard scattering and prior to exit from the medium respectively. 
The model of the medium is essentially contained in the space-time profile chosen for the 
initial temperature $T$ appearing in the transition rates. 

\paragraph{Model comparison}

The four energy-loss formalisms discussed above can be roughly divided into two groups. 
Those calculating the radiated gluon spectrum i.e. the energy lost by the initial parton (BDMPS/ASW and GLV) 
and those determining directly the change in the final distribution of the traversing partons (Higher Twist and AMY). 
Each approach has its advantages and disadvantages:
\begin{itemize}
\item {\it ASW} and {\it GLV}: Both are applicable to thin and thick media (confined or deconfined), but 
they do not account for the energy flow {\it into} the medium.
\item {\it Higher Twist}: It can directly compute the medium-modified fragmentation functions, and allows for
the study of multi-hadron correlations, but the formalism is more appropriate for thin than thick media.
\item {\it AMY}: It is the only framework that accounts for processes where a thermal gluon or quark is
absorbed by a hard parton, and elastic losses can be included in a simple way, 
but it does not take into account vacuum radiation (nor vacuum-medium interference) 
and its application to non-thermalised media is questionable.
\end{itemize}

\begin{figure}[htbp]
\centering
\includegraphics[height=5.9cm,width=0.8\linewidth]{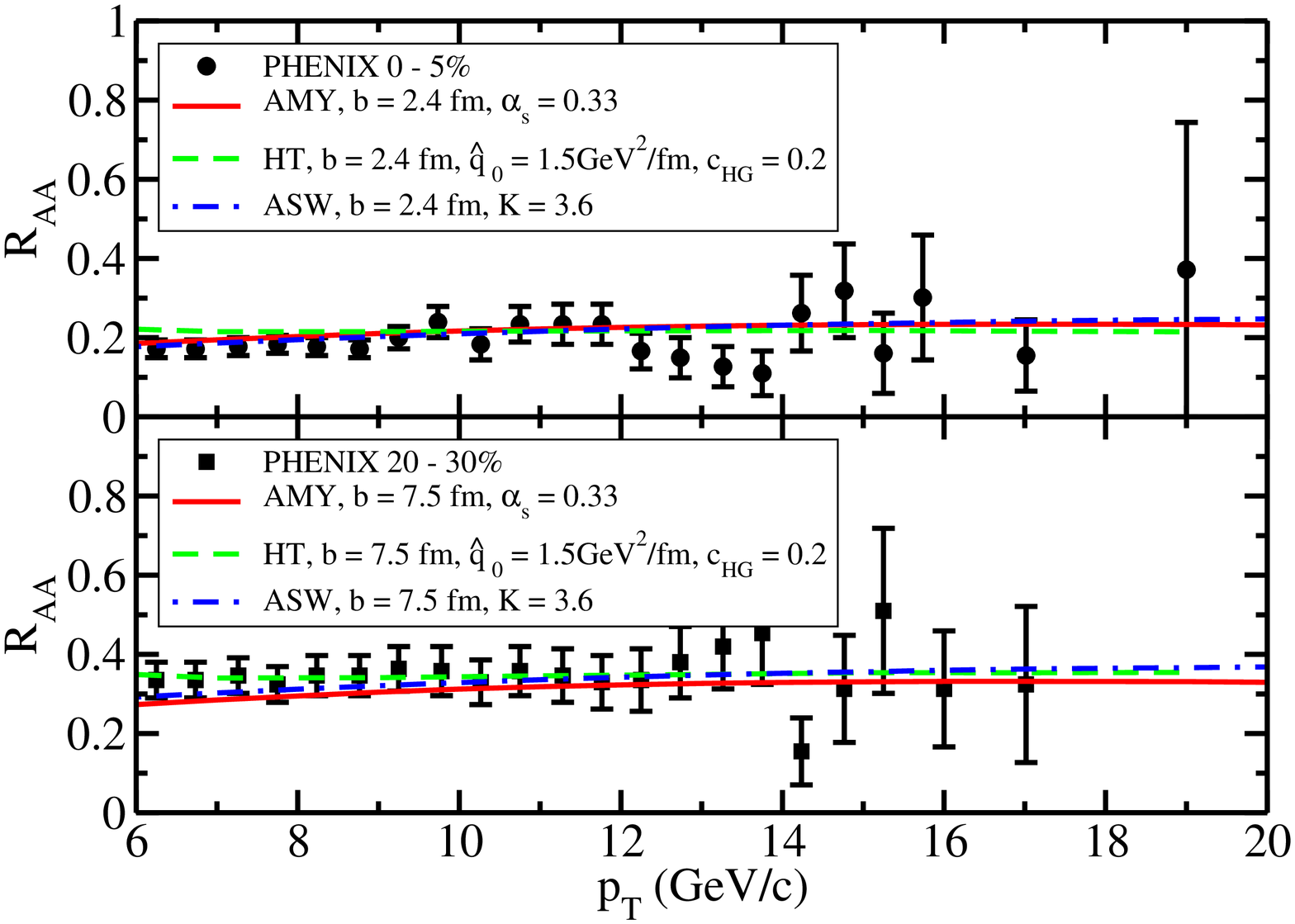}
\caption{Suppression factor for high-$p_T$ pions in central (top) and semicentral (
bottom) \AuAu\ collisions at RHIC~\cite{Adare:2008qa} compared to AMY, HT and ASW
energy loss calculations~\cite{Bass:2008ch}.}
\label{fig:RAA_Bass}
\end{figure}

All four schemes have independently made successful comparisons to the available data 
(see Fig.~\ref{fig:RAA_Bass} and coming Sections). 
The outcome of the models is one parameter tuned to ideally fit all experimental observables: 
$\mean{\qhat}$ in the BDMPS/ASW scheme, the initial $dN^g/dy$ density in GLV, the $C$ correlator 
normalization (or energy loss $\epsilon_0$) in HT, and the temperature $T$ in AMY. All jet-quenching 
observables in \AuAu\ collisions at $\sqrtsnn$~=~200~GeV can only be reproduced with medium 
parameters consistent with a QGP at temperatures above the QCD phase transition ($\Tcrit\approx$~0.2~GeV).
The analytical results of the different schemes under ``controlled'' situations are in principle equivalent, 
see e.g.~\cite{Arnold:2008iy}. Yet, the detailed comparison of the phenomenological results of the 
models is not always straightforward as they 
\begin{itemize}
\item use different approximations in their calculations (e.g. running-coupling, or fixed to $\alpha_s$~=~0.3, 0.5),
\item do not always include the same list of physics processes (e.g. $\Delta E_{coll}$ is neglected in some cases),
\item choose different fitting parameters to characterise the medium (see above), and
\item the space-time profile of the quenching medium is not always equivalent (e.g. static-plasma with 
average path-length vs. 1D Bjorken expansion or vs. full 3D hydrodynamical evolutions with varying 
thermalisation times $\tau_0$).
\end{itemize}

The quantitative consistency of the different schemes has been investigated within a 
3-dimensional hydrodynamics approach (Fig.~\ref{fig:RAA_Bass})~\cite{Bass:2008ch}
linking the various medium properties via thermodynamical relations and using the same 
space-time evolution. Yet, the extracted $\qhat$ values still differ by factors of 2~--~3 (see Section~\ref{sec:AAhighpT_hadrons}a).
At least part of the uncertainty is due to the relative insensitivity of the $\qhat$ parameter
to the irreducible presence of hadron from (unquenched) partons emitted from the surface of the plasma~\cite{Eskola:2004cr}.
Additional constraints on $\qhat$ can be placed by requiring also the model reproduction of the
suppressed {\it dihadron} azimuthal correlations (see Section~\ref{sec:IAA}).

\subsection{Jet quenching Monte Carlo's}
\label{sec:MCs}

Ultimately, the energy loss schemes discussed in the previous Section are all based on a final 
energy-rescaling of the vacuum parton-to-hadron fragmentation functions (Fig.~\ref{fig:FF_erescaling}). 
Recently, attempts to reformulate parton energy loss as a 
medium-modification of the perturbative {\it evolution} of the fragmentation functions have been 
implemented in Monte Carlo (MC) models~\cite{Armesto:2007dt,Zapp:2008gi,Domdey:2008gp,Renk:2008pp}. 
Such MC approaches allow one to address more detailed experimental observables such as the particle and energy flows within a jet.
The DGLAP scale-dependence ($Q^2$-evolution) equation of the FFs reads
\begin{equation}
\frac{\partial \mathcal{D}_{i\to h}(x,Q^2)}{\partial \log Q^2} = \sum_j \int_x^1 \frac{dz}{z}\frac{\alpha_s}{2\pi}\, \; P_{i \to j}(z)\;\mathcal{D}_{j\to h}(x/z,Q^2)\;,
\label{eq:FF_dglap}
\end{equation}
with splitting functions $P_{i \to j}(z)$, Eq.~(\ref{eq:splitt}).
The probabilistic nature of parton showering -- with a {\it Sudakov factor} $\exp[- \int dQ/Q^2\; \int \alpha_s/2\pi \; P_{i \to j}(z,Q^2)\,dz\,]$
giving the probability that a parton evolves from virtualities $Q_1$ to $Q_2$ without branching --
can be easily implemented in MC codes 
which compute the virtuality and energy fraction of a parton at each branching point with proper energy-momentum 
conservation. Parton showers are a basic ingredient of event generators such as \pythia~\cite{Sjostrand:2006za}
or \herwig~\cite{Corcella:2002jc} often used to compare the experimental jet data to the 
details of the underlying QCD radiation pattern. 
Medium effects can be easily included by e.g. modifying the splitting functions in Eq.~(\ref{eq:FF_dglap}).
\hydjet~\cite{Lokhtin:2005px,Lokhtin:2008wg} was the first MC code which incorporated medium effects  
via a {\sc pyquen} routine which modifies the standard \pythia\ branching algorithm to include radiative 
and elastic energy losses. More recent developments like \qpythia\ and \qherwig~\cite{Armesto:2008qh}
change the DGLAP evolution of these two parton-shower MCs. 
The \jewel\ MC~\cite{Zapp:2008gi} implements elastic scattering in DGLAP evolution 
plus radiative energy loss through a multiplicative constant in the infrared part of
the splitting functions~\cite{Borghini:2005em}, whereas {\sc YaJEM}~\cite{Renk:2008pp} increases
the virtuality of the traversing partons (i.e. their probabilities to radiate) in \pythia\ according to the medium $\qhat$.



\subsection{Parton energy loss in AdS/CFT}

So far, we have discussed perturbative calculations of parton energy loss in an ideal QGP. Yet, the medium produced at RHIC 
has temperatures $\mathscr{O}(2\,\Tcrit)$ in a domain where lattice QCD~\cite{latt}
still predicts large deviations with respect to the asymptotic ideal-gas behaviour. 
Many experimental signals at RHIC are consistent with the formation of a {\it strongly-coupled} 
plasma (sQGP)~\cite{Shuryak:2003xe,tdlee04,gyulassy_mclerran04}. Such a 
regime is theoretically treatable via the Anti-de-Sitter/Conformal-Field-Theory (AdS/CFT) 
correspondence between weakly-coupled gravity and strongly-coupled gauge theories~\cite{adscft}. 

The AdS/CFT correspondence conjectures that string theories described in an Anti--de--Sitter 
space\footnote{$AdS_5$ is a 5-dimensional space with constant and negative curvature.}
times a 5-dimensional sphere ($AdS_5 \times S_5$) are equivalent to a conformal field theory (CFT), 
defined on the 4-dimensional boundary of this space. 
A particularly useful case is ${\cal N}=4$ Super--Symmetric Yang Mills (SYM)\footnote{SYM 
is a quantum-field $SU(N_c)$ theory like QCD (${\cal N}=4$ indicates 4 additional
super-charges) but  dissimilar from QCD in many aspects: extra SUSY degrees of freedom, no running 
coupling, no confinement, ... Yet, such differences ``wash out'' 
at finite-$T$~\cite{Kajantie:2006hv}.} at strong coupling $g_{YM}$ and large number of colours 
$N_c$ (i.e. at large 't~Hooft coupling $\lambda = g_{YM}^2\,N_c\gg 1$) which is dual to supergravity in a curved space-time.
The string coupling $g_s$, the curvature radius $R$ of the $AdS$ metric, and the string tension $(2\pi\alpha')^{-1}$ are 
related to the SYM quantities via: $R^2/\alpha' = \sqrt{\lambda}$, and $4\pi\, g_s =  g_{YM}^2 = \lambda/N_c$. 
Essentially, taking the large $N_c$ limit at fixed $\lambda$ (i.e. weakly coupled gravity: $g_s\rightarrow 0$) and the 
large $\lambda$ limit (i.e. weakly curved space and large string tension), the SYM theory can be described by 
classical gravity in a 5-D space. By virtue of such a duality, one can carry out analytical calculations of 
gravity, which can then be mapped out ``holographically'' to the {\it non-perturbative} dynamics of the gauge (QCD-like) theory. 

One can further exploit the AdS/CFT correspondence for theories at {\it finite-temperature}, by replacing the $AdS_5$ space
by an $AdS$ Schwarzchild black-hole. The temperature of the gauge theory is then equal to the black-hole Hawking 
temperature: $T=r_0/(\pi\,R^2)$, where $r_0$ is the coordinate of the black-hole horizon.
Recent applications of this formalism in the context of heavy-ions physics have led to the determination of 
transport properties of strongly-coupled (SYM) plasmas -- such as its viscosity~\cite{kovtun04}, the $\qhat$ parameter~\cite{wiedem06}, 
and the heavy-quark diffusion coefficients~\cite{heavyQ_adscft,CasalderreySolana:2006rq,CasalderreySolana:2007qw,Gubser:2006bz} 
-- from simpler black hole thermodynamics calculations. 

In the case of jet quenching calculations~\cite{Liu:2006he,CasalderreySolana:2007zz}, 
one can express the propagation of a parton through a medium in terms of Wilson lines.
The $\hat{q}$ parameter is identified with the coefficient in the exponential of an adjoint Wilson loop 
averaged over the medium length: $\mean{W^A(C)}\propto \exp\left[\hat{q}L\right]$~\cite{wiedem06}.
One then evaluates the gravity dual of this Wilson loop
given by the classical action of a string stretching in an $AdS_5 \times S_5$ space with
a Schwarzschild black hole background. 
After solving the equations of motion of the string, the  $\hat{q}$ parameter is found to be 
\begin{eqnarray}
\qhat_{sym} & = & \frac{\pi^{3/2} \Gamma(\frac{3}{4})}{\Gamma(\frac{5}{4})}\;\sqrt{g^2 N_c} T^3 .
\label{eq:qhat_ads}
\end{eqnarray}
Though this result is computed in the infinite coupling and number of colours limits, 
typical values of $\alpha_s=0.5$ and $N_c=3$ 
lead to $\hat{q}$ = 4.5 -- 20.7~GeV$^2$/fm for $T$~=~0.3~--~0.5~GeV~\cite{CasalderreySolana:2007zz},
consistent with phenomenological fits of the RHIC data~\cite{Eskola:2004cr}.
There have been also AdS/CFT-based calculations~\cite{heavyQ_adscft,CasalderreySolana:2006rq,CasalderreySolana:2007qw,Gubser:2006bz} 
of the transport properties of a {\it heavy} quark, described by a semiclassical string in the gravity theory,
such as its diffusion constant in a ${\cal N}=4$~SYM plasma~\cite{CasalderreySolana:2006rq}
\begin{eqnarray}
D \approx \frac{0.9}{2\,\pi T}\left(\frac{1.5}{\alpha_s N_c}\right)^{1/2},
\end{eqnarray}
which agrees with the drag coefficient, see Eq.~(\ref{eq:diff_eq}), computed independently. 



\section{High-$p_T$ leading hadron suppression: data vs. theory}

The simplest empirically testable (and theoretically computable) consequence of jet quenching
in heavy-ion collisions 
is the suppression of the single inclusive high-$p_T$ hadron spectrum relative to that in 
proton-proton collisions. Since most of the energy of the fragmenting parton goes into a single 
{\it leading} hadron, 
QCD energy loss was predicted to result in a significantly suppressed production of high-$p_T$ hadrons 
($R_{AA}\ll 1$)~\cite{gyulassy90}. We compare in this Section the existing measurements 
of large-$p_T$ hadroproduction in \pp\ and \AaAa\ collisions, and discuss their agreement 
with jet quenching models.


\subsection{High-$p_T$ hadron spectra in proton-proton and proton-nucleus collisions}
\label{sec:highpT_hadrons}

\begin{figure}[htbp]
\centering
\includegraphics[width=9.5cm,height=7.5cm]{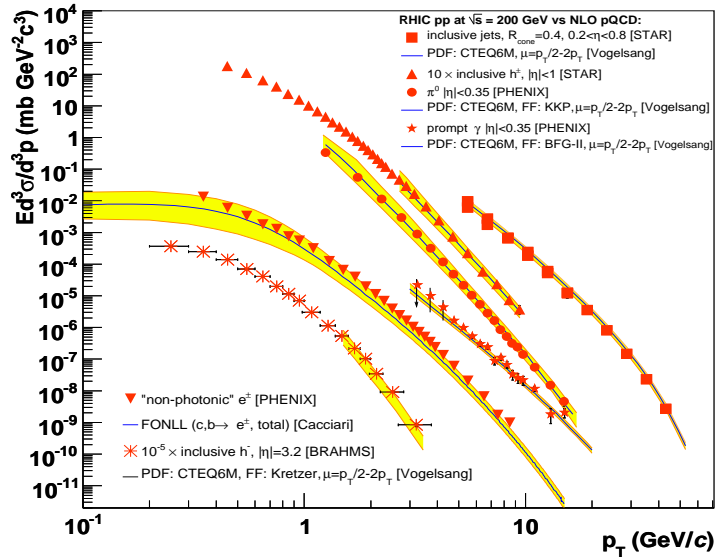}
\caption{Compilation of hard cross sections in \pp\ 
at $\sqrts$ = 200 GeV measured by STAR~\cite{star_pp_jets_200,star_hipt_200}, 
PHENIX~\cite{phnx_pp_pi0_200,phnx_pp_gamma_200,phnx_pp_nonphot_elec_200}, 
and BRAHMS~\cite{brahms_pp_dAu} (10\%-30\% syst. uncertainties not shown for clarity) 
compared to NLO~\cite{vogelsang_pi0,vogelsang_gamma} and NLL~\cite{cacciari05} pQCD 
predictions (yellow bands).}
\label{fig:pp_rhic_vs_nlo}
\end{figure}

Figure~\ref{fig:pp_rhic_vs_nlo} collects several $p_T$-differential inclusive cross sections
measured at RHIC in $pp$ collisions at $\sqrts$ = 200 GeV: jets~\cite{star_pp_jets_200},
charged hadrons~\cite{star_hipt_200}, neutral pions~\cite{phnx_pp_pi0_200},
direct photons~\cite{phnx_pp_gamma_200}, and $D, B$ mesons (indirectly measured via inclusive $e^\pm$
from their semileptonic decays)~\cite{phnx_pp_nonphot_elec_200} at central rapidities ($y=0$), and 
negative hadrons at forward pseudorapidities ($\eta$ = 3.2)~\cite{brahms_pp_dAu}. The existing 
measurements cover 9 orders of magnitude in cross section (from 10~mb/GeV$^2$ down to 1~pb/GeV$^2$), 
and broad ranges in transverse momentum (from zero for $D,B$ mesons up to 45 GeV/c, a half
of the kinematical limit, for jets) and rapidity ($\eta$ = 0 -- 3.2).

Standard next-to-leading-order (NLO)~\cite{vogelsang_pi0,vogelsang_gamma} or resummed 
next-to-leading log (NLL)~\cite{cacciari05} pQCD calculations (yellow bands in Fig.~\ref{fig:pp_rhic_vs_nlo})
with recent proton PDFs~\cite{cteq}, fragmentation functions~\cite{kkp,kretzer}, and with 
varying factorisation-renormalisation scales ($\mu=p_T/2-2p_T$) reproduce well the \pp\ data.
This is true even in the semi-hard range $p_T\approx 1- 4$~GeV/c, where a perturbative description 
would be expected to give a poorer description of the spectra. These results indicate that the hard QCD 
cross sections at RHIC energies are well under control both experimentally and theoretically in their full
kinematic domain.

\begin{figure}[htbp]
\centering
\includegraphics[width=0.85\linewidth]{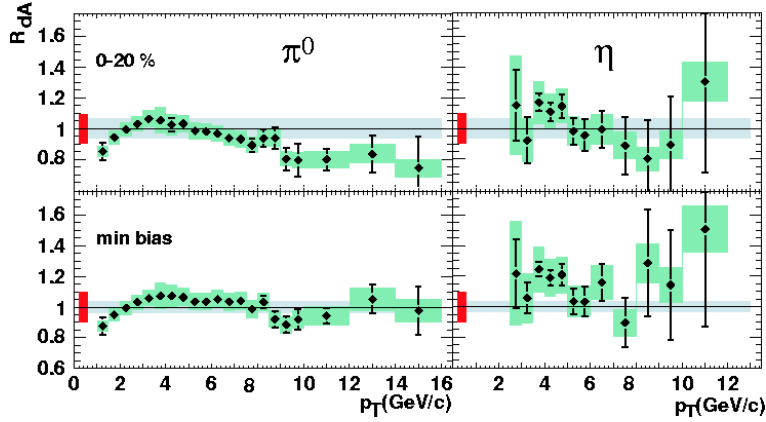}
\caption{Nuclear modification factors for high-$p_T$ $\pi^{0}$ (left) and $\eta$ (right) mesons 
at midrapidity in $dAu$ collisions at $\sqrtsnn$~=~200~GeV~\cite{phnx_dAu_pi0,Adler:2006wg}.}
\label{fig:dAu_rhic_vs_nlo}
\end{figure}

Not only the proton-proton hard cross sections are well under theoretical control at RHIC but the 
hard yields measured in deuteron-gold collisions do not show either any significant deviation from the perturbative 
expectations. Figure~\ref{fig:dAu_rhic_vs_nlo} shows  the nuclear modification factors measured in 
$dAu$ collisions at $\sqrtsnn$~=~200~GeV for high-$p_T$ $\pi^{0}$ at $y=0$~\cite{phnx_dAu_pi0,Adler:2006wg}. 
The maximum deviation from the $R_{dAu}$~=~1 expectation 
is of the order of  $\sim$10\%, well accounted for by standard pQCD calculations~\cite{vogt05,guzey04} 
that include DGLAP-based parametrisations of nuclear PDFs~\cite{eks98} and/or a mild amount of 
initial-state $p_T$ broadening~\cite{accardi09} to account for a modest ``Cronin enhancement''\footnote{The 
``Cronin effect'' is the observation of enhanced hadron production at $p_T\approx$~1~--7~GeV/c 
in proton-nucleus compared to proton-proton collisions (see~\cite{accardi09} for a recent review).}~\cite{cronin}. 
[The only exception to this behaviour is baryon (in particular, proton) production, which shows a 
{\it large} Cronin enhancement: $R_{dAu}$~=~1.5-2.0~\cite{Adams:2006nd}.]
These data clearly confirm that around {\it midrapidity} at RHIC energies, the parton flux of the incident gold nucleus can be 
basically obtained by geometric superposition of the nucleon PDFs, and that the nuclear $(x,Q^2)$ 
modifications of the PDFs are very modest\footnote{The same is not true at {\it forward} rapidities
where gluon saturation effects in the $Au$ PDFs play an important role (see e.g.~\cite{d'Enterria:2006nb} and references therein).}. 
Since no dense and hot system is expected to be produced in $dAu$ collisions, such results 
indicate that any value of $R_{AA}$ different than $R_{dAu}^2\approx~1 \pm 0.2$ potentially 
observed for hard probes in  \AuAu\ collisions (at central rapidities) can only be due to {\it final-state} 
effects in the medium produced in the latter reactions.


\subsection{High-$p_T$ hadron spectra in nucleus-nucleus collisions}
\label{sec:AAhighpT_hadrons}

Among the most exciting results from RHIC is the large high-$p_T$ hadron suppression ($R_{AA}\ll$ 1) 
observed in central \AuAu\ compared to \pp\ or \dAu\ reactions. 
We discuss here the properties of the measured suppression factor and compare it to detailed predictions 
of parton energy loss models.

\begin{figure}[htbp]
   \includegraphics[width=0.5\linewidth,height=6.8cm]{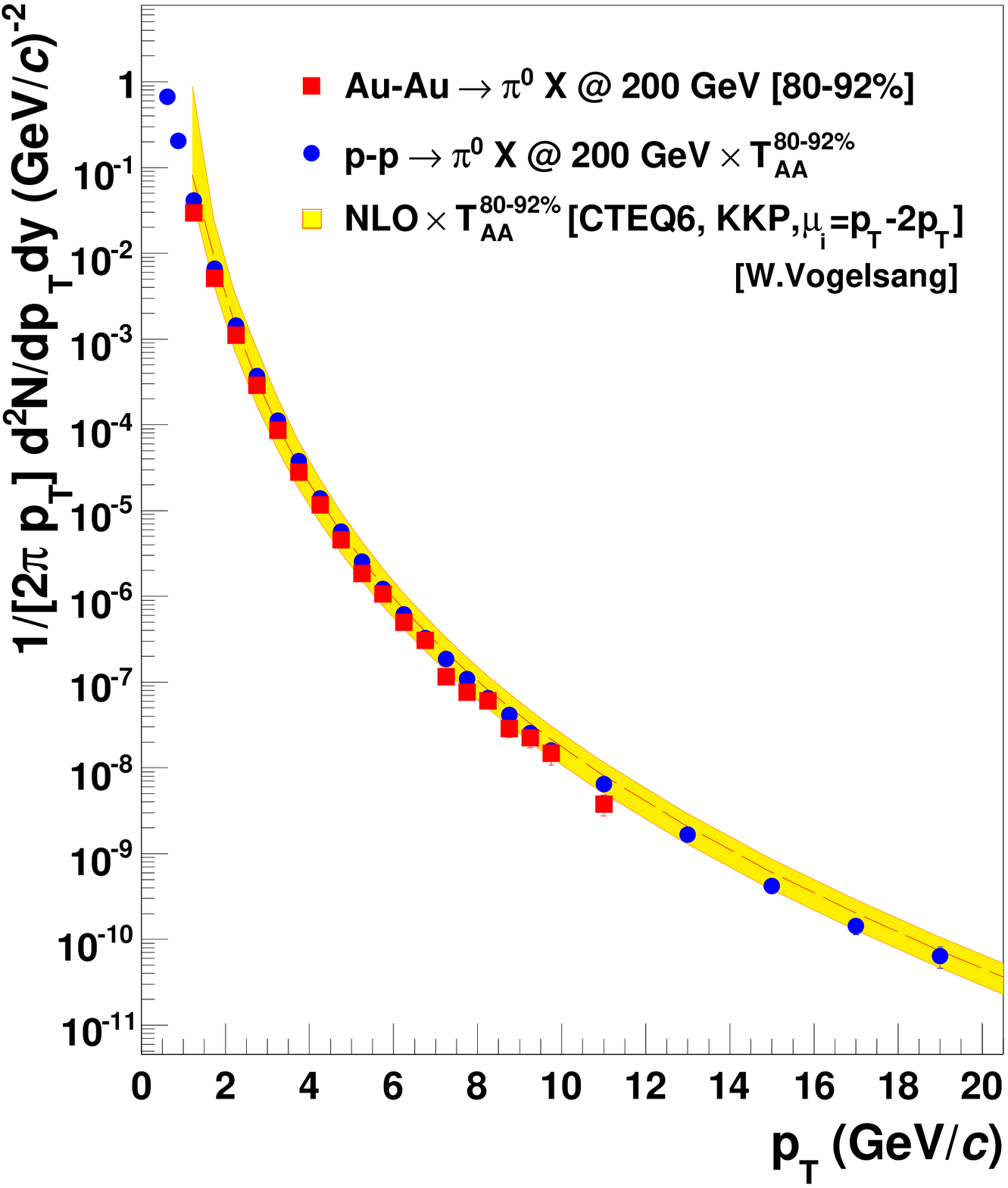}
   \includegraphics[width=0.5\linewidth,height=6.8cm]{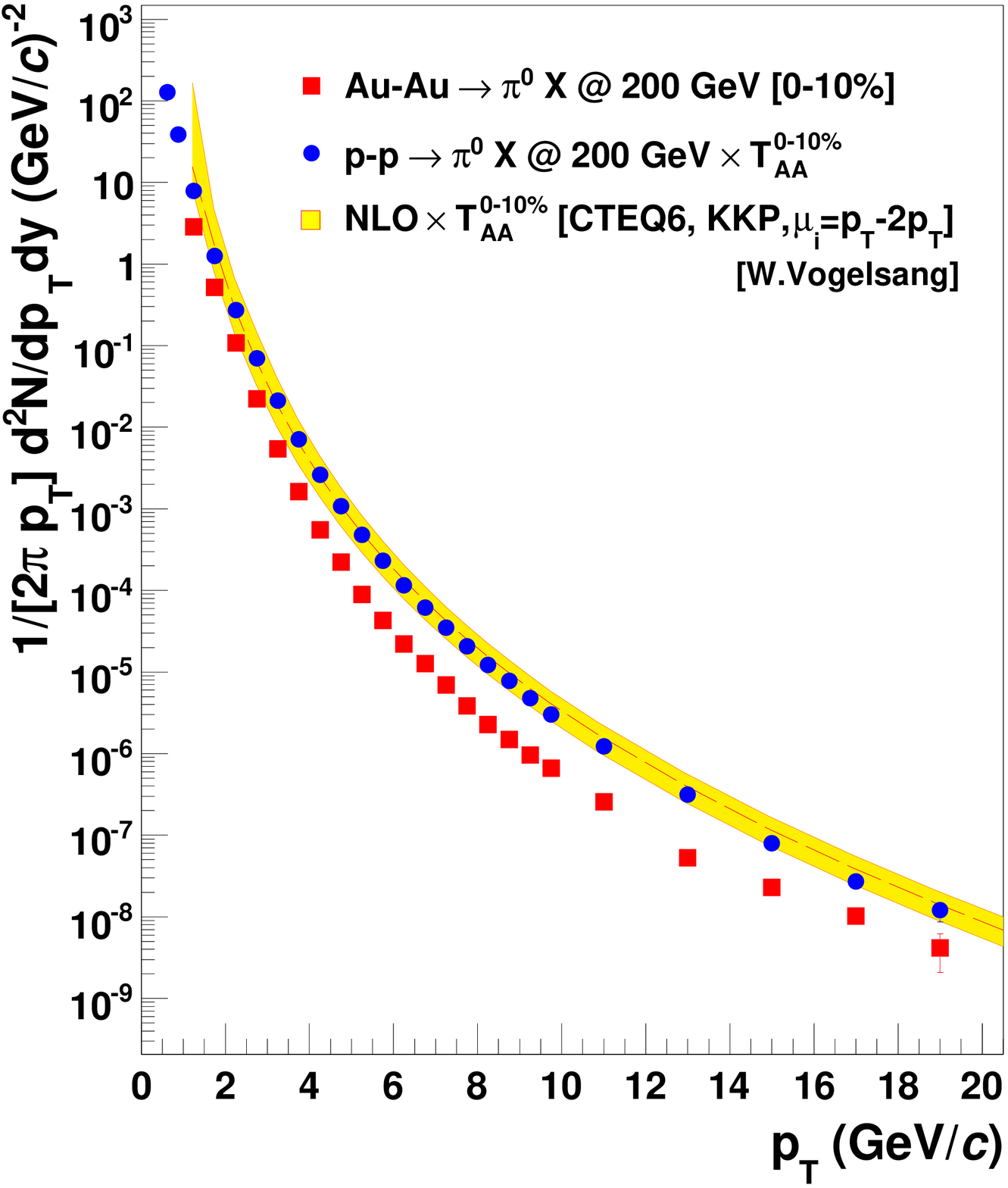}
\caption{Invariant $\pi^0$ yields measured by PHENIX in peripheral (left) and central (right)
\AuAu\ collisions (squares)~\protect\cite{Adare:2008qa}, compared to the ($T_{AA}$-scaled) $pp \rightarrow \pi^0+X$ cross section
(circles)~\protect\cite{Adare:2007dg} and to a NLO pQCD calculation (curves and yellow band)~\protect\cite{vogelsang_pi0}.}
\label{fig:phenix_pi0_pp_AuAu}
\end{figure}


\subsubsection{(a) Magnitude of the suppression: medium properties}

\begin{figure}[htbp]
\centering
\includegraphics[width=0.85\linewidth,height=6.3cm]{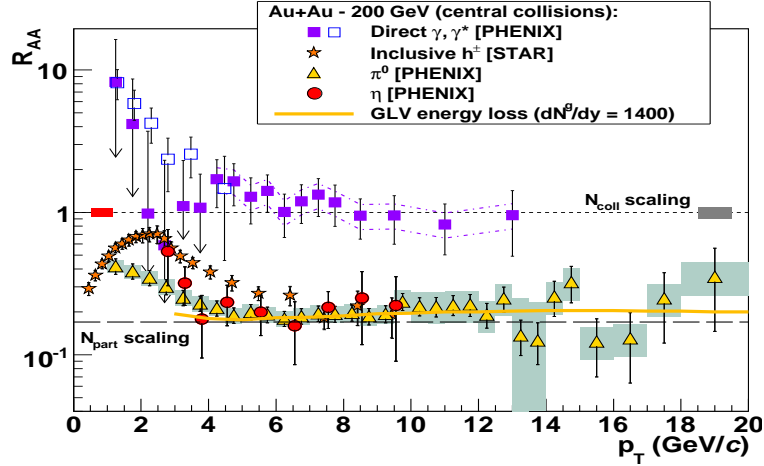}
\caption{$R_{AA}(p_T)$ measured in central \AuAu\ at 200 GeV for $\pi^0$~\protect\cite{Adare:2008qa} and 
$\eta$~\protect\cite{phenix_hipt_pi0_eta_200} mesons, charged hadrons~\protect\cite{star_hipt_200},
and direct photons~\protect\cite{phnx_gamma_AuAu200,Adare:2008fqa} compared to theoretical predictions for 
parton energy loss in a dense medium with $dN^g/dy=$~1400 (yellow curve)~\protect\cite{vitev_gyulassy}.}
\label{fig:R_AA_RHIC_200}
\end{figure}

Figure~\ref{fig:phenix_pi0_pp_AuAu} shows the $\pi^0$ spectrum measured in \pp\ collisions~\cite{Adare:2007dg} compared
to peripheral (left) and central (right) \AuAu\ spectra~\cite{Adare:2008qa} at 200 GeV, as well as to NLO pQCD calculations~\cite{vogelsang_pi0}. 
Whereas the peripheral \AuAu\ spectrum is consistent with a simple superposition of individual 
$NN$ collisions, the data in central \AuAu\ show a suppression factor of 4 -- 5 with respect to this expectation.
The amount of suppression is better quantified taking the ratio of both spectra in the nuclear modification factor, 
Eq.~(\ref{eq:R_AA}). Figure~\ref{fig:R_AA_RHIC_200} compiles the measured $R_{AA}(p_T)$ 
for various hadron species 
and for direct  $\gamma$ in central \AuAu\ collisions at $\sqrtsnn$~=~200~GeV. 

Above $p_T\approx$~5~GeV/c, $\pi^0$~\cite{phenix_hipt_pi0_200}, $\eta$~\cite{phenix_hipt_pi0_eta_200}, 
and charged hadrons~\cite{star_hipt_200,phenix_hipt_200} 
(dominated by $\pi^\pm$~\cite{phenix_hipt_200}) show all a common factor 
of $\sim$5 suppression relative to the $R_{AA}$~=~1 expectation that holds for hard probes, 
such as direct photons, which do no interact with the medium~\cite{phnx_gamma_AuAu200}. 
The fact that $R_{AA}\approx$~0.2 irrespective of the nature of the finally produced 
hadron is consistent with a scenario where final-state energy loss of the {\it parent} parton 
takes place {\it prior} to its fragmentation into hadrons in the vacuum 
according to {\it universal} (but energy-rescaled) FFs.
The suppression factor at top RHIC energies is very close to the ``participant scaling'', 
$(N_{part}/2)/N_{coll}\approx$~0.17, expected in the strong quenching limit 
where only hadrons coming from partons produced at the {\it surface} of the medium show
no final-state modifications in their spectra~\cite{Muller:2002fa}. From the $R_{AA}$ one 
can approximately obtain the fraction of energy lost, $\epsilon_{loss} = \Delta p_{T}/p_{T}$, via
\begin{equation}
\epsilon_{loss} \approx 1 - R_{AA}^{1/(n-2)}\;,
\label{eq:eloss}
\end{equation}
when the \AuAu\ and \pp\ {\it invariant} spectra are both a power-law with 
exponent $n$, i.e. $1/p_{T}\,{d}N/{d}p_{T} \propto p_{T}^{-n}$~\cite{Adler:2006bw}. 
At RHIC ($n\approx$~8, $R_{AA}\approx$~0.2), one finds $\epsilon_{loss}\approx$~0.2.

\begin{figure}[htbp]
\centering
\includegraphics[height=5.5cm,width=12.cm,clip]{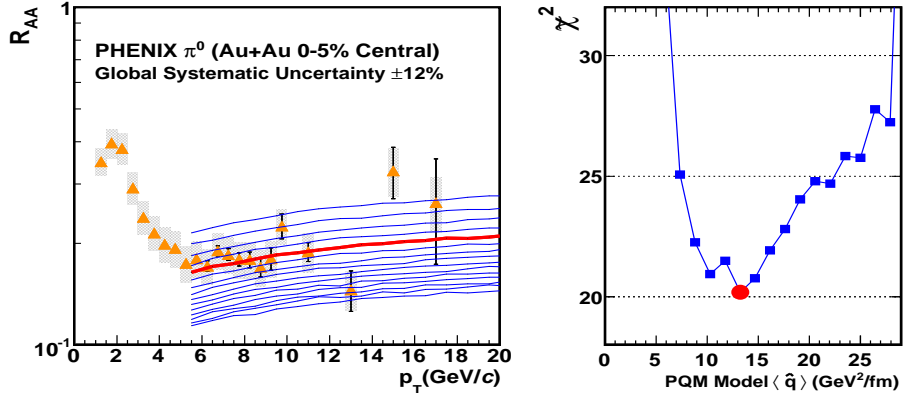}
\caption{Left: $R_{AA}(p_T)$ for neutral pions in central \AuAu\ collisions (triangles)~\cite{Adare:2008qa}
compared to PQM predictions~\cite{pqm} for varying values of the $\qhat$ coefficient 
(red curve, best fit for $\mean{\hat{q}}$~=~13.2\,GeV$^2$/fm).
Right: Corresponding (data vs. theory) $\chi^2$ values for the PQM $\qhat$ parameters that fit
the data points on the left plot~\cite{Adare:2008cg}.}
\label{fig:RAA_vs_PQM}
\end{figure}

The high-$p_T$ \AuAu\ suppression can be well reproduced by parton energy loss models that 
assume the formation of a very dense system with initial gluon rapidity densities $dN^g/dy\approx$~1400 
(yellow line in Fig.~\ref{fig:R_AA_RHIC_200})~\cite{vitev_gyulassy}, transport coefficients
$\mqhat\approx$~13~GeV$^2$/fm (red line in Fig.~\ref{fig:RAA_vs_PQM}, left)~\cite{pqm}, 
or plasma temperatures $T\approx$~0.4~GeV~\cite{Turbide:2005fk}. 
The quality of agreement between the theory and data has been studied in detail 
in~\cite{Adare:2008qa,Adare:2008cg} taking into account the experimental 
(though not theoretical) uncertainties. 
The PHENIX $\pi^0$ suppression data allows one to constrain the  transport coefficient 
of the PQM model~\cite{pqm} $\mean{\hat{q}}$ as 13.2 $^{+2.1}_{-3.2}$ and $^{+6.3}_{-5.2}$\,GeV$^2$/fm 
at the one and two standard-deviation levels (Fig.~\ref{fig:RAA_vs_PQM}, right).

The consistency between the extracted $\qhat$, $dN^g/dy$ and $T$ values in the various models 
can be cross-checked considering the simple case of a gluon traversing a thermalised gluon plasma.
The transport coefficient, Eq.~(\ref{eq:qhat}), is the product of the medium particle density, the medium Debye-mass, 
and the parton-medium cross section. Taking $\sigma_{gg}=9\pi\alpha_s^2/(2\,m_D^2)$ with 
$\alpha_s$~=~0.5 for the latter, one has a simple relation\footnote{Conversion between units is done
multiplying by suitable powers of $\hslash\,c\,=$~0.197~GeV\,fm.} 
between $\qhat$ and $\rho$:
\begin{equation}
\qhat \mbox{\small [GeV$^2$/fm]}= m_D^2 \times \sigma \times \rho = m_D^2 \times 9\pi\alpha_s^2/(2\,m_D^2)\times \rho 
\approx 0.14 \,K\; \rho\mbox{\small [fm$^{-3}$]}\;,
\label{eq:qhat_rho}
\end{equation}
where we introduce a $K$-factor to account for possible higher-order scatterings not included in the LO perturbative
expression for $\sigma_{gg}$. For an ideal ultrarelativistic gas, the particle density scales with the cube of the temperature as $\rho \approx$~ndf$/9\cdot T^{3}$.
For a pure gluon plasma (with ndf = 16  degrees of freedom), $\rho\mbox{\small [fm$^{-3}$]} 
\approx 260\cdot (T\mbox{\small [GeV]})^{3}$, 
and one can write Eq.~(\ref{eq:qhat_rho}) as:
\begin{equation}
\qhat \mbox{\small [GeV$^2$/fm]} \approx 36 \,K \cdot (T\mbox{\small [GeV]})^{3}
\label{eq:qhat_T}
\end{equation}
In addition, from the relation $\rho${\small [fm$^{-3}$]}~$\approx 1.9\cdot(\varepsilon\mbox{\small [GeV/fm$^3$]})^{3/4}$
between particle and energy densities,
one can also express Eq.~(\ref{eq:qhat_rho}) as: 
\begin{equation}
\hat{q}\mbox{\small [GeV$^2$/fm]}\approx 0.27\,K\cdot(\varepsilon\mbox{\small [GeV/fm$^3$]})^{3/4}\,.
\label{eq:qhat_eps}
\end{equation}
In an expanding plasma, the density follows a power-law evolution as a function of time, $\rho=\rho_0\,(\tau_0/\tau)^\alpha$, 
and thus so does the transport coefficient (\ref{eq:qhat_rho}):
\begin{equation}
\hat{q}(\tau)\mbox{\small [GeV$^2$/fm]}\approx 0.14\,K \cdot\rho_0\left(\frac{\tau_0}{\tau}\right)^\alpha = 
0.14\,K \cdot\frac{dN^{g}}{dV}\left(\frac{\tau_0}{\tau}\right) \approx 0.14 \,K\cdot\frac{1}{A_T}\,\frac{dN^{g}}{dy}\,\frac{1}{\tau} \, ,
\label{eq:qhat_dNgdy}
\end{equation}
where for the two last equalities we have assumed a 1-dimensional (aka Bjorken) longitudinal expansion 
i.e. $\alpha = 1$ and 
$dV = A_T\,\tau_0\,dy$, where $A_T${\small [fm$^2$]} is the transverse area of the system. 
Combining Eq.~(\ref{eq:qhat_dNgdy}) with Eq.~(\ref{eq:qhat_time}) that relates the {\it time-averaged} $\hat{q}(\tau)$ 
to that of a static medium with effective length $L_{\ensuremath{\it eff}}$, 
we finally get
\begin{equation}
\mean{\hat{q}}\mbox{\small [GeV$^2$/fm]}\approx 0.14\,K\cdot \frac{2}{L_{\ensuremath{\it eff}}\mbox{\small [fm]}\,A_T\mbox{\small [fm$^2$]}}
\cdot\frac{dN^{g}}{dy} \approx 1.4\,10^{-3}\cdot K\cdot\frac{dN^{g}}{dy}\;,
\label{eq:avqhat_dNgdy}
\end{equation}
where, for the last equality, we use $L_{\ensuremath{\it eff}}\approx$~2~fm and $\langle A_T\rangle\approx$~100~fm$^2$ 
for the overlap area in 0-10\% most central \AuAu. 
This approximate relation between the average transport coefficient and the original gluon density is only
well fulfilled by the data (see Table~\ref{tab:dNdy_qhat_sqrts} below) for very large $K \approx$~7 factors.
The fact that the jet-quenching data favours an effective elastic parton-medium cross-section much larger 
than the LO perturbative estimate of $\sigma_{gg}\approx$~1.5~mb, see Eq.~(\ref{eq:qhat_rho}), has been 
discussed many times in the literature -- e.g. in the context of the strong partonic elliptic flow seen in the 
data~\cite{Molnar:2001ux} -- and supports the strongly-coupled nature of the QGP produced at RHIC~\cite{gyulassy_mclerran04}.\\

Equation~(\ref{eq:avqhat_dNgdy}) is just a simple order-of-magnitude estimate based on simplifying assumptions.
A more detailed comparison of different energy-loss schemes within a realistic {\it 3-dimensional}
hydrodynamics evolution has been carried out in~\cite{Bass:2008rv}. The extraction of a common 
$\qhat$ parameter from the different model predictions relies on the use of (thermo)dynamical relationships
such as Eqs.~(\ref{eq:qhat_T}) or (\ref{eq:qhat_eps}). The results for the ASW, AMY and HT schemes 
are shown in Table~\ref{tab:qhats}. 
The ASW calculations consistently predict a higher $\qhat$ than AMY or HT. Seemingly, 
as of today, comparisons of model predictions to RHIC results for $R_{AA}(p_T)$ can only 
constrain $\qhat$ within a factor of 2~--~3. The origin of such a large variability can be traced to a 
combination of (i) the relative insensitivity of using just a single-inclusive observable\footnote{Irreducible 
parton production from the outer corona of the medium -- which remains unsuppressed even for extreme 
densities in the centre -- makes of $R_{AA}(p_T)$ a ``fragile'' observable~\cite{Eskola:2004cr}.}, 
$R_{AA}(p_T)$, in the data--model comparisons~\cite{Eskola:2004cr} (additional independent measurements place 
extra constraints on $\qhat$ as discussed in Section~\ref{sec:IAA}), and (ii) the assumptions about 
the equation-of-state of the medium (and its time evolution) and the corresponding approximations 
relating its thermodynamical and transport properties. 
Genuine model differences (e.g. AMY accounts for collisional loses which are neglected in the purely radiative ASW 
approach) play also a role. A working group~\cite{techqm} has been recently created to clarify discrepancies 
among the formalisms. 

\begin{table}[htbp]
\caption{Transport coefficients $\hat{q}$ derived in a 3-D hydro simulation of an 
expanding QGP with initial temperature $T_0$~=~0.4 GeV (at $\tau_0$~=~0.6~fm/c)~\cite{Bass:2008rv} 
with different parton energy loss implementations (ASW, HT and AMY schemes) that reproduce 
the high-$p_T$ $\pi^0$ suppression observed in central \AuAu\ at RHIC~\protect\cite{Adare:2008qa}. 
The $a,b$ exponents indicate two choices of scaling of $\hat{q} ({\bf r},\tau)$ with the initial
plasma temperature or energy-density: $(a)\;\hat{q}_0\propto T_0^3({\bf r}, \tau)$,
and $(b)\;\hat{q}_0\propto \epsilon_0^{3/4}({\bf r}, \tau)$. The PQM/ASW result 
(Fig.~\ref{fig:RAA_vs_PQM}, $\mqhat$ for a {\it static} plasma) is also listed for comparison.}
\vspace{0.2cm}
\label{tab:qhats}
\centering
\begin{tabular}{c | c | c | c } \hline\hline\noalign{\smallskip}
\hspace{2mm} & \hspace{2mm}ASW \hspace{2mm} & \hspace{2mm}HT \hspace{2mm} & \hspace{5mm}AMY \hspace{5mm}\\\hline \noalign{\smallskip}
\hspace{2mm}$\hat{q}$ (GeV$^2$/fm) \hspace{2mm} & \hspace{2mm} 10$^{(a)}$ -- 18.5$^{(b)}$, 13.2$^{(PQM)}$\hspace{2mm} & \hspace{2mm} 2.3$^{(a)}$ -- 4.3$^{(b)}$ \hspace{2mm}&  \hspace{2mm} 4.1$^{(a)}$ \hspace{5mm}\\\noalign{\smallskip}\hline\hline
\end{tabular}
\end{table}



\subsubsection{(b) Centre-of-mass energy dependence}

As one increases the centre-of-mass energy in nucleus-nucleus collisions, the produced plasma
reaches higher energy and particle densities and the system stays longer in the QGP phase.
Since $\Delta E_{\ensuremath{\it loss}}\propto dN^g/dy \propto dN_{ch}/d\eta$,
and since the charged particle multiplicity in \AaAa\ at midrapidity increases with 
collision-energy as~\cite{ppg019}
\begin{equation}
dN_{ch}/d\eta \approx 0.75\cdot(N_{part}/2)\cdot \ln(\sqrtsnn\,\mbox{\small [GeV]}/1.5),
\label{eq:dNchdy}
\end{equation}
(where $N_{part}$ is the number of nucleons participating in the collision),
one naturally expects the hadron quenching (at a given centrality) to increase logarithmically 
with $\sqrtsnn$. 
The actual ``excitation function'' of $R_{AA}$ does not follow exactly the same c.m.-energy
dependence of Eq.~(\ref{eq:dNchdy}) because for increasing energies 
other factors play counteracting roles: (i) 
the lifetime of the quenching medium becomes longer (which enhances the energy loss), (ii) 
the parton spectrum becomes flatter (which leads to a comparatively 
{\it smaller} suppression for the same value of $\Delta E_{\it loss}$, see below), and (iii) 
the relative fraction of quarks and gluons produced at a given $p_T$ changes and so does the quenching factor
(see Fig.~\ref{fig:q_g_RAA_fraction} below and the discussion on the colour-factor dependence of the suppression).

\begin{figure}[htbp]
\centering
\includegraphics[width=0.85\linewidth,height=6.2cm]{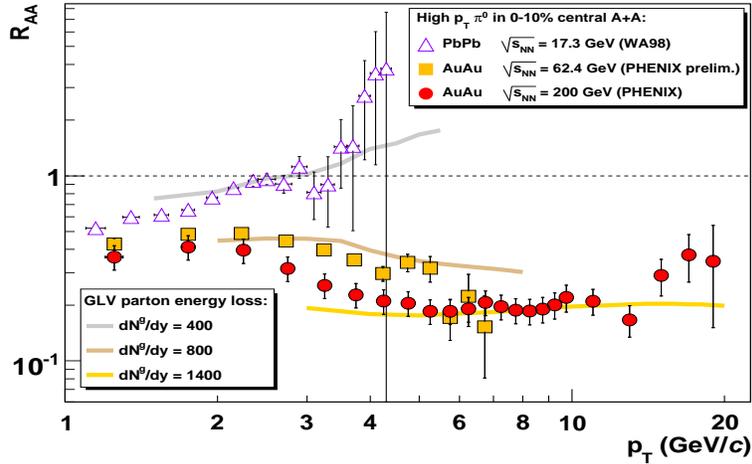}
\caption{Nuclear modification factor, $R_{AA}(p_T)$, for neutral 
pions in central \PbPb\ at $\sqrtsnn$~=~17.3~GeV~\protect\cite{Aggarwal:2001gn,d'Enterria:2004ig}
and \AuAu\ at $\sqrtsnn$~=~62.4~GeV~\protect\cite{phenix_hipt_62}, 
200~GeV~\protect\cite{phenix_hipt_pi0_200}; 
compared to GLV  energy loss calculations for initial gluon densities: 
$dN^g/dy$ = 400, 800, 1400 \protect\cite{vitev_gyulassy,Vitev:2004gn} respectively.
Experimental normalisation errors, $\mathscr{O}$(10\%--25\%), not shown.}
\label{fig:RAA_compilation}
\end{figure}

Figure~\ref{fig:RAA_compilation} compiles the measured $R_{AA}(p_T)$ for 
high-$p_T$ $\pi^0$ measured in central \AaAa\ collisions in the range 
$\sqrtsnn\approx$ 20 -- 200 GeV compared to 
parton energy loss calculations that assume the formation of a QGP with initial gluon densities 
per unit rapidity in the range $dN^g/dy\approx$ 400 -- 1400~\cite{vitev_gyulassy,Vitev:2004gn} 
or, equivalently, averaged transport coefficients 
$\mean{\hat{q}}\approx$~3.5~--~13~GeV$^2$/fm~\cite{pqm} (Table~\ref{tab:dNdy_qhat_sqrts}). 
The theoretical predictions reproduce very well the experimental data.
The SPS data show an $R_{AA}$ for central \PbPb\ which, 
though consistent with unity~\cite{d'Enterria:2004ig}, is significantly suppressed compared to the 
``Cronin enhancement'' observed for peripheral \PbPb\ and for $pPb$ collisions~\cite{Aggarwal:2007gw}.
The onset of the suppression must lie close to the highest energies reached at SPS, in a domain of
c.m. energies that would be worth studying detailedly (including proper high-$p_T$ \pp\ reference measurements) in coming RHIC runs.
At $\sqrtsnn$~=~62~GeV, the suppression is already comparatively large because, as can be seen from Eq.~(\ref{eq:eloss}), 
$R_{AA}$ not only depends on $\Delta E_{loss}$ but also on the steepness (power-law exponent $n$) of the parton $p_{T}$ 
spectrum: with decreasing $\sqrtsnn$, the $p_{T}$ spectra become steeper effectively leading to a comparatively 
{\it larger} suppression (i.e. smaller $R_{AA}$) for the {\it same} value of $\Delta E_{loss}$.\\

In any case, it is interesting to remark that for each collision energy the values for 
$dN^g/dy$ derived from the jet-quenching models 
are consistent with the final charged hadron density $dN_{ch}/d\eta$ measured in the reactions.  
This is expected in an isentropic\footnote{Namely, expanding at constant entropy i.e. without extra 
particle production.} expansion process, where all the hadrons produced at midrapidity in a 
\AaAa\ collision come directly from the original gluons released
\footnote{We use: $N_{tot}/N_{ch}$ = 3/2 and the Jacobian 
$|d\eta/dy| = E/m_T \approx$~1.2 for a mostly pionic system.} in the collision:
\begin{equation}
\frac{dN^g}{dy}\approx\frac{N_{tot}}{N_{ch}}\,\left|\frac{d\eta}{dy}\right|\,\frac{dN_{ch}}{d\eta}
\approx 1.8\cdot\frac{dN_{ch}}{d\eta}.
\label{eq:dNgdy}
\end{equation}
This relationship is relatively well fulfilled by the data as can be seen by comparing
the fourth and fifth columns of Table~\ref{tab:dNdy_qhat_sqrts}.

\begin{table}[htbp]
\caption{Initial gluon densities $dN^g/dy$~\protect\cite{vitev_gyulassy,Vitev:2004gn}, and 
transport coefficients $\mean{\hat{q}}$~\protect\cite{pqm,dde_hp04} 
for the dense media produced in central \AaAa\ collisions at SPS and RHIC 
obtained from parton energy loss calculations reproducing the observed high-$p_T$ 
$\pi^0$ suppression at each $\sqrtsnn$. The measured charged particle densities at 
midrapidity, $dN_{ch}^{exp}/d\eta$~\protect\cite{ppg019}, are also quoted.}
\label{tab:dNdy_qhat_sqrts}
\centering
\begin{tabular}{lcccc}\hline\hline\noalign{\smallskip}
\hspace{2mm} & \hspace{2mm}$\sqrtsnn$ \hspace{2mm} & \hspace{2mm} $\mean{\hat{q}}$ \hspace{2mm}  &  \hspace{2mm} $dN^g/dy$  \hspace{2mm} &  \hspace{2mm} $dN_{ch}^{exp}/d\eta$  \hspace{2mm} \\
\hspace{2mm} & \hspace{2mm} (GeV) \hspace{2mm} &  \hspace{2mm}(GeV$^2$/fm)  \hspace{2mm} & \hspace{2mm} & \\
\noalign{\smallskip}\hline\noalign{\smallskip}
SPS  &  17.3 & 3.5 & 400  & 312 $\pm$ 21 \\
RHIC &  62.4 & 7. & 800  & 475 $\pm$ 33 \\
RHIC &  130. &  $\sim$11  & $\sim$1000   & 602 $\pm$ 28 \\
RHIC &  200. & 13 & 1400 & 687 $\pm$ 37 \\ \noalign{\smallskip}\hline\hline
\end{tabular}
\end{table}
\vspace{-0.6cm}



\subsubsection{(c) $p_T$-dependence of the suppression}

\begin{figure}[htbp]
\centering
\includegraphics[width=0.85\textwidth,height=5.7cm]{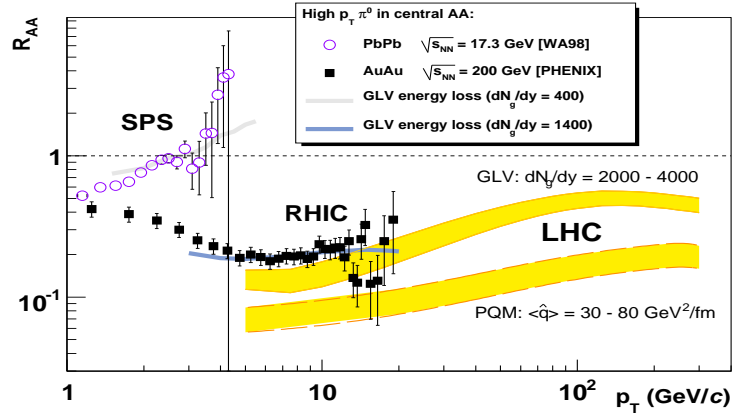}
\caption{$R_{AA}(p_{T})$ for neutral pions at SPS~\cite{Aggarwal:2001gn,d'Enterria:2004ig} 
and RHIC~\cite{Adare:2008qa} compared to the suppression of charged hadrons in central 
\PbPb\ at the LHC ($\sqrtsnn$~=~5.5~TeV) predicted by the GLV ($dN^g/dy$ = 2000 -- 4000)~\protect\cite{vitev_gyulassy,Abreu:2007kv} 
and PQM ($\mqhat\approx$  30 -- 80 GeV$^2$/fm)~\protect\cite{pqm,Abreu:2007kv} models.}
\label{fig:RAA_SPS_RHIC_LHC}
\end{figure}

At RHIC top energies, the hadron quenching factor remains relatively constant from 5~GeV/c up to the highest 
transverse momenta measured so far, $p_T\approx$~20~GeV/c (see Figs.~\ref{fig:R_AA_RHIC_200},~\ref{fig:RAA_SPS_RHIC_LHC}). 
On rather general grounds~\cite{Renk:2006pk}, one expects a rise of $R_{AA}$ with $p_T$ 
for any model in which the energy loss probability does not strongly depend on the initial parton
energy as more of the shift in energy becomes accessible. The detailed form of the rise is sensitive 
to the energy loss probability distribution, Eq.~(\ref{eq:qw}). 
The measured flatness of $R_{AA}(p_T)$ 
was not expected in various original analytical QCD energy-loss calculations including the LPM effect 
(see e.g.~\cite{Jeon:2002dv}) which instead predicted an $R_{AA}$ slowly (logarithmically) increasing with $p_T$. 
However, the combined effect of (i) kinematics constraints (which modify the asymptotic $\Delta E_{loss}$ 
formulas), (ii) the steeply falling $p_T$ spectrum of the scattered partons, and (iii) $\mathscr{O}($20\%) 
$p_T$-dependent (anti)shadowing differences between the proton and nuclear PDFs included in the various 
models~\cite{vitev_gyulassy,pqm,Jeon:2003gi,Eskola:2004cr}, do result in an effectively flat $R_{AA}(p_T)$ 
as found in the data. 


The much larger kinematical range opened at LHC energies~\cite{Abreu:2007kv} will allow one to 
test the $p_T$-dependence of parton energy loss over a much wider domain than at RHIC. 
The GLV and PQM predictions for the charged hadron suppression in \PbPb\ at 5.5 TeV 
are shown in Fig.~\ref{fig:RAA_SPS_RHIC_LHC}. Apart from differences in the {\it absolute}
quenching factor, PQM seemingly predicts a {\it slower} rise of $R_{AA}(p_T)$ than GLV.
The large $p_T$ reaches of the ALICE~\cite{Alessandro:2006yt}, ATLAS~\cite{Grau:2008ef} 
and CMS~\cite{D'Enterria:2007xr} experiments (up to 300~GeV/c for the nominal luminosities) 
will allow them to test such level of model details.


\subsubsection{(d) Centrality (system-size) dependence}

\begin{figure}[!Hhtb]
\includegraphics[width=\textwidth,height=5.1cm]{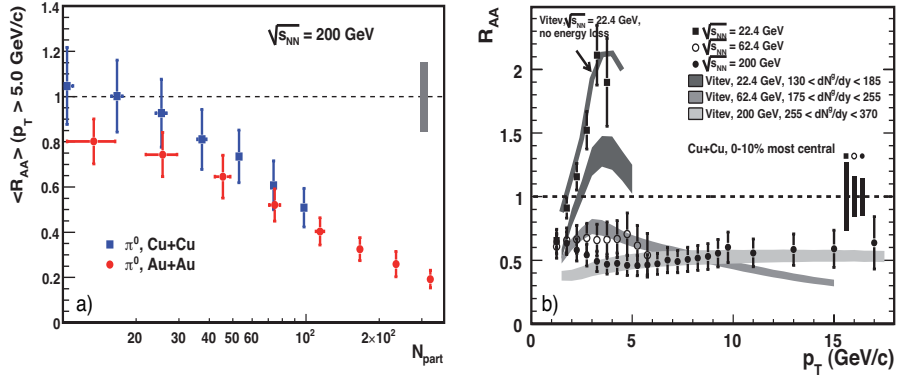}
\caption{Left: Centrality ($N_{part}$) dependence of the high-$p_T$ $\pi^0$ suppression
in \CuCu\ and \AuAu\ at 200~GeV~\cite{Reygers:2008pq}. 
Right: $R_{AA}(p_T)$ for $\pi^0$ in central \CuCu\ collisions at 
22.4, 62.4 and 200~GeV compared to GLV calculations with initial gluon 
densities $dN^g/dy\approx$~100~--~370~\cite{Adare:2008cx}.}
\label{fig:RAA_vs_Npart}
\end{figure}

The volume of the overlap zone in a heavy-ion collision can be ``dialed'' either by selecting 
a given impact-parameter $b$ -- i.e. by choosing more central or peripheral reactions -- or by 
colliding larger or smaller nuclei. 
From Eq.~(\ref{eq:glv}), the relative amount of suppression depends\footnote{Since 
${dN^g}/{dy} \, \propto \,  {dN_{ch}}/{dy} \propto  A_{\rm eff} \,  \propto \,  N_{\rm part}$, 
$\;L \propto  A_{\rm eff}^{1/3} \, \propto \,  N_{\rm part}^{1/3}$, and
$A_{\perp} \propto A_{\rm eff}^{2/3} \, \propto \, N_{\rm part}^{2/3}$.} on the effective mass 
number $A_{\rm eff}$ or, equivalently, on the number of participant nucleons in the collision $N_{\rm part}$,
as: $\epsilon = { \Delta E }/{E} \; \propto \; A_{\rm eff}^{2/3} \; \propto \; N_{\rm part}^{2/3}$.
Combining this expression with Eq.~(\ref{eq:eloss}) yields~\cite{Vitev:2005he}
\begin{equation}
R_{AA} = (1 - \kappa\; N_{part}^{\alpha})^{n-2}\;, \mbox{ with $\alpha\approx$~2/3, and $\kappa$ an arbitrary constant.}
\label{eq:RAA_Npart}
\end{equation}
Figure~\ref{fig:RAA_vs_Npart} (left) compares the measured high-$p_T$ pion suppression in 
\CuCu\ and \AuAu\ at $\sqrtsnn$~=~200~GeV~\cite{Adare:2008cx,Reygers:2008pq}. 
Because of the large difference in the $Cu$ ($A$~=~63) and $Au$ ($A$~=~197) mass numbers, the same $N_{part}$ 
value (i.e. the same overlap volume) implies very different collision geometries: a thin, elongated collision zone in \AuAu, 
and a thicker, more spherical one in the \CuCu\ case. Yet, interestingly the average suppression in the two 
systems depends only on $N_{part}$. Fitting this dependence with expression (\ref{eq:RAA_Npart}) 
yields $\alpha = 0.56 \pm 0.10$, consistent with $\alpha \approx 0.6$ expected 
in detailed parton energy loss calculations~\cite{Adare:2008qa,Vitev:2005he}.


The right plot of figure~\ref{fig:RAA_vs_Npart} shows the $R_{AA}(p_T)$ measured 
in \CuCu\ at 22.4, 62.4, and 200\,GeV~\cite{Adare:2008cx}. The observed amount of suppression is 
roughly a factor of $(A_{Au}/A_{Cu})^{2/3} \approx$~2 lower than in \AuAu\
at the same energies (Fig.~\ref{fig:RAA_compilation}).
The $R_{AA}(p_T)$ can be described by the GLV model with initial gluon densities 
$dN^g/dy\approx$~100~--~370 (the \CuCu\ enhancement at 22.4\,GeV is actually 
consistent with a scenario {\it without} parton energy loss). 


\subsubsection{(e) Path-length dependence}

\begin{figure}[!Hhtb]
\includegraphics[width=6.cm,height=4.5cm]{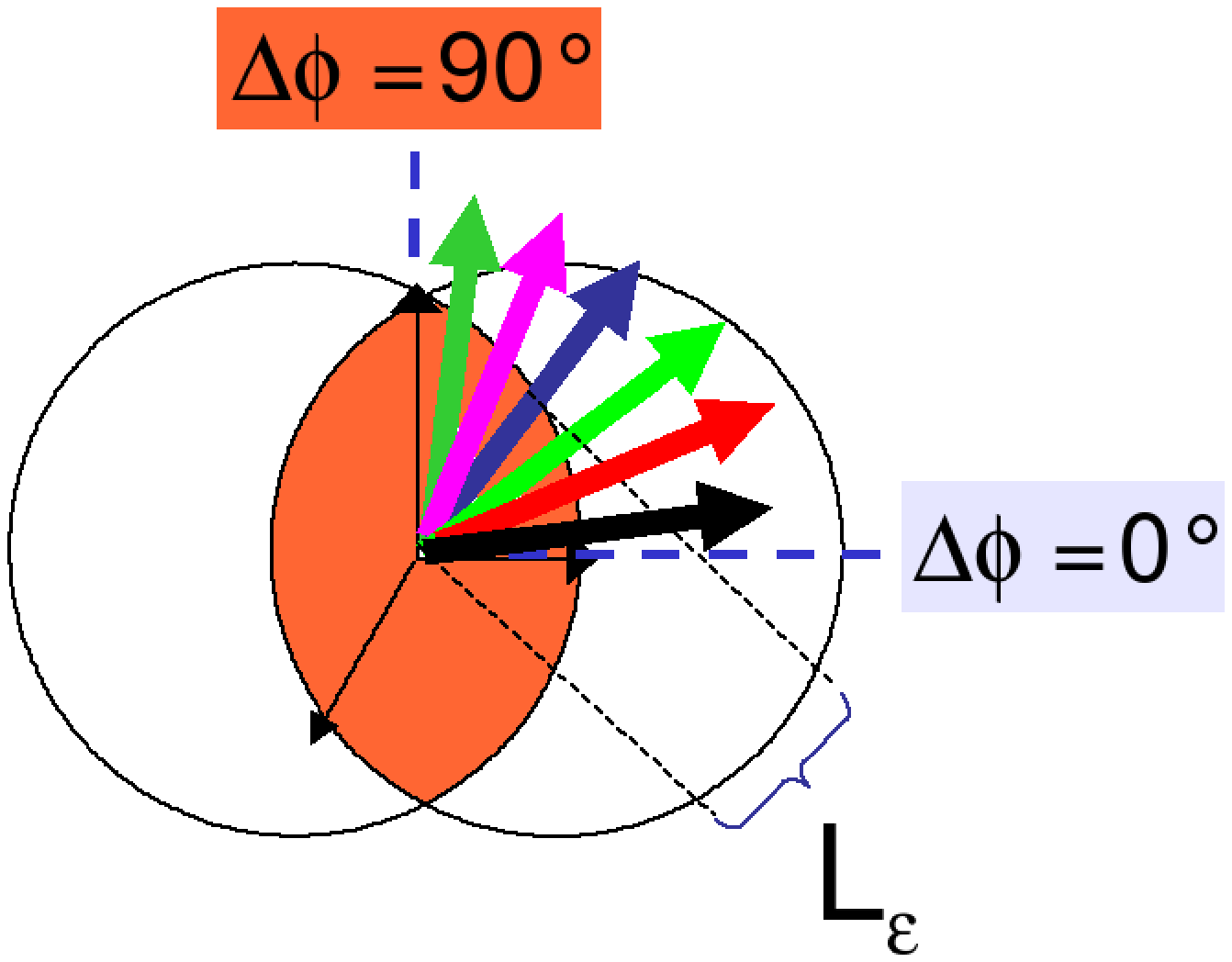}
\includegraphics[width=5.8cm,height=4.5cm]{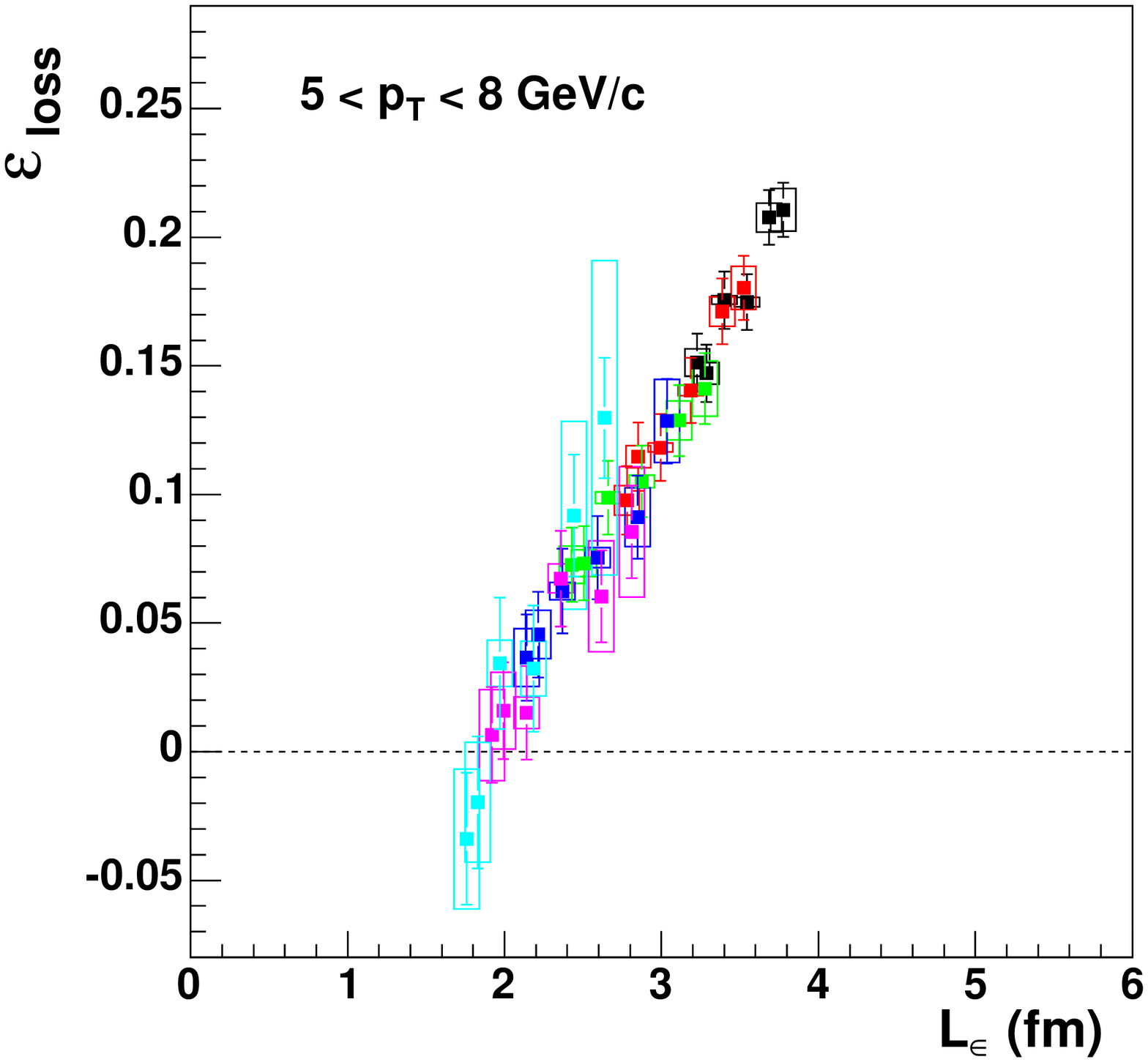}
\caption{Left: Effective thicknesses along various azimuthal directions with respect to the reaction
plane in the overlap region of two heavy-ions. Right: Fraction of energy loss $\epsilon_{loss}$ versus 
effective path-length $L_\epsilon$ measured for high-$p_{\rm T}$ neutral pions in \AuAu\ at 
200~GeV~\cite{Adler:2006bw}.}
\label{fig:Sloss_L}
\end{figure}

The analytical {\it quadratic} dependence of the energy loss on the thickness of a {\it static} medium $L$,
Eq.~(\ref{eq:LPM_QCD}), becomes effectively a {\it linear} dependence on the {\it initial} value of $L$
when one takes into account the expansion of the plasma, see Eq.~(\ref{eq:glv}). Experimentally, one can test the 
$L$-dependence of parton suppression by exploiting the spatial asymmetry of the  system produced 
in non-central nuclear collisions (Fig.~\ref{fig:Sloss_L}, left). Partons produced ``in plane'' (``out-of-plane'') 
i.e. along the short (long) direction of the ellipsoid matter with eccentricity $\epsilon$ will comparatively 
traverse a shorter (longer) thickness. 

PHENIX~\cite{Adler:2006bw,Adare:2009iv} has measured the high-$p_T$ neutral pion suppression as a function 
of the angle with respect to the reaction plane, $R_{AA}(p_T,\phi)$.
Each azimuthal angle $\phi$ can be associated with an average medium path-length $L_{\epsilon}$ 
via a Glauber model. Figure~\ref{fig:Sloss_L} (right) shows the measured fractional energy loss 
$\epsilon_{loss}(\phi)$, obtained via Eq.~(\ref{eq:eloss}), as a function of $L_{\epsilon}$ for pions
in the range $p_T =$~5~--~8~GeV/c (markers of different colours correspond to varying centralities, 
i.e. eccentricities $\epsilon$). The energy loss is found to satisfy the expected $\Delta E_{loss} \propto L$
dependence above a minimum length of $L\approx$~2~fm. The absence of suppression in the surface of
the medium is explained as due to a geometric ``corona'' effect~\cite{Pantuev:2005jt}. 


\subsubsection{(f) Non-Abelian (colour factor) dependence}

The amount of energy lost by a parton in a medium is proportional to its colour Casimir factor 
$C_R$, i.e. $C_A =$~3 for gluons, $C_F$~=~4/3 for quarks. Asymptotically, the probability 
for a gluon to radiate another gluon is $C_A/C_F$ = 9/4 times larger than for a quark and, thus, 
$g$-jets are expected to be more quenched than $q$-jets in a QGP. One can test such a genuine 
{\it non-Abelian} property of QCD energy loss  in two ways:
\begin{description}
\item (1) by measuring hadron suppression at a {\it fixed} $p_T$ for {\it increasing} $\sqrt{s}$~\cite{dde_hp04,wang05},
\item (2) by comparing the suppression of high-$p_T$ {\it (anti)protons} (coming mostly from gluon fragmentation)
to that of {\it pions} (which come from both $g$ and $q,\bar{q}$).
\end{description}

\begin{figure}[htbp]
\includegraphics[width=6.cm,height=4.7cm,clip]{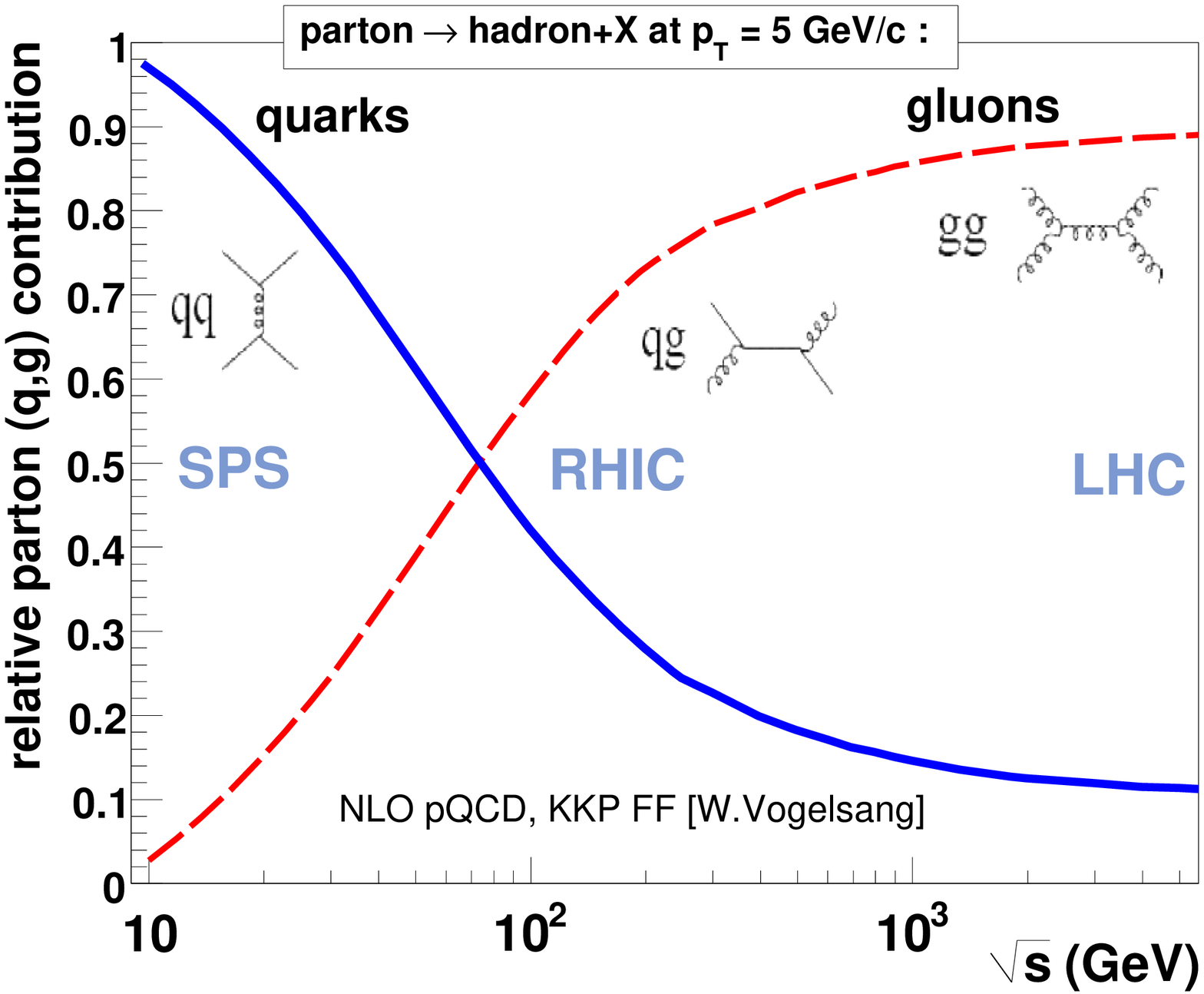}
\includegraphics[width=6.cm,height=4.8cm,clip]{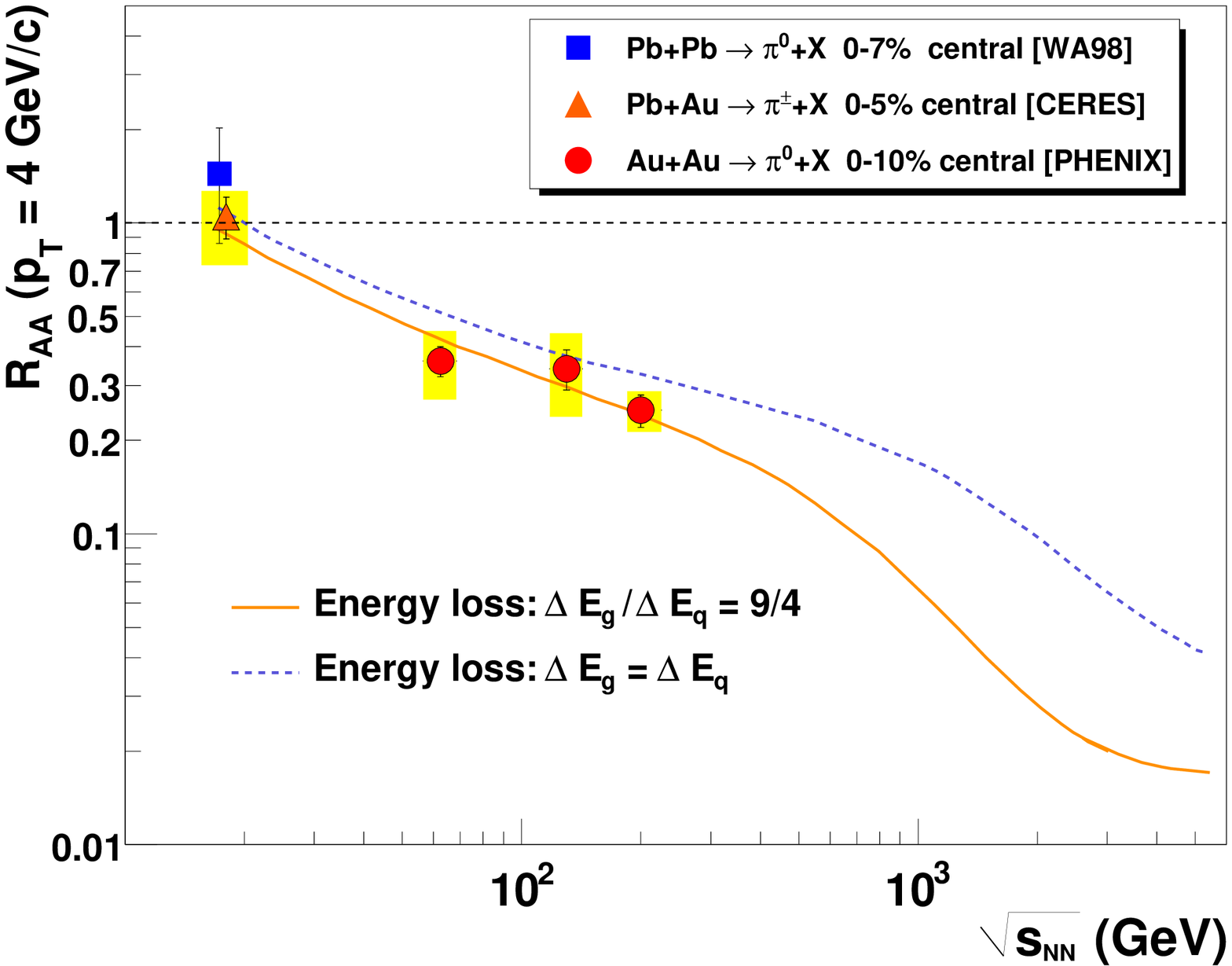}
\caption{Left: Relative fraction of quarks and gluons fragmenting into a hadron at $p_T$~=~5~GeV/c 
in \pp\ collisions in the range $\sqrt{s}$ = 10 -- 5500 GeV given by NLO pQCD~\protect\cite{vogelsang_pi0}.
Right: $R_{AA}(p_T$~=~4~GeV/c) for $\pi^0$  in central \AaAa\ collisions as function of collision energy
compared to non-Abelian (solid) and ``non-QCD'' (dotted) energy loss curves~\cite{dde_hp04,wang05}.}
\label{fig:q_g_RAA_fraction}
\end{figure}

The motivation for (1) is based on the fact that the fraction of quarks and gluons scattered 
at midrapidity in a \pp\ or \AaAa\ collision 
at a {\it fixed} $p_T$ varies with $\sqrtsnn$ 
in a proportion given\footnote{The different ``hardness'' of quarks and gluons fragmenting into
a given hadron at the corresponding $z=p_{hadron}/p_{parton}$ plays also a (smaller) role.} by the relative density 
of  $q,\bar{q}$ and $g$ at the corresponding Bjorken $x = 2p_T/\sqrts$ in the proton/nucleus. 
At large (small) $x$, the hadronic PDFs are dominated by valence-quarks (by ``wee'' gluons) and 
consequently hadroproduction is dominated by quark (gluon) scatterings. 
A full NLO calculation~\cite{vogelsang_pi0} (Fig.~\ref{fig:q_g_RAA_fraction}, left) predicts that
hadrons with $p_T\approx$~5~GeV/c at SPS (LHC) energies are $\sim$100\% produced by 
quarks (gluons), whereas at RHIC they come 50\%-50\% from both species.

Figure~\ref{fig:q_g_RAA_fraction} (right) shows the $R_{AA}$ for 4-GeV/c pions measured 
at SPS and RHIC compared to two parton energy loss curves, both normalised at the $R_{AA}\approx$~1 
measured at SPS and extrapolated all the way up to LHC energies~\cite{wang05}. 
The lower curve shows the expected $R_{AA}$ assuming the normal non-Abelian behaviour 
($\Delta E_g/\Delta E_q$~=~9/4).
The upper (dotted) curve shows an arbitrary prescription in which quarks and gluons lose the same 
energy ($\Delta E_g = \Delta E_q$). 
Above $\sqrtsnn\approx$~100~GeV, gluons take over as the dominant parent parton 
of hadrons with $p_T\approx$~5~GeV/c and, consequently, the $R_{AA}$ values 
drop faster in the canonical non-Abelian scenario. The experimental high-$p_T$ $\pi^0$ data 
thus supports the expected colour-factor dependence of $R_{AA}(\sqrtsnn)$~\cite{dde_hp04}.\\

\begin{figure}[htbp]
\hspace{-0.4cm}
\includegraphics[width=6.cm,height=4.9cm,clip]{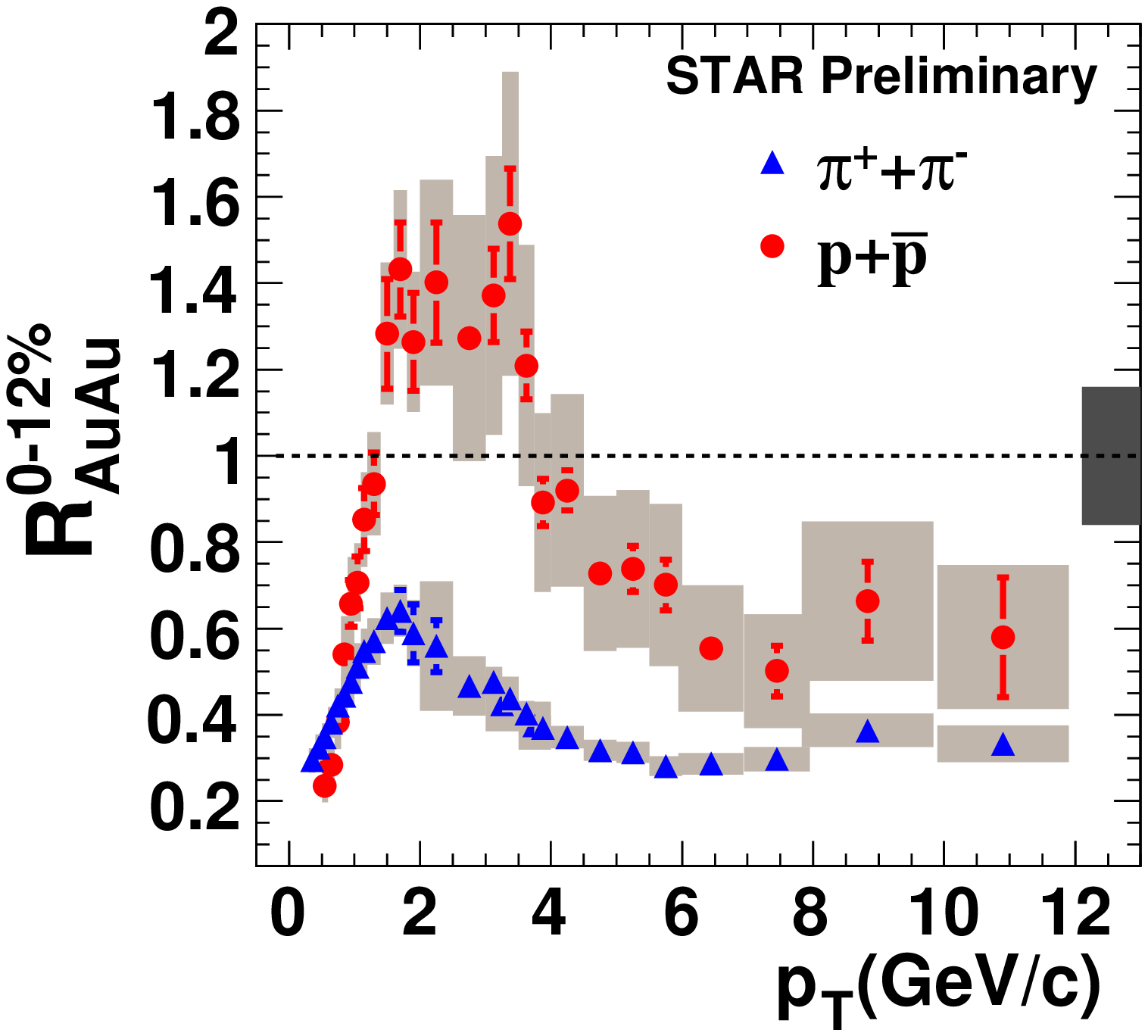}\hspace{0.2cm}
\includegraphics[width=6.cm,height=4.9cm]{figs/denterria_qg_Eloss_ivitev08.eps}\vspace{-0.2cm}
\caption{Left: $R_{AA}(p_T)$ for pions and (anti)protons in central \AuAu\ at $\sqrtsnn$~=~200~GeV~\protect\cite{Mohanty:2008tw}.
Right: Comparison between $\Delta E_g$ and $\Delta E_q$ in central  collisions of 
large nuclei at RHIC and LHC showing large deviations from $\Delta E_g = 2.25 \Delta E_q$ 
for finite parton energies~\cite{Vitev:2008jh}.}
\label{fig:hipt_ppbar}
\end{figure}

The second test of the colour charge dependence of hadron suppression is based on the fact 
that gluons fragment comparatively more into (anti)protons than quarks do. One would thus
naively expect $R_{AA}^{p,\bar{p}}<R_{AA}^{\pi}$.
The STAR results (Fig.~\ref{fig:hipt_ppbar}, left) are however seemingly at variance 
with this  expectation: pions appear more suppressed than protons at high-$p_T$~\cite{Mohanty:2008tw}. 
The use of (anti)protons as a reference for perturbative particle production is however questionable: 
$p,\bar{p}$ are already found to be enhanced in \dAu\ compared to \pp\ collisions by a factor 
$\sim 50$\% -- 100\% for $p_T$'s as large as 7~GeV/c~\cite{Adams:2006nd}. It is likely that there 
is an extra mechanism of baryon production, based e.g. on in-medium quark coalescence~\cite{reco},
which compensates for the energy loss suffered by the parent partons. It is also
important to stress that the $\Delta E_g/\Delta E_q$~=~9/4 expectation holds only for
asymptotic parton energies. Finite energy constraints yield 
$\Delta E_g/\Delta E_q\approx$~1.5 for realistic kinematics (Fig.~\ref{fig:hipt_ppbar}, right)~\cite{pqm,Vitev:2008jh}.


\subsubsection{(g) Heavy-quark mass dependence}

\begin{figure}[htbp]
\includegraphics[width=0.49\linewidth,height=5.cm,clip]{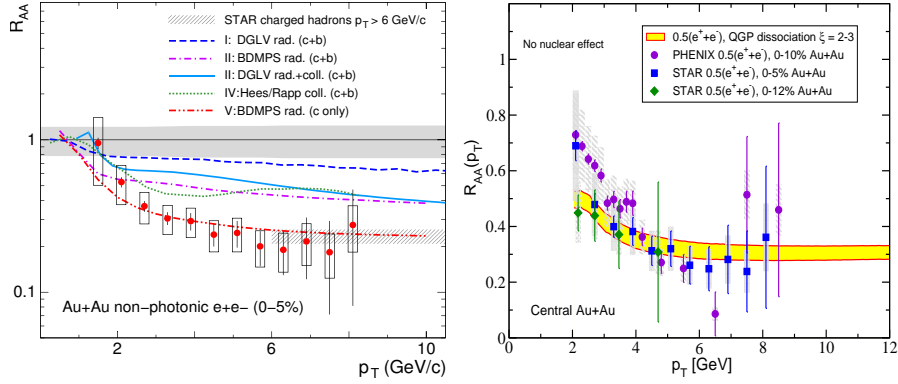}\hspace{0.15cm}
\includegraphics[width=0.49\linewidth,height=5.cm,clip]{figs/denterria_RAA_nonphot_elec_adilvitev.eps}
\caption{
$R_{AA}(p_T)$ for decay electrons from $D$ and $B$ mesons in central \AuAu\ at 
$\sqrt{s_{NN}}=200$~GeV~\cite{Adler:2005xv,Adare:2006nq,Abelev:2006db} 
compared to various radiative+elastic energy loss models for $c$ and $b$ quarks (left) 
and to a model of $D$ and $B$ meson dissociation in the plasma~\cite{Adil:2006ra} (right).}
\label{fig:heavyflavourRAA}
\end{figure}

A robust prediction of QCD energy loss models is the hierarchy $\Delta E_{Q} < \Delta E_{q} <  \Delta E_{g}$.
Due to the dead-cone effect, the radiative energy loss for a charm (bottom) quark is $\sim$25\% (75\%)
less than for a light-quark (see Section~\ref{sec:intro_theory}). Surprisingly, PHENIX and STAR measurements of high-$p_T$
electrons from the semi-leptonic decays of $D$- and $B$-mesons (Fig.~\ref{fig:heavyflavourRAA}) indicate that their suppression 
is comparable to that of light mesons: $R_{AA}(Q)\sim R_{AA}(q,g)\approx$~0.2~\cite{Adler:2005xv,Adare:2006nq,Abelev:2006db}. 
Such a low $R_{AA}$ cannot be described by radiative energy loss calculations with the same initial gluon 
densities or transport coefficients needed to reproduce the quenched light hadron spectra~\cite{djordj04,Armesto:2005mz}.

Various explanations have been proposed to solve such a `heavy flavor puzzle':
\begin{itemize}
\item First, if only $c$ quarks (three times more suppressed than the heavier $b$ quarks) actually contributed 
to the measured high-$p_T$ decay electron spectrum, then one would indeed expect $R_{AA}(c)\approx$~0.2~\cite{armesto05}. 
Yet, indirect measurements from PHENIX~\cite{Adare:2009ic} and STAR~\cite{Mischke:2008qj} have confirmed 
the similar production yields of $e^\pm$ from $D$ and $B$ decays above $p_T\approx$~5~GeV/c predicted by NLL pQCD~\cite{Cacciari:2005rk}.
\item The heavy-quark suppression has revived the interest of computing {\it elastic} energy loss
in a QGP~\cite{Mustafa:1997pm,Mustafa:2004dr,Peshier:2006hi,Djordjevic:2006tw,Peigne:2008nd}. As discussed in Sect.~\ref{sec:intro_theory}, 
$\Delta E_{coll}$ can indeed be a significant contribution for heavy quarks (see `rad.+coll.' curves in Fig.~\ref{fig:heavyflavourRAA}, left). 
\item 
The strongly-coupled nature of the plasma at RHIC would lead, according to AdS/CFT
calculations~\cite{heavyQ_adscft,CasalderreySolana:2006rq,CasalderreySolana:2007qw,Gubser:2006bz,CaronHuot:2008uh}, 
to a larger heavy-quark momentum diffusion parameter than expected in perturbative approaches~\cite{vanHees:2007me}. 
This could explain the larger charm/bottom quenching observed in the data.
\item Two works~\cite{Sorensen:2005sm,MartinezGarcia:2007hf} have argued that the large charm-quark coalescence into 
$\Lambda_c$ baryons (with a small semileptonic decay branching ratio) in the plasma would deplete the number of 
open-charm mesons, and correspondingly reduce the number of decay electrons, compared to \pp\ collisions.
\item The assumption of vacuum hadronisation (after in-medium radiation) implicit in all parton energy loss
formalisms may well not hold in the case of a heavy quark. All existing quark-hadronisation time estimates~\cite{accardi09} 
are inversely proportional to the mass $m_h$ of the final produced hadron: the heavier the hadron, the fastest it is
formed. In the rest frame\footnote{Note that in the laboratory system there is an extra Lorentz boost factor: 
$\tau_{_{\ensuremath{\it lab}}} = \gamma_Q\cdot\tau_{_{form}}$.} of the fragmenting heavy-Q, the formation time of $D$- and $B$-mesons~\cite{Adil:2006ra}
\begin{equation}
\tau_{_{form}} = \frac{1}{ 1+\beta_Q} \frac{ 2z(1-z)p^+}{  {\bf k}^2 + (1-z)m_h^2 - z(1-z)m_Q^2   } \;,  \;\mbox{ where $\beta_Q = {p_Q}/{E_Q}$,}
\label{tfrag}
\end{equation}
is of order $\tau_{_{form}}\approx$~0.4 -- 1 fm/c respectively. Thus, theoretically, one needs to account for both
the energy loss of the heavy-quark as well as the possible dissociation of the heavy-quark {\it meson} 
inside the QGP. The expected amount of suppression in that case is larger and consistent with the data (Fig.~\ref{fig:heavyflavourRAA}, right).
\end{itemize}


\section{High-$p_T$ di-hadron $\phi,\eta$ correlations: data vs. theory}
\label{sec:highpTcorrs}

Beyond the leading hadron spectra discussed in the previous Section, detailed studies of the
modifications of the jet structure in heavy-ion collisions have been addressed via high-$p_T$ 
multi-particle (mostly di-hadron) $\phi,\eta$ correlations at RHIC (and to a lesser extent at SPS). 
Jet-like correlations are measured on a {\it statistical} basis by selecting high-$p_T$ {\it trigger} 
particles and measuring the azimuthal ($\Delta\phi = \phi - \phi_{trig}$) and pseudorapidity 
($\Delta\eta = \eta - \eta_{trig}$) distributions of {\it associated} hadrons 
($p_{T}^{assoc}<p_{T}^{trig}$) relative to the trigger:
\begin{equation}
C(\Delta\phi,\Delta\eta) = \frac{1}{N_{trig}}\frac{d^2N_{pair}}{d\Delta\phi d\Delta\eta}.
\end{equation}
Combinatorial background contributions, corrections for finite pair acceptance, and
the superimposed effects of {\it collective} azimuthal modulations (elliptic flow)
can be taken care of with different techniques~\cite{star_hipt_awayside,phenix_machcone,star_hipt_etaphi}.
A commonly-used $C(\Delta\phi)$ background-subtraction method  is the so-called ``zero yield at minimum'' (ZYAM)~\cite{zyam}.
\begin{figure}[htbp]
\centering
\includegraphics[width=10.2cm]{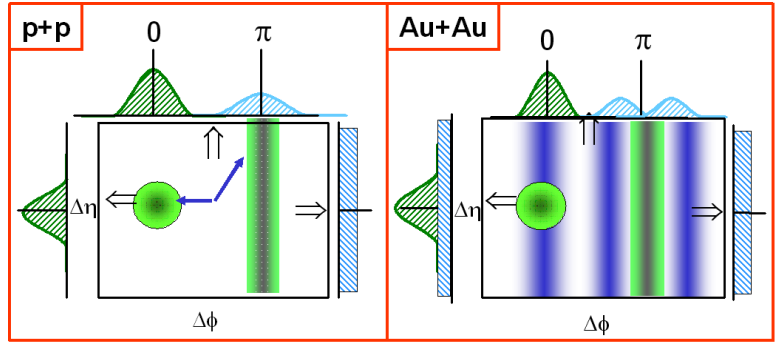}
\caption{Schematic illustration summarising the jet-induced di-hadron correlation signals in $\Delta\phi$ and $\Delta\eta$ 
observed in \pp\ (left) and central \AuAu\ (right) at $\sqrtsnn$~=~200~GeV~\protect\cite{Jia:2008kf}.}
\label{fig:dNdphideta}
\end{figure}

A schematic representation of the di-hadron azimuthal-pseudorapidity correlations
$dN_{pair}/d\Delta\phi d\Delta\eta$ measured in \pp\ and central \AuAu\ collisions is shown in
Fig.~\ref{fig:dNdphideta}. In the \pp\ case, without significant initial- or final- state interactions, 
a dijet signal appears clearly as two distinct back-to-back Gaussian-like peaks at $\Delta\phi\approx$ 0, $\Delta\eta\approx$ 0 
(near-side) and at $\Delta\phi\approx\pi$ (away-side). Note that the away-side peak is naturally broader in 
$\Delta\eta$ (up to $\Delta\eta \approx 2$) than the near-side peak due to the {\it longitudinal}
momentum  imbalance between the two colliding partons with different $x_1, x_2$ momentum 
fractions (the collision is boosted in $\eta$ by an amount $\ln(x_1/x_2)$ in the direction of the larger $x_{1,2}$).
At variance with such a standard dijet topology, the di-hadron correlations in \AuAu\ reactions 
at RHIC show several striking features, discussed in detail below:
\begin{itemize}
\item  The {\bf away-side} azimuthal peak at $\Delta\phi\approx\pi$  is {\bf strongly suppressed} with 
increasing centrality 
for hadrons with $p_{T}^{assoc} \gtrsim 2$~GeV/c, consistent with strong 
suppression of the leading fragments of the recoiling jet traversing the medium~\cite{star_hipt_awayside}.
\item The vanishing of the away-side peak 
is accompanied with an {\it enhanced} production of {\it lower} $p_T$ hadrons 
($p_{T}^{assoc} \lesssim$~2~GeV/c)~\cite{phenix_machcone}
with a characteristic {\bf double-peak} structure at $\Delta\phi\approx\pi\,\pm$~1.1 -- 1.3~\cite{phenix_machcone,star_hipt_etaphi}. 
\item One observes a large {\bf broadening}  (``ridge''), out to $\Delta\eta\approx$~4, 
of the {\bf near-side pseudo-rapidity} $dN_{pair}/d\Delta\eta$ correlations~\cite{star_hipt_etaphi}.
\end{itemize}


\subsection{Azimuthal correlations: away-side quenching and energy loss}
\label{sec:IAA}

\begin{figure}[htbp]
\centering
\includegraphics[width=12.cm]{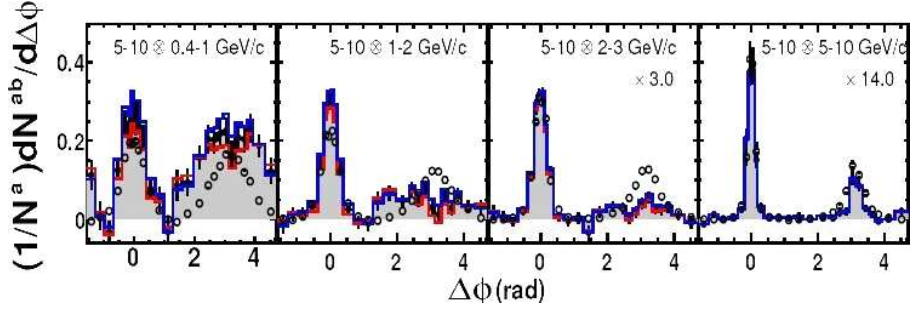}
\caption{Comparison of the azimuthal di-hadron correlation $dN_{pair}/d\Delta\phi d\eta$ for \pp\ (open symbols) 
and central \AuAu\ (histograms) at $\sqrtsnn$~=~200~GeV for $p_T^{trig}=$~5--10~GeV/c and increasingly
smaller (right to left) values of $p_T^{assoc}$~\protect\cite{Adare:2008cq}.}
\label{fig:dNdphi}
\end{figure}

Figure~\ref{fig:dNdphi} shows the increasingly distorted back-to-back azimuthal correlations 
in high-$p_T$ triggered central \AuAu\ events as one decreases the $p_T$ of the associated hadrons
(right to left). Whereas, the \AuAu\ and \pp\ near-side peaks are similar for all $p_T$'s, 
the away-side peak is only present for the highest partner $p_T$'s but progressively disappears 
for less energetic partners~\cite{Adare:2007vu,Adare:2008cq}.
Early STAR results~\cite{star_hipt_awayside} showed a monojet-like topology with a complete 
disappearance of the opposite-side peak for $p_{T}^{assoc}\approx$~2~--~4~GeV/c.

\begin{figure}[htbp]
\includegraphics[width=6.1cm,height=5.7cm,clip]{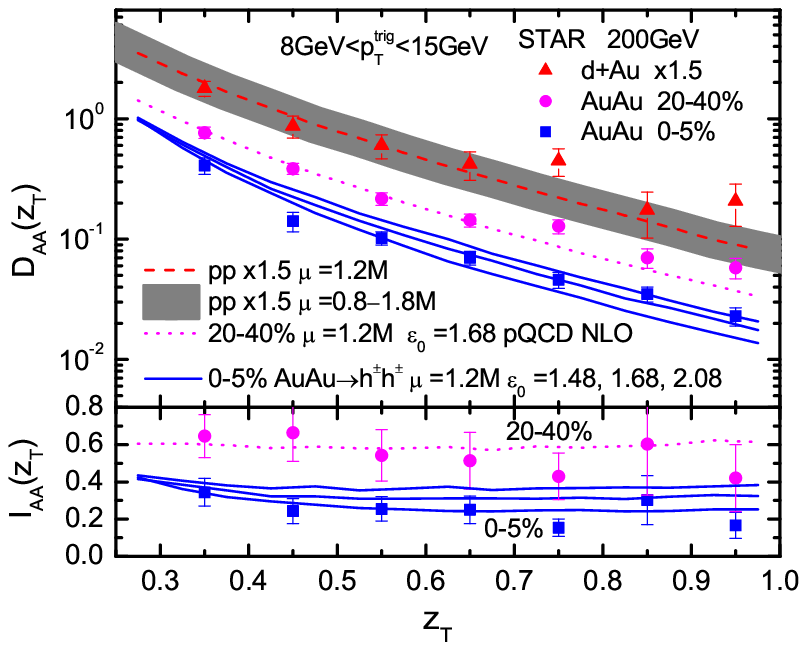}
\includegraphics[width=6.1cm,height=5.8cm,clip]{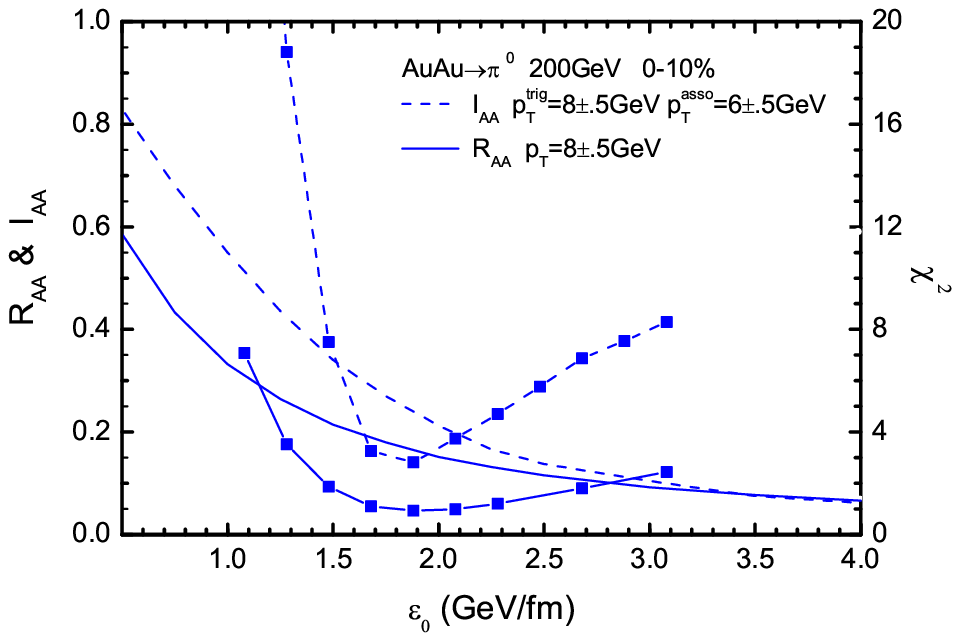}
\caption{Left: $D^{away}_{AA}(z_T)$ distributions for \dAu\ and \AuAu\ collisions at 200~GeV
and $I_{AA}(z_T)$ ratio (for central \AuAu)~\cite{star_jet_punchthrough}, compared to HT 
calculations~\cite{Zhang:2007ja} for varying $\epsilon_0$ energy loss parameter. 
Right: Corresponding (data vs. theory) $\chi^2$ values for the fitted $\epsilon_0$ 
parameters~\cite{Zhang:2007ja}.}
\label{fig:IAA_qhat}
\end{figure}

For any range of trigger $p_{T}^{trig}$  and associated $p_{T}^{assoc}$ intervals, the correlation strength 
over an azimuthal range $\Delta\phi$ between a trigger hadron $h_{t}$ and a partner hadron 
$h_{a}$ in the opposite azimuthal direction can be constructed as a function of the momentum fraction 
$z_T=p_{T}^{assoc}/p_{T}^{trig}$ via a ``pseudo-fragmentation function''~\cite{Tannenbaum:2007qx}:
\begin{equation}
D^{away}_{AA} (z_T) =
\int_{p^{trig}_{T,min}}^{p^{trig}_{T,max}}dp_{T}^{trig}
\int_{p^{assoc}_{T,min}}^{p^{assoc}_{T,max}}dp_{T}^{assoc}
\int_{away} d\Delta\phi\; \frac{d^3\sigma_{AA}^{h_t h_a}/d p_{T}^{trig}d p_{T}^{assoc}
d\Delta\phi}{d\sigma_{AA}^{h_t}/d p_{T}^{trig}}\,.
\label{eq:D_AA}
\end{equation}
Figure~\ref{fig:IAA_qhat} (left) shows the measured $D^{away}_{AA}$ distributions 
for \pp\ and \AuAu\ collisions 
as a function of $z_T$ compared to predictions of the HT jet-quenching model for various 
values of the $\epsilon_0$ parameter quantifying the amount of parton energy loss~\cite{Zhang:2007ja}.
Similarly to $R_{AA}$, the magnitude of the suppression of back-to-back jet-like two-particle 
correlations can be quantified with the ratio $\IAA(z_T) = D_{AA}(z_T)/D_{pp}(z_T)$. 
$\IAAa(z_T)$ (bottom-left panel of Fig.~\ref{fig:IAA_qhat}) is found to decrease with increasing centrality, 
down to about 0.2~--~0.3 for the most central events~\cite{star_hipt_awayside,star_jet_punchthrough}. 
The right plot of Fig.~\ref{fig:IAA_qhat} shows the best $\epsilon_0\approx$~1.9~GeV/fm value that fits the 
measured $R_{AA}$ and $I_{AA}$ factors. Due to the irreducible presence of (unquenched) partons emitted 
from the surface of the plasma, the leading-hadron quenching factor $R_{AA}(p_T)$ is in general 
less sensitive to the value of $\epsilon_0$ than the dihadron modification ratio $I_{AA}(z_T)$. 
The combination of $R_{AA}(p_T)$ and $I_{AA}(z_T)$ provides robust quantitative information on 
the medium properties.


\subsection{Azimuthal correlations: away-side broadening and ``conical'' emission}

\begin{figure}[htbp]
\includegraphics[width=7.cm,height=4.6cm,clip]{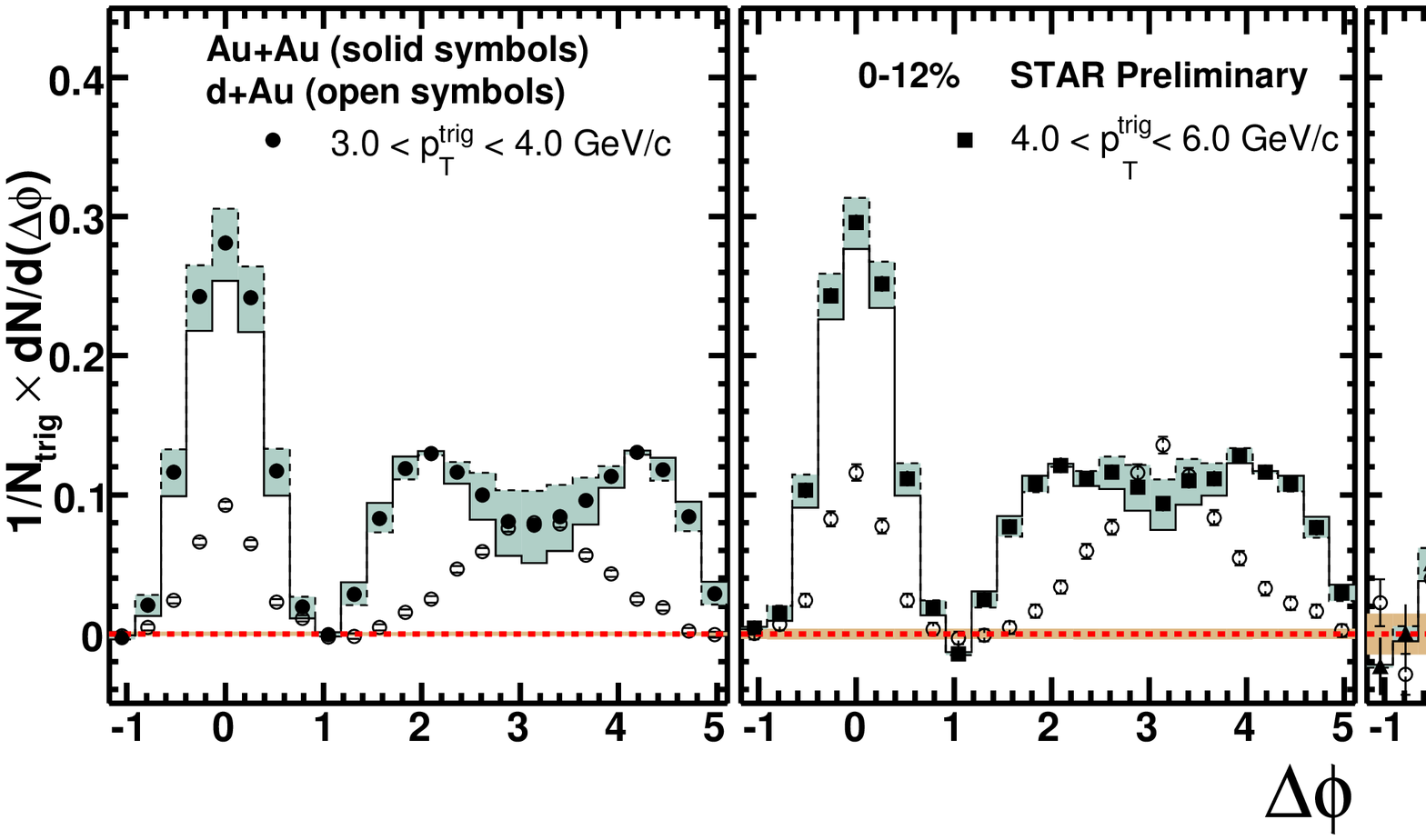}
\includegraphics[width=4.7cm,height=4.9cm,clip]{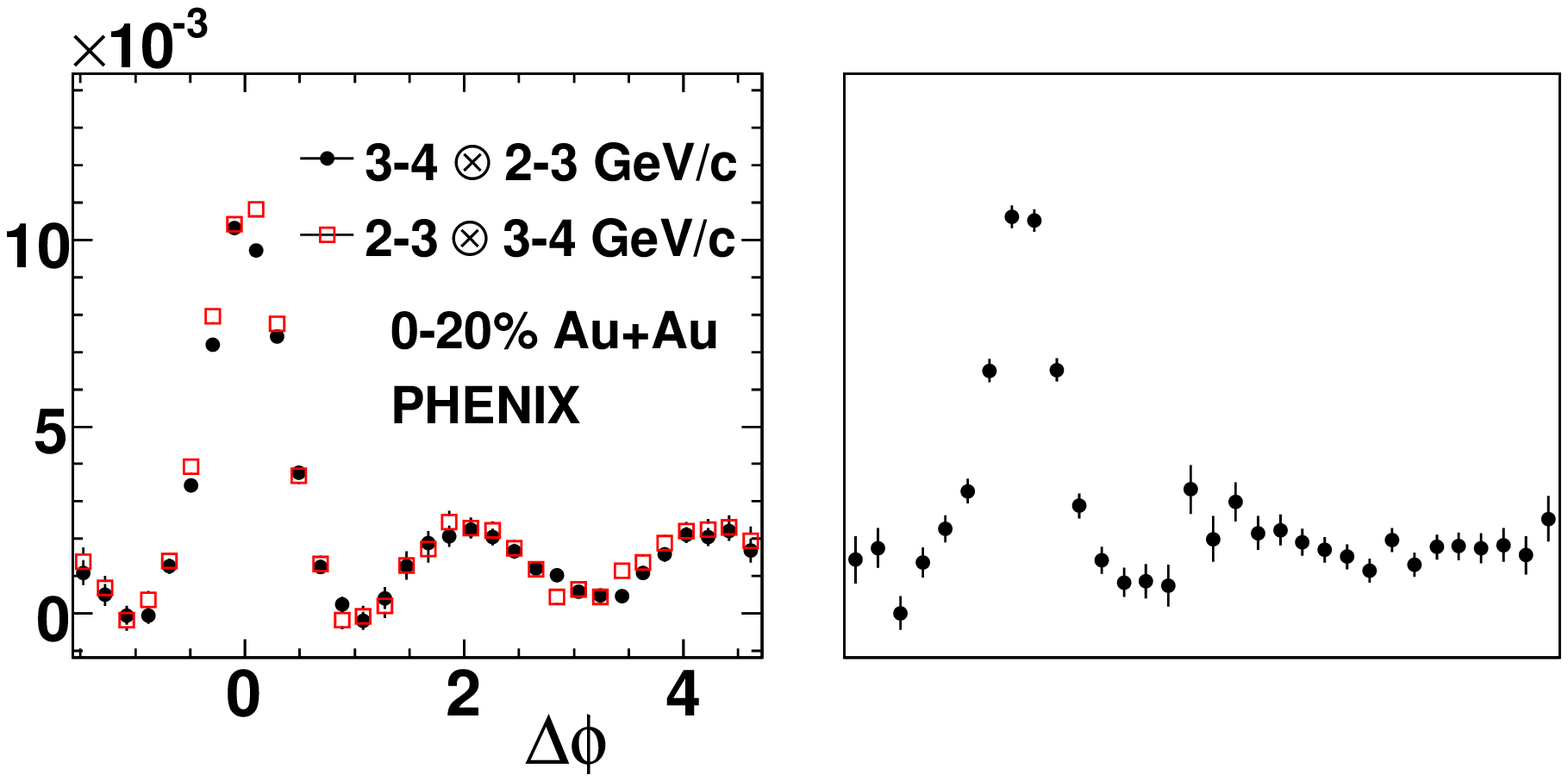}
\caption{Normalised azimuthal dihadron distributions, $1/N_{trig}\,dN_{pair}/d\Delta\phi$.
Left: STAR data in central $AuAu$ (squares) and $dAu$ (circles) for $p_{T}^{assoc}$~=~1.3 -- 1.8~GeV/c and 
two ranges of $p_{T}^{trig}$~\protect\cite{star_machcone1}.
Right: PHENIX results in central \AuAu\ for various $p_{T}^{trig,assoc}$ ranges~\protect\cite{Adare:2008cq}.}
\label{fig:dNdphi_cone}
\end{figure}

Since energy and momentum are conserved, the ``missing''  fragments of the away-side (quenched) parton 
at intermediate $p_T$'s should be either shifted to lower energy ($p_T\lesssim$~2~GeV/c) and/or 
scattered into a broadened angular distribution. Both, softening and broadening, 
are seen in the data when the $p_T$ of the away-side associated hadrons is {\it lowered} (see two
leftmost panels of Fig.~\ref{fig:dNdphi}). Figure~\ref{fig:dNdphi_cone} shows in detail the dihadron 
azimuthal correlations $dN_{pair}/d\Delta\phi$ 
in central $AuAu$ collisions~\cite{Adare:2008cq,star_machcone1}: 
the away-side hemisphere shows a very unconventional angular distribution with a ``dip'' 
at $\Delta\phi \approx \pi$ and two neighbouring local maxima at 
$\Delta\phi \approx \pi\,\pm$~1.1~--~1.3. 

Such a ``volcano''-like profile has been interpreted as due to the preferential emission of energy 
from the quenched parton at a finite angle with respect to the jet axis. This could happen in a purely radiative 
energy loss scenario due to large-angle radiation~\cite{Polosa:2006hb}, but more intriguing explanations 
have been put forward based on the dissipation of the lost energy into a {\it collective} mode 
of the medium in the form of a wake of lower energy gluons with Mach-~\cite{mach1,mach2,rupp05} or 
\v{C}erenkov-like~\cite{rupp05,cerenkov1,cerenkov2} angular emissions. 

\begin{figure}[htbp]
\centering
\includegraphics[width=5.cm,height=4.5cm]{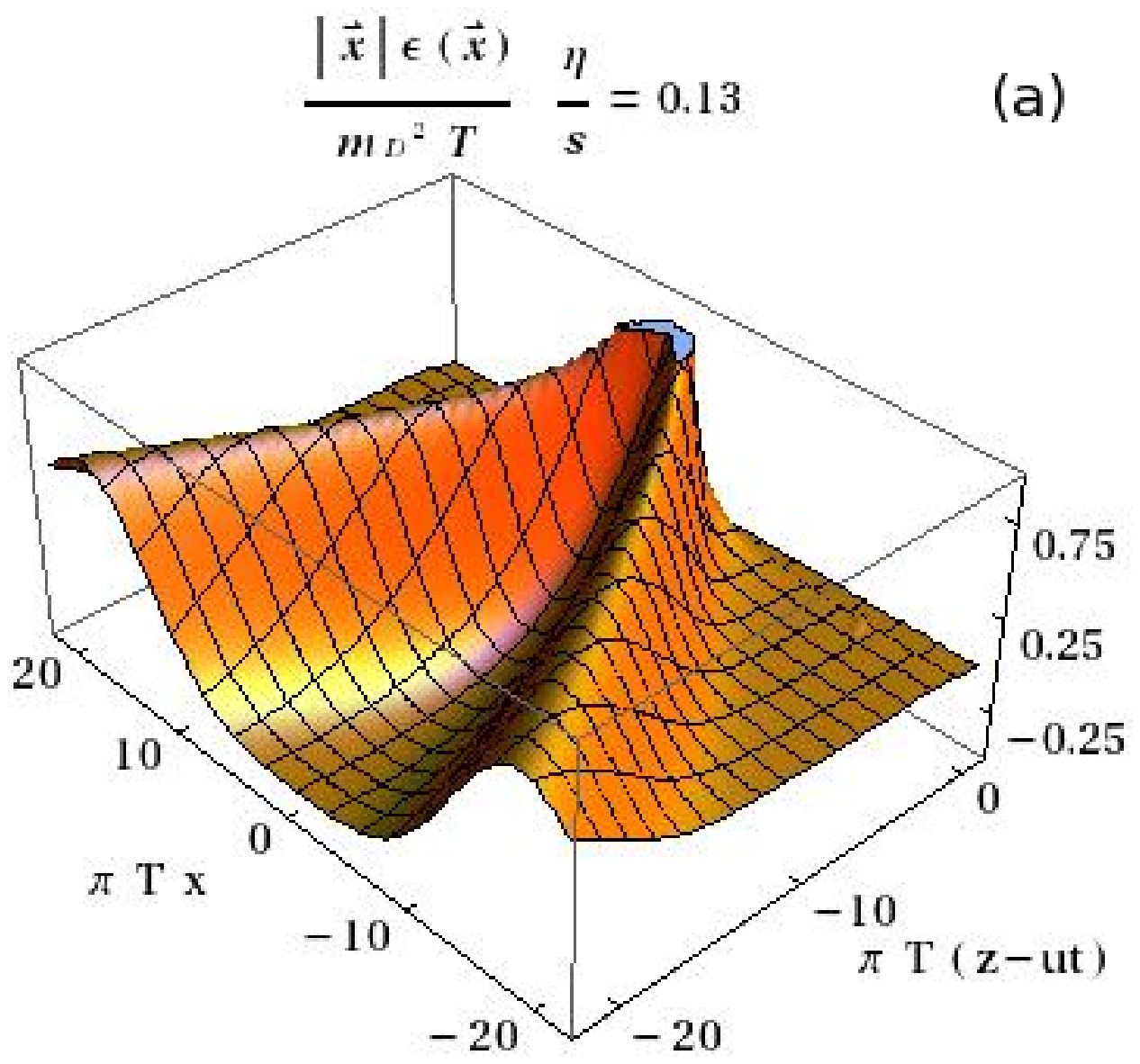}
\includegraphics[width=5.cm,height=4.5cm]{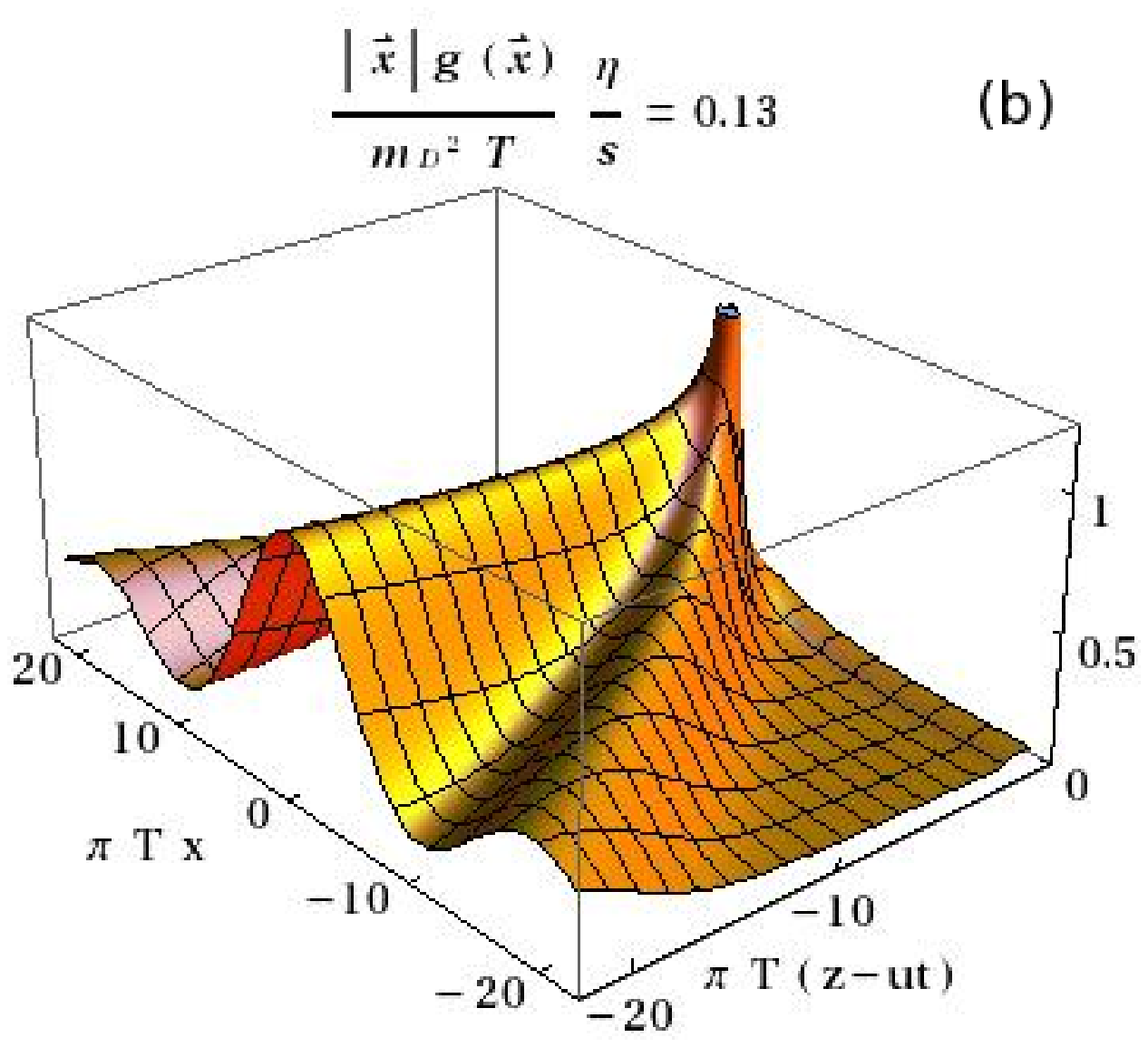}
\includegraphics[width=10.cm,height=3.7cm]{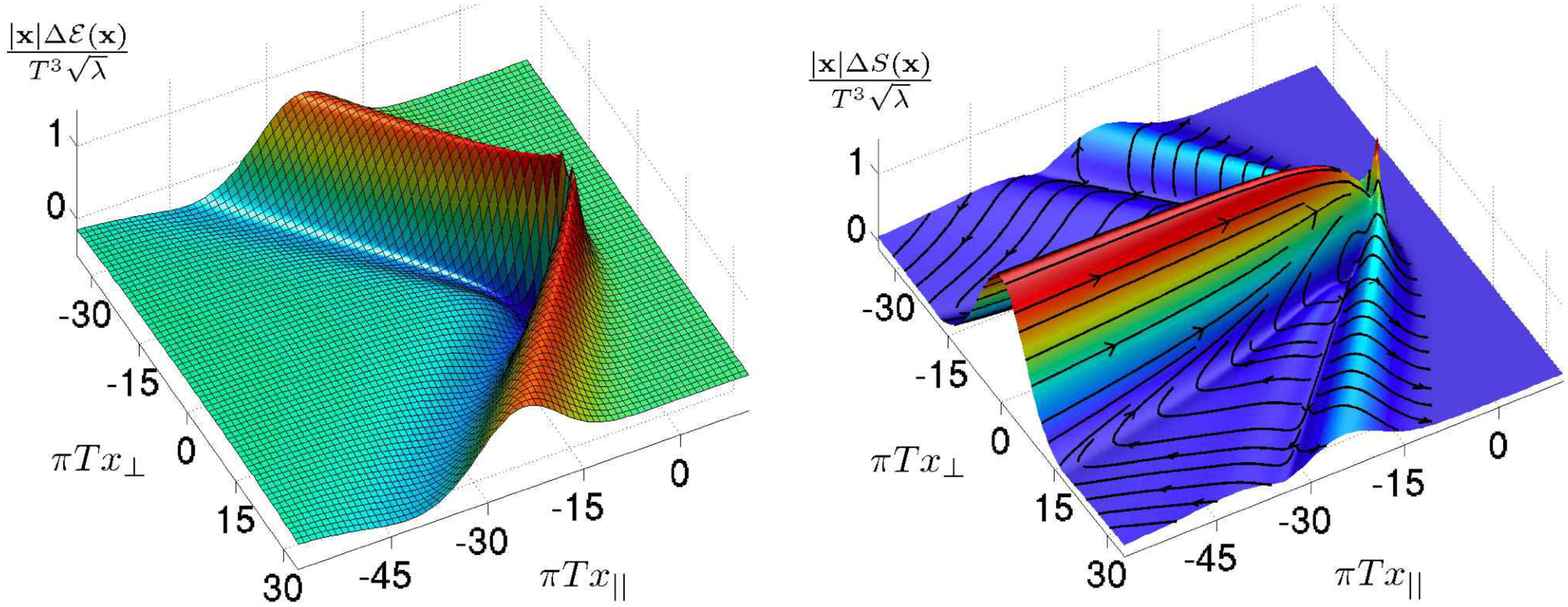}
\caption{Top: Perturbed energy (left) and momentum (right) densities for a gluon moving 
with $\beta=0.99955$ in a perturbative QGP~\protect{\cite{Neufeld1}}. Bottom: Perturbed energy density (left) 
and energy flux (Pointing vector, right) for a jet with $\beta=0.75$ from an AdS/CFT model~\protect\cite{Chesler:2007sv}.}
\label{fig:pQCD_AdSCFT}
\end{figure}
In the Mach cone scenario~\cite{mach1,mach2,rupp05}, the local maxima in central \AuAu\ 
are caused by the Mach shock of the supersonic recoiling parton traversing the medium 
with a resulting preferential emission of secondary partons from the plasma at an angle $\theta_{M}$ 
(Fig.~\ref{fig:pQCD_AdSCFT}). 
Such a mechanism would give access, via Eq.~(\ref{eq:mach}), to the speed sound $c_{s}$ of the traversed matter.
In an expanding plasma, the speed of sound changes from $c_s=1/\sqrt{3}$ (QGP) to $c_s\approx \sqrt{0.2}$ 
(hadron gas) through $c_s=0$ (mixed phase) in the case of a first-order phase transition. 
The time-averaged value is $\mean{c_s} =\frac{1}{\tau}\int_0^\tau dt \,c_s(t)\,\approx 0.3$~\cite{mach2}
with a resulting Mach angle $\theta_M$~=~arccos$(c_s)\;\approx$~1.3, see Eq.~(\ref{eq:mach}), 
in rough agreement with the experimental data.\\

In the \v{C}erenkov picture~\cite{rupp05,cerenkov1,cerenkov2}, it is argued that the combination 
of the LPM gluonstrahlung interference and a medium with a {\it large} dielectric 
constant ($n\approx$ 2.75 is needed in Eq.~(\ref{eq:cerenkov}) to reproduce the location of the 
experimental peaks), would also result in the emission of QCD \v{C}erenkov radiation with the 
double-hump structure observed in the data. However, at variance with the Mach angle 
which is constant in the fluid, 
the \v{C}erenkov angle {\it decreases} with the momentum of the radiated gluon. 
Such a trend is seemingly in disagreement with the fact that 
the measured $\theta_{c}$  remains relatively constant as a function of 
$p_{T}^{assoc}$~\cite{Adare:2008cq,star_machcone2}. In addition, STAR~\cite{star_3particles_conical} 
and PHENIX~\cite{Ajitanand:2006is} {\it three}-particle correlations studies, seem to clearly favour the 
{\it conical} over deflected-jets interpretation. \\ 





Theoretically, the disturbance of the energy-momentum tensor caused by a light-quark has been studied
in a perturbative plasma~\cite{Neufeld1} as well as for heavy-quark in a $\mathcal{N}=4$ SYM plasma~\cite{gubsermach,Chesler:2007sv}. 
In  both cases a clear conical structure 
as well as a strong flow generated along the path of the jet (diffusion wake~\cite{mach1,Betz:2008ka})
are observed (Fig.~\ref{fig:pQCD_AdSCFT}). The results are sensitive to the viscosity of the medium. 
Yet, it is unclear if phenomenologically such partonic collective wake(s) and cone survive both hadronisation and the final hadronic 
freeze-out~\cite{Betz:2008ka,Noronha,Betz_pQCD_AdSCFT,Neufeld2}. 
Results for a pQCD plasma~\cite{Betz_pQCD_AdSCFT} 
indicate that the conical signal does not survive freeze-out: a peak at $\Delta\phi$~=~$\pi$ 
appears due to the strong 
diffusion wake. More involved studies, accounting for e.g. the plasma expansion and the 
hadronic phase evolution, are needed before a final conclusion can be reached. 


\subsection{Pseudo-rapidity correlations: near-side ``ridge''}

\begin{figure}[htb]
\includegraphics[width=6.0cm,clip]{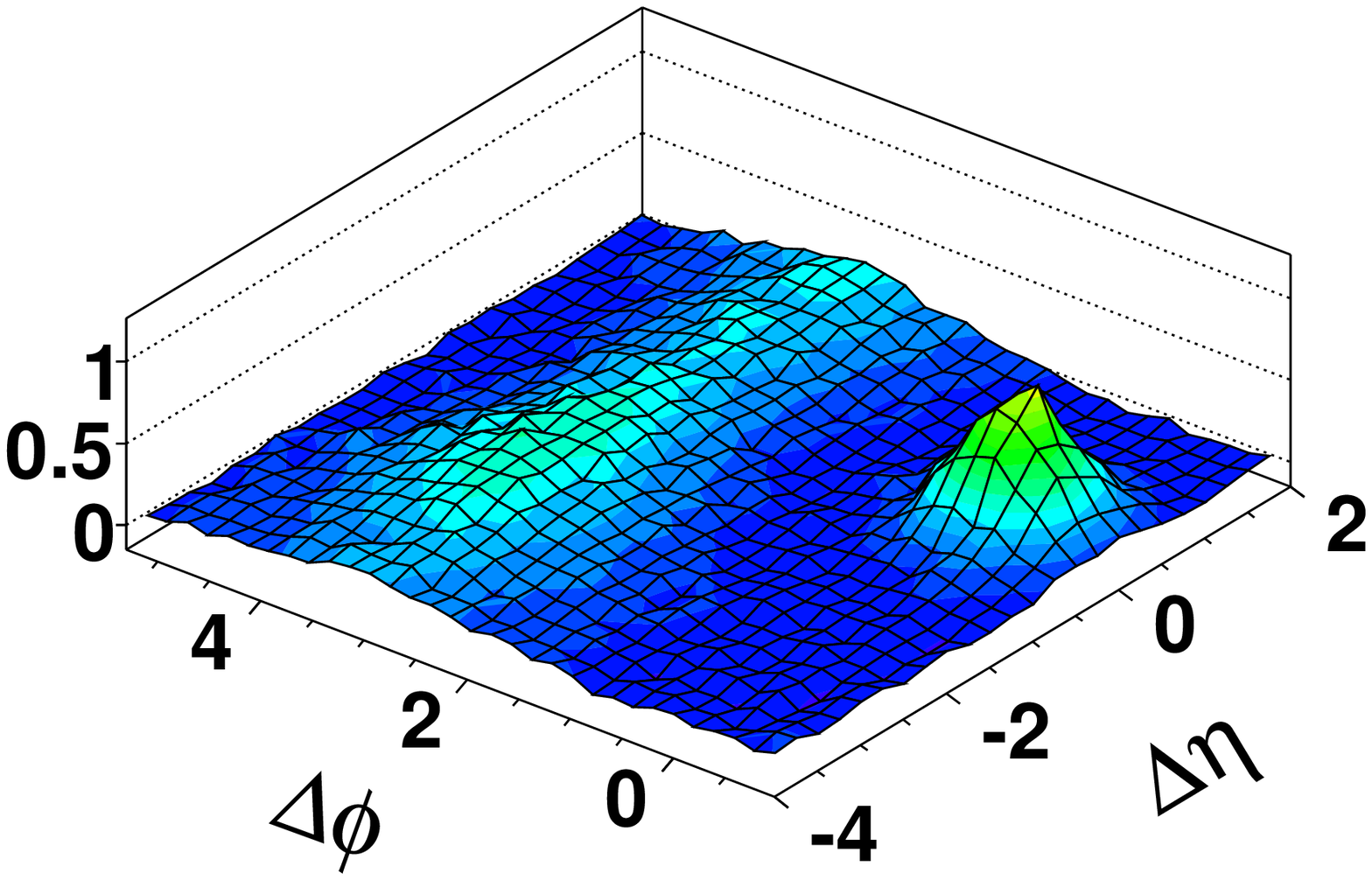} 
\includegraphics[width=6.0cm,clip]{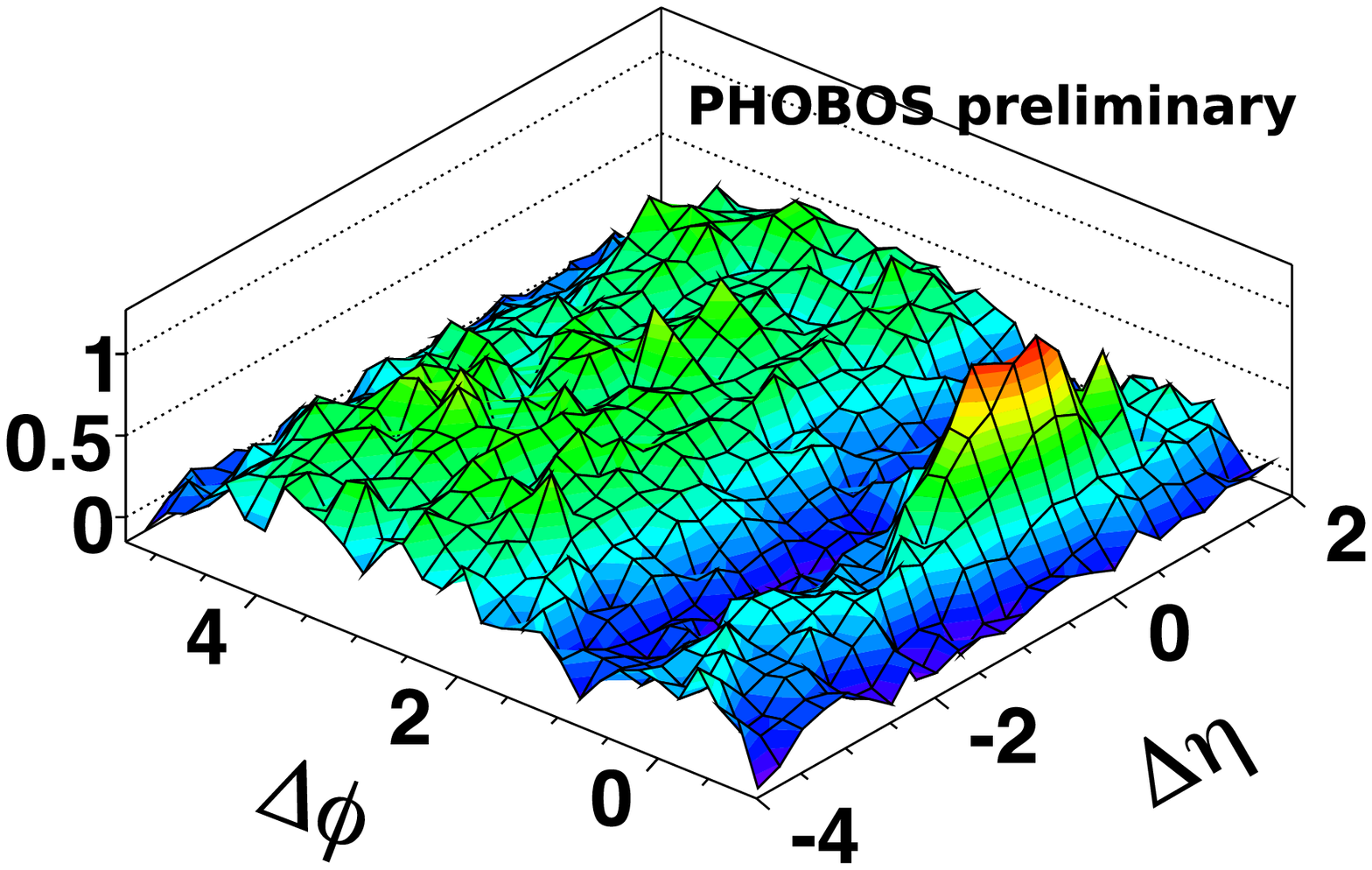} 
\caption{Per-trigger associated hadron yield for $p^{trig}_{T} >$~2.5~GeV/c as a function 
of $\Delta\eta$ and $\Delta\phi$ for \pp\  (\pythia, left) and 0-30\% central \AuAu\ 
(PHOBOS, right) collisions at 200~GeV~\cite{Wenger:2008ts}.}
\label{fig:ridge_phobos}
\end{figure}

Figure~\ref{fig:ridge_phobos} shows the associated $\Delta\eta$-$\Delta\phi$ particle yield 
(down to very low $p_{T}^{assoc} \gtrsim$~20~MeV/c) for trigger hadrons 
$p^{trig}_{T} >$~2.5~GeV/c in \pp\ (\pythia\ simulations) and in central \AuAu\ (PHOBOS data)
at 200~GeV. Both distributions show a clear peak at $(\Delta\eta,\Delta\phi) \approx  (0,0)$ 
as expected from jet fragmentation, but the near-side peak in heavy-ion collisions features a wide 
associated yield out to $\Delta\eta\approx$~4, referred to as the ``ridge''~\cite{Putschke:2007mi}. 
The existence of such unique long-range rapidity correlations in the near-side of the {\it trigger} parton 
which is, by construction, the least affected by the medium, is puzzling. 
The properties (particle composition, $p_T$ slope, intra-particle correlations) of this structure
are very similar to those of the soft underlying event in the collision~\cite{vanLeeuwen:2008pn}. 
This clearly suggests that the ridge is formed from bulk matter and not from jet fragments.
Though many different interpretations have been put forward (see e.g.~\cite{Pruneau:2007ua} for a summary), 
models  that do {\it not} require jet triggers for the effect to appear 
-- such as e.g.``glasma'' flux-tubes~\cite{Dumitru:2008wn} or ``trivial'' modifications of the 
2- and 3-particle correlations due to  radial flow~\cite{Pruneau:2007ua} -- seem favoured by the data.


\section{Jet observables in \AaAa\ collisions}
\label{sec:jets}

The measurement of the leading fragments of a jet (single-hadron spectra and/or di-hadron azimuthal 
correlations at high-$p_T$) in \AaAa\ collisions has been covered in detail in the previous Sections. 
In this last Chapter,  we focus on {\it full jet} reconstruction in nuclear reactions.
The study of the energy and particle-multiplicity distributions within a jet issuing from the fragmentation 
of a quenched parton is a powerful tool to study the response of hot and dense QCD matter to fast quark and gluons. 

\subsection{Full jet reconstruction in \AaAa\ collisions}

Experimental reconstruction of jets in hadronic collisions is an involved exercise~\cite{Ellis:2007ib,Buttar:2008jx}
that requires at least three steps:
\begin{description}
\item {\it (1) Clustering algorithm}: 
Hadrons belonging to a given jet are measured in the detector (usually in the cells of hadronic and 
electromagnetic calorimeters) and are clustered together, according to relative ``distances'' in momentum 
and/or space, following an {\it infrared- and collinear-safe} procedure (see below) that can be also 
appropriately applied to ``theoretical'' (Monte Carlo) jets. 
The algorithm needs to be {\it fast} enough to be run over events with very high multiplicities. 
Various jet-finders exist presently that fulfill all such conditions such as e.g. the $k_T$~\cite{Catani:1993hr} 
and SISCone~\cite{Salam:2007xv} algorithms implemented in the \fastjet\ package~\cite{Cacciari:2005hq}.\\
\item {\it (2) Background subtraction}: Jets are produced on top of a large ``underlying event'' (UE) of hadrons 
coming from other (softer) parton-parton collisions in the same interaction. At LHC energies, extrapolating from 
$dE_T/d\eta|_{\eta=0}=$~0.6~TeV measured at RHIC~\cite{ppg019}, one expects a total transverse energy 
of $\sim$1~TeV in 1-unit rapidity at midrapidity in \PbPb. Jet reconstruction is usually carried out with small cone radius 
$R=\sqrt{\Delta\eta^2+\Delta\phi^2}$~=~0.3~--~0.5 (or similar $k_T$-distances, $D$) to minimise the
UE contributions. Indeed, at the LHC in a $R = 0.4$ cone one expects 
$\Delta E_T = \pi \times R^2 \times 1/(2\pi) \times \dETdeta \approx$
~80~GeV, with large fluctuations, 
making it challenging to reconstruct jets below $E_T\approx$~50~GeV. Various UE subtraction techniques 
have been developed in combination with the $k_T$~\cite{Cacciari:2007fd,Cacciari:2008gn}, 
UA1-cone~\cite{Morsch:2008zz,Blyth:2006vb} or iterative-cone~\cite{Kodolova:2007hd} algorithms.\\
\item {\it (3) Jet corrections}: The energy of the reconstructed and background-subtracted jets has to be corrected 
for various experimental and model-dependent uncertainties before comparing it to theoretical predictions.
Experimentally, the {\it jet energy-scale} (JES) is the most important source of systematic uncertainties in the 
jet yield and requires careful data-driven studies (e.g. via dijet and $\gamma$-,$Z$-jet $E_T$-balancing in proton-proton
collisions). 
In addition, before a given ``parton-level'' pQCD calculation can be compared to a measured ``hadron-level''
jet spectrum, 
one needs to estimate the non-perturbative effects introduced by the underlying-event and hadronisation 
corrections. In \pp\ collisions, this final step is carried out usually comparing the results from two Monte Carlo's
(e.g. \pythia\ and \herwig) with different models for the UE multiparton-interactions as well as for the 
hadronisation (string- and cluster- fragmentation respectively).
\end{description}


\subsubsection{(1) Jet clustering algorithms}

\begin{figure}[htbp]
\includegraphics[height=7.5cm,width=5.8cm,clip]{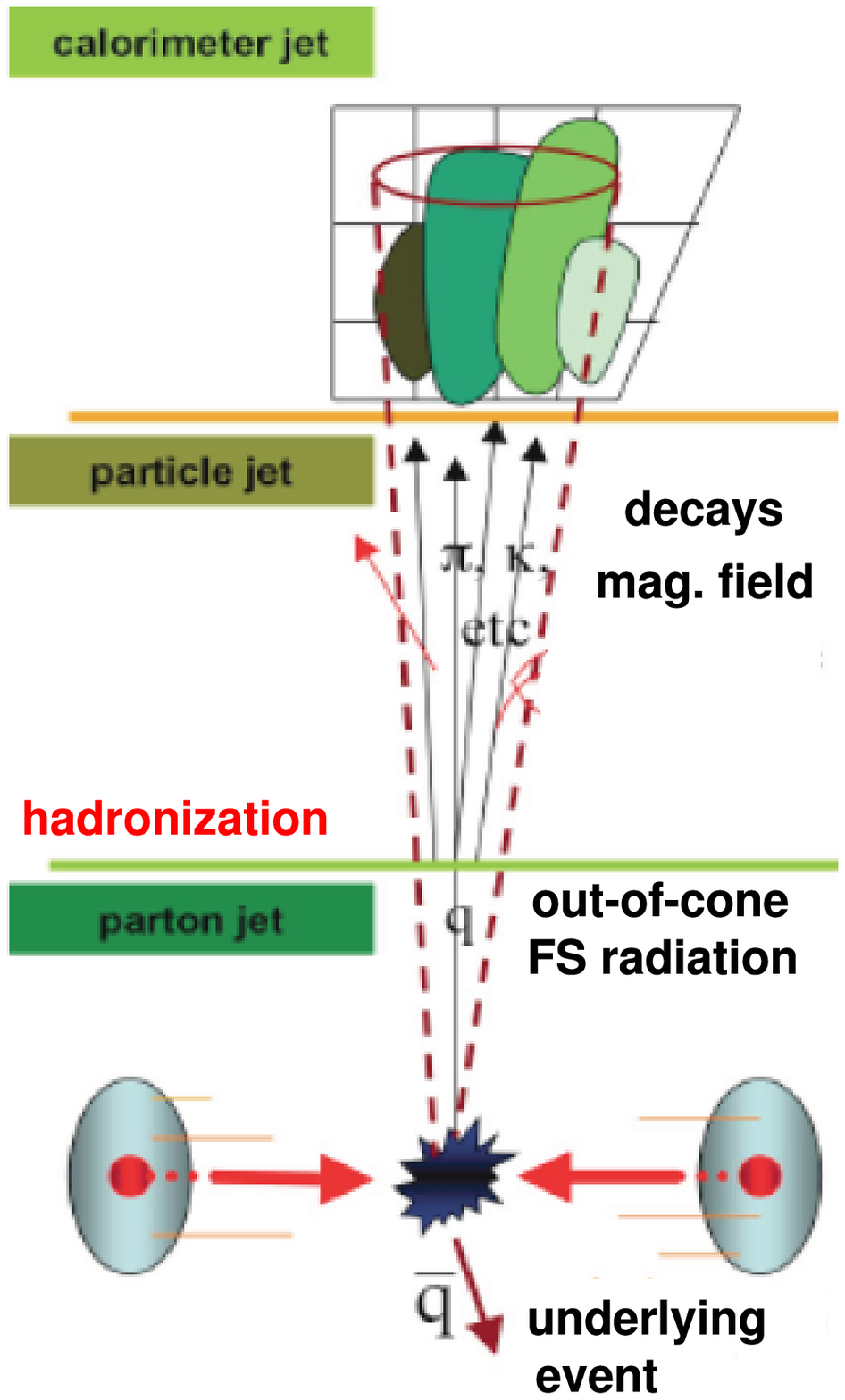}
\includegraphics[height=7.5cm,width=6.25cm,clip]{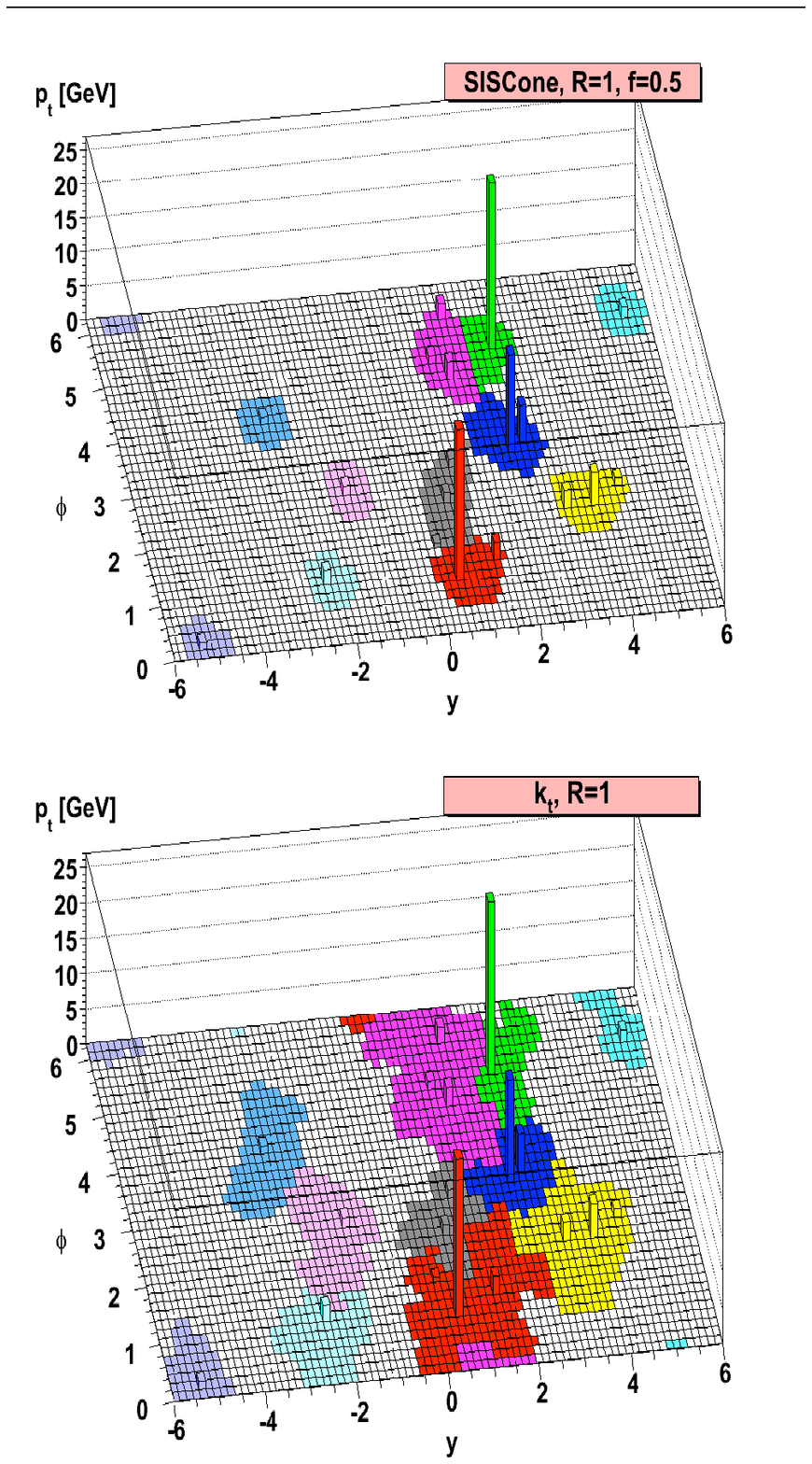} 
\caption{Left: Schema of jet production and measurement~\cite{Ellis:2007ib}.
Right: Reconstructed jets in $\eta$-$\phi$ space with the SISCone (top) and $k_T$ (bottom)
algorithms for a simulated \pp\ event at the LHC~\cite{Cacciari:2008gp}.}
\label{fig:jet_algos}
\end{figure}

In practical terms one usually deals with three types of ``jets''  (Fig.~\ref{fig:jet_algos}, left).
Experimentally, a {\it calorimeter jet} (aka ``CaloJet'') is a collection of 4-vectors based on the energy 
deposited in calorimeter towers clustered in pseudorapidity-azimuth according to a given algorithm. 
[Often nowadays, the experiments use also the {\it momentum} of the (low $p_T$) charged hadrons measured by the 
tracking system to reconstruct the jet energy with improved resolution, so the name ``Calo'' is not fully justified.].
At the Monte-Carlo generator level, a {\it hadron or particle jet} 
(aka ``GenJet'') is a collection of hadrons issuing from the (non-perturbative) hadronisation of a 
given parton. Theoretically, a {\it parton-level jet} is what one actually calculates in pQCD. 
The (non-unique) method of linking an initial parton to a set of final-state particles (or other objects
with four-vector like properties) relies on a procedure known as ``jet algorithm''. 

The goal of a clustering algorithm is to combine hadrons into jets according to a given ``distance'' (radius).
Theoretically, such a procedure must be {\it infrared-} and {\it collinear-safe} -- i.e. adding a soft 
gluon or splitting a given parton must {\it not} change the final number of reconstructed jets.
In addition, the jet-finder must not be too sensitive to nonperturbative effects -- hadronisation, 
underlying-event (and pileup\footnote{{\it Pile-up} refers to the additional \pp\ collisions occurring
in the same bunch-crossing at high luminosity (at $\mathscr{L}$~=~10$^{34}$~cm$^{-2}$s$^{-1}$ 
one expects $\sim$25 simultaneous collisions at the LHC).}
in \pp) -- and it must be realistically applicable at detector level 
(e.g. not too slow). There are two broad classes of jet algorithms~\cite{Ellis:2007ib,Buttar:2008jx}:
\begin{itemize}
\item  {\bf Cone-type} algorithms are ``top-down'' approaches i.e. they identify energy flow into 
pre-defined cones of a given radius. One sums the momenta of all particles $j$ within a cone 
of radius $R$ around a seed particle $i$ (often the particle or calorimeter tower with the largest 
transverse momentum) in azimuthal angle $\phi$ and pseudorapidity $\eta$, i.e. taking all $j$ 
such that
\begin{equation}
  \Delta_{ij}^2 = (\eta_i - \eta_j)^2 + (\phi_i-\phi_j)^2 < R^2\,.
\label{eq:deltaij}
\end{equation}
The direction of the resulting sum is then used as a new seed direction, and one iterates 
the procedure until the direction of the resulting cone is stable. 
There exist various cone jet-finders: JetClu, ILCA/MidPoint, ICone,  SISCone, ...
which have been mainly employed at hadron colliders (see e.g. Fig.~\ref{fig:jet_algos}, top-right). 
Their main advantages are their speed, 
which makes them easy to implement in triggers, and the simplicity of the UE corrections. 
On the other hand, their particular implementations can be messy (seeding, split-merge, ``ratcheting'',
missed or ``dark'' towers, ...) and infrared/collinear safety is not guaranteed in many cases.\\

\item {\bf Sequential clustering} algorithms are ``bottom-up'' approaches that rely on 
pair-wise successive recombinations of the closest hadrons in momentum up to a given 
(predefined) distance $D$. One introduces distances $d_{ij}$ between entities 
(particles, protojets) $i$ and $j$, and $d_{iB}$ between entity $i$ and the beam (B). 
The clustering proceeds by identifying the smallest of the distances and if it is a $d_{ij}$ 
recombining entities $i$ and $j$, while if it is $d_{iB}$ calling $i$ a jet and removing 
it from the list. 
The distances are recalculated and the procedure repeated 
until no entities are left. The distance measures for several algorithms are of the form
\begin{equation}
    d_{ij} = \min(k_{T,i}^{2p}, k_{T,j}^{2p}) \frac{\Delta_{ij}^2}{D^2}\;\;,\;\;  d_{iB} = k_{T,i}^{2p}\,,
\label{eq:genkt}
\end{equation}
where $\Delta_{ij}^2$ is defined in Eq.~(\ref{eq:deltaij}), $k_{T,i}$  is the transverse 
momentum of particle $i$, $D$ is the jet-radius parameter (equivalent to $R$ in the cone finders), 
and $p$ parameterises the type of algorithm:  $k_T$ ($p$~=~1)~\cite{Ellis:1993tq},  Cambridge/Aachen 
($p$~=~0)~\cite{Wobisch:1998wt}, anti-$k_T$ ($p$~=~-1)~\cite{Cacciari:2008gp}
(Fig.~\ref{fig:jet_algos}, bottom-right).
On the positive side, these algorithms -- widely used at LEP and HERA -- are explicitly infrared 
and collinear safe and more ``realistic'' than the cone-based ones as they mimic (backwards) 
the QCD shower branching dynamics. On the other hand, they used to be slow and lead
to jets with irregular shapes\footnote{Yet, {\it with large multiplicities}, the $k_T$ algorithm
(often labelled a ``vacuum cleaner'') has actually an average area $\sim\pi R^2$, whereas 
modern versions of the cone finder (assumed to have always an area $\pi R^2$)
with split-merge steps such as SISCone, turn out to be quite non-conical, with 
\emph{small} areas $\sim \pi R^2/2$ that renders them efficient in noisy environments~\cite{Salam:2008qq}.}
which complicated  the UE subtraction compared to the cone jet-finders (this has been 
now solved as discussed in the next subsection), 
making them not competitive in a heavy-ion environment with very large hadron multiplicities. 
Recently, the time taken to cluster $N$ particles has been significantly improved in the 
\fastjet~\cite{Cacciari:2005hq} package, based on Voronoi diagrams,
going down for the default $k_T$ jet-finder from $\mathscr{O}(N^3)$  to
$\mathscr{O}(N\ln N)$. Jet clustering in nucleus-nucleus collisions is now routinely 
performed at sub-second times.
\end{itemize}


\subsubsection{(2) Underlying event subtraction}

\begin{figure}[htbp]
\hspace{-0.2cm}\includegraphics[width=5.8cm]{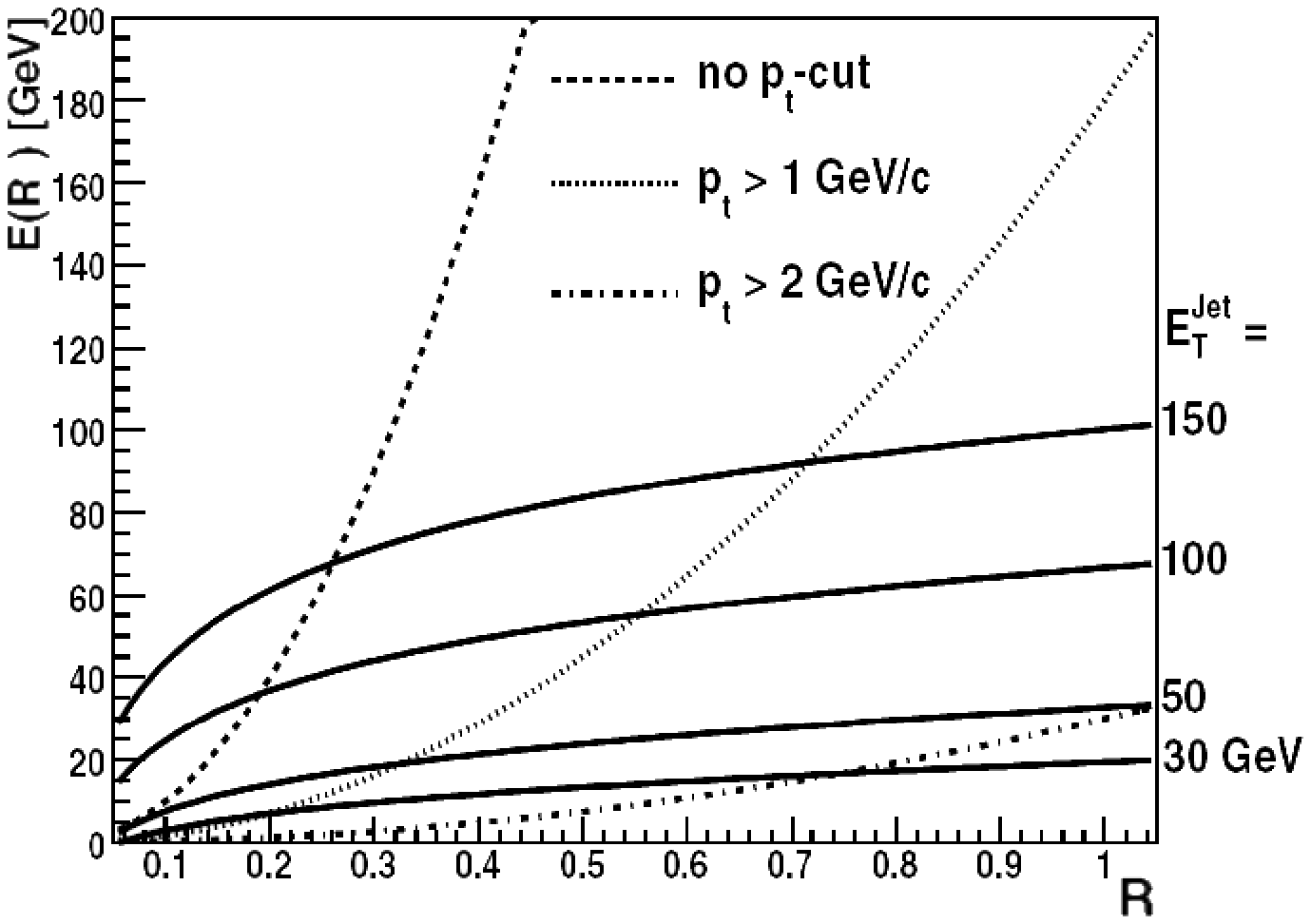} 
\includegraphics[width=6.cm]{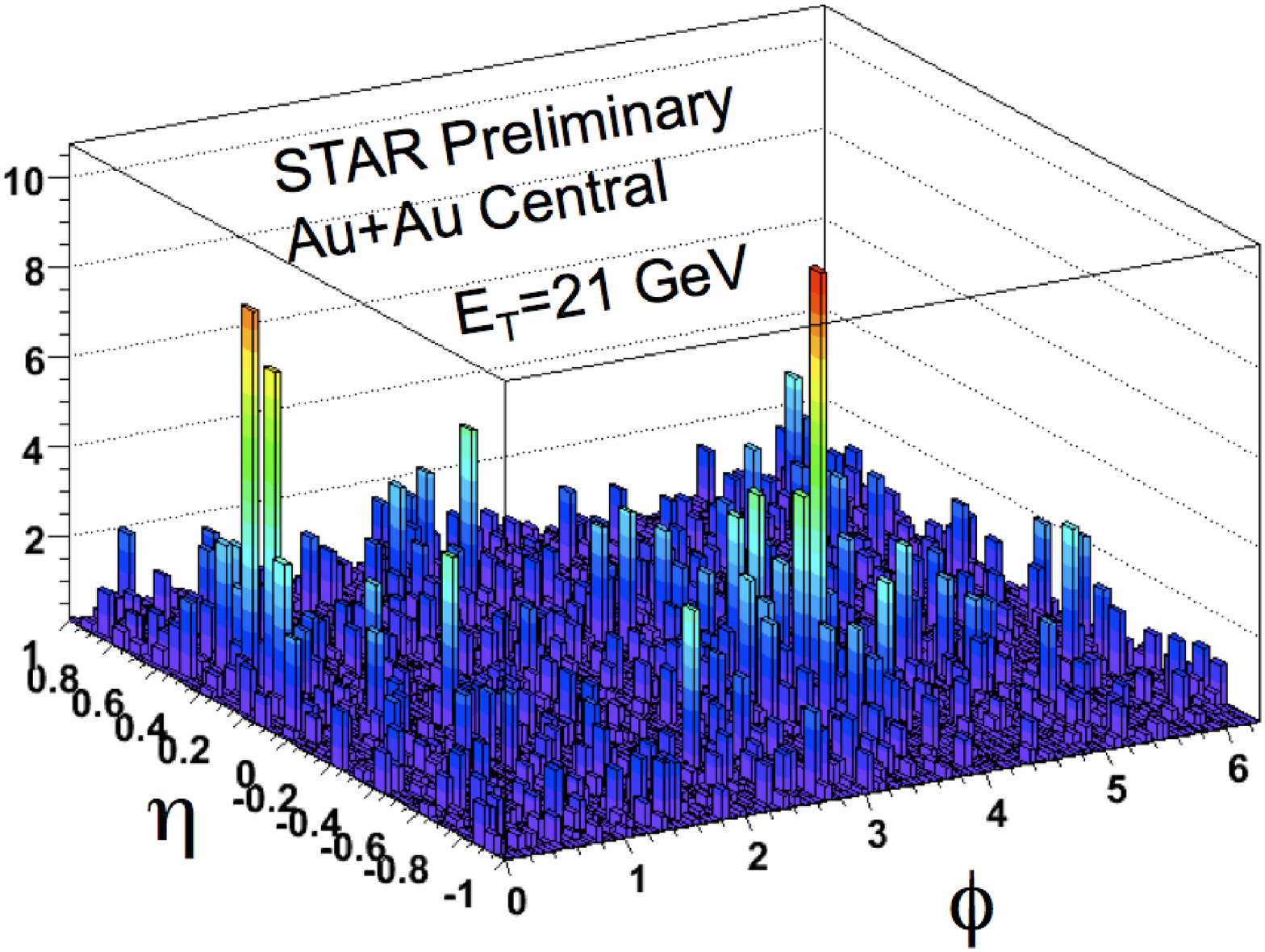}
\caption{Left: Charged jet energy in a cone of radius $R$ (full lines) in ALICE 
compared to the background energy from a \hijing~\cite{hijing} \PbPb\ simulation for 
different cuts in the particles $p_T$ (dashed lines)~\cite{Alessandro:2006yt}.
Right: STAR \AuAu\ dijet event after background subtraction~\cite{Putschke:2008wn,Salur:2008vm}.}
\label{fig:STARdijet}
\end{figure}

Background energy in a jet cone of size $R$ is $\mathscr{O}(R^2)$ and background fluctuations 
are $\mathscr{O}(R)$. As aforementioned, the soft background from the underlying event in a cone 
of $R$~=~0.4 in central nucleus-nucleus collisions at RHIC (LHC) is about 40 (80)~GeV. 
Fig.~\ref{fig:STARdijet} (left) shows the (charged) jet and background energies as a function 
of the cone radius $R$ in ALICE~\cite{Alessandro:2006yt,Morsch:2008zz}. 
Jets can only be identified if the background energy within the cone is smaller than the signal energy. 
This can be achieved by using small cone sizes ($E_T^{bgd}\propto R^2$), $R$~=~0.3~--~0.5, 
and/or by applying $p_T$ or energy cuts on the charged hadrons or calorimeter towers. The latter option is 
not optimal since it also introduces potential biases in the investigation of jet-quenching effects. 
STAR~\cite{Putschke:2008wn,Salur:2008vm} (Fig.~\ref{fig:STARdijet}, right) uses a seeded-cone 
algorithm with $R$~=~0.4 and $p_T^{cut}$~=~0.1~--~2~GeV/c, and estimates the UE background 
event-by-event from the average energy in cones {\it without} seeds which is then subtracted 
from the reconstructed jets. ALICE uses a modified version of the UA1-cone algorithm ($R$~=~0.4)
where the mean cell energy from cells outside a jet cone is recalculated after each iteration of the cone 
jet finder and subtracted from all cells~\cite{Alessandro:2006yt,Morsch:2008zz}.

Similarly, CMS~\cite{D'Enterria:2007xr,Kodolova:2007hd} 
subtracts the UE on an event-by-event basis with a variant of the iterative ``noise/pedestal subtraction'' 
for \pp\ collisions~\cite{Ball:2007zza}. Initially, the mean value and dispersion 
of the energies in the calorimeter cells are calculated for rings of constant pseudorapidity, $\eta$. The value of 
this pedestal function, P($\eta$), is subtracted from all cells (the cell energy is set to zero in case of negative 
values) and the jets are reconstructed with the default ICone finder. In a second iteration, the pedestal function 
is recalculated using only calorimeter cells outside the area covered by  jets with $E_{T}>$~30~GeV. The cell 
energies are updated with the new pedestal function and the jets are reconstructed again, using the updated calorimeter cells. 

Alternatively, \fastjet~\cite{Cacciari:2007fd} proposes a background-subtraction procedure {\it after} 
running any infrared-safe algorithm. The method is based on the concept of a `jet area' $A$ constructed by 
adding infinitely soft particles (``ghosts'') and identifying the region in $\eta$,$\phi$ where those ghosts 
are clustered within each jet~\cite{Cacciari:2008gn}. 
Each reconstructed jet $p_T$ is then corrected by subtracting the median value of the noise distribution in the event, 
$\rho\,=\, \mbox{median}\left[ \left\{p_{T}/A \right\} \right]$, in the jet-area $A$, via
$p^{sub}_{T} = p_{T} - A\cdot\rho$. 
In practical terms, one fits the measured $p_T(\eta)/A$ background distribution for each event 
with a parabola form, $\rho(\eta) = a+b\,\eta^2$ (which excludes any jet peak) and corrects then
the jet $p_{T}$ using the $p^{sub}_{T}$ formula above.


\subsubsection{(3) Jet energy corrections}

\begin{figure}[htbp]
\centering
\includegraphics[width=\textwidth,clip]{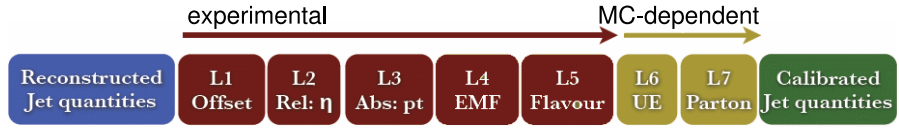} 
\caption{List of typical factorised jet-energy corrections (CMS analysis)~\cite{Ball:2007zza}.}
\label{fig:JetCorrs}
\end{figure}

The last step of any jet analysis consists in correcting the $p_T$ of any measured {\it CaloJet} 
to match closely that of the associated {\it GenJet} and/or {\it PartonJet}, so that it can
be compared to theoretical expectations. In principle, the different corrections can be decomposed 
as shown in Fig.~\ref{fig:JetCorrs}.
The experimental corrections (labelled Level  1~--~5 in the plot) can be extracted from the data 
themselves. For example, the correction L1 (noise offset) can be obtained from minimum bias events {\it without}
jet activity, and the L2 (flattening of relative, $\eta$-dependent, $p_T$ responses of the calorimeters) 
and L3 (absolute $p_T$ calibration) can be derived using $p_T$-balancing techniques in back-to-back di-jet 
and $\gamma$-, $Z$-jet events in \pp\ collisions. A precise calibration of the jet energy scale (JES) is 
essential. Given the steep (power law) fall-off of the jet cross section as a function of energy and the 
relatively large binning\footnote{The bin-width is not constant in the whole spectrum 
but given by the absolute $p_T$ resolution at each bin: relative jet $p_T$ resolutions are in the 25\%-10\% range
for \PbPb\ at the LHC~\cite{Alessandro:2006yt,Grau:2008ef,D'Enterria:2007xr}.} 
of the jet spectra of $\mathscr{O}(5$~--~30~GeV/c$)$, an uncertainty 
of 10\% in the JES can propagate into uncertainties as large as 50\% (!) in the jet yield at a given $p_T$ bin.
The L3 correction is thus the most important source of experimental uncertainty in any jet measurement.
The two last corrections, L4 (fraction EMF of energy deposited by hadrons in the EM calorimeter), 
and L5 (flavour correction accounting for the different characteristics of
-- and therefore detector responses to -- gluon, light-quark and heavy-quark jets), can be e.g. obtained 
in back-to-back $\gamma$-jet and $b$,$c$-identified dijet events in \pp. 

The two ``theoretical'' corrections (L6--UE and L7--parton) aim at bringing the $p_T$ of a {\it CaloJet} 
as close as possible to that of its originating parton. They can only be obtained from MC simulations
that model the effects of final-state-radiation (FSR), hadronisation and underlying-event.
FSR and hadronisation tend to remove energy {\it out} of the jet, whereas the UE has the contrary
``splash-in'' effect. In \pp\ collisions, the total shift on a jet $p_T$ due to these effects can be approximated by the uncorrelated sum
$\mean{\delta p_T^2 } \approx \mean{\delta p_T}^2_{_{FSR}} +  \mean{\delta p_T}^2_{_{hadr}} + \mean{\delta p_T}^2_{_{UE}}$~\cite{Dasgupta:2007wa}.
 The way these effects modify the jet energy as a 
function of the parton $p_T$, flavour and the used cone radius $R$ are summarised in Table~\ref{tab:jet_effects}.
Whereas the effect of FSR can be in principle computed perturbatively, the UE and hadronisation corrections
rely on model-dependent descriptions of 
multi-parton interactions (MPI) and parton-to-hadron fragmentation. 
In \pp\ collisions, one usually compares the result of \pythia\ and \herwig\ 
-- which have different MPI and different (string vs. cluster) fragmentation models -- 
to gauge the dependence of the measured jet observables on these non-perturbative phenomena.

\begin{table}
  \caption{Main physical effects contributing to a shift $\mean{\delta p_T}$ 
   in the transverse momentum of a jet compared to that of its parent parton in \pp\ collisions
   (cases with `--' do not have any dependence at LO)~\cite{Dasgupta:2007wa}.}
  \label{tab:jet_effects}
  \centering 
  \begin{tabular}{l|c|c|c}\hline\hline\noalign{\smallskip}
    & \multicolumn{3}{c}{Dependence of jet $\mean{\delta p_T}$ shift on }\\[3pt] 
    & \hspace{0.5cm} parton $p_T$ \hspace{0.5cm} & \hspace{0.2cm} colour factor \hspace{0.2cm} & \hspace{0.5cm} radius $R$ \hspace{0.5cm} \\\hline\noalign{\smallskip}
    \hspace{0.1cm} final-state radiation \hspace{0.3cm}& $\sim\alpha_s(p_T)\, p_T$ & $C_i$ & $\ln R + \mathscr{O}(1)$ \\[3pt] 
    \hspace{0.1cm} hadronisation  \hspace{0.3cm} & -- & $C_i$ & $-1/R + \mathscr{O}(R)$ \\[3pt] 
    \hspace{0.1cm} underlying event \hspace{0.3cm} & -- & -- & $R^2/2 + \mathscr{O}(R^4)$ \\\noalign{\smallskip}\hline\hline 
  \end{tabular}
\end{table}

In heavy-ion collisions, in-medium FSR and UE are significantly enhanced compared to \pp\ jets (see 
cartoon in Fig.~\ref{fig:jet_quench}), but at high enough $p_T$ the final parton-to-hadron non-perturbative fragmentation 
occurs in the vacuum and should be the same as in \pp. Ideally, the effects of the UE can be controlled 
embedding MC jets in real events, and the influence of hadronisation can be gauged comparing the results of 
e.g. \qpythia\ and \qherwig~\cite{Armesto:2008qh} (adding also eventually medium-induced modifications 
of the colour structure of the jet shower evolution). Effects on jet quenching observables -- which are 
the ultimate goal of our studies -- can then be isolated comparing the results of different 
parton energy-loss MCs such as e.g. {\sc pyquen} (with large out-of-cone elastic 
energy-loss) and \qpythia\ (with its embedded BDMPS radiative energy loss).


\subsection{Jet spectra}

\begin{figure}[htb]
\hspace{-0.2cm}
\includegraphics[width=6.cm,height=6.cm,clip]{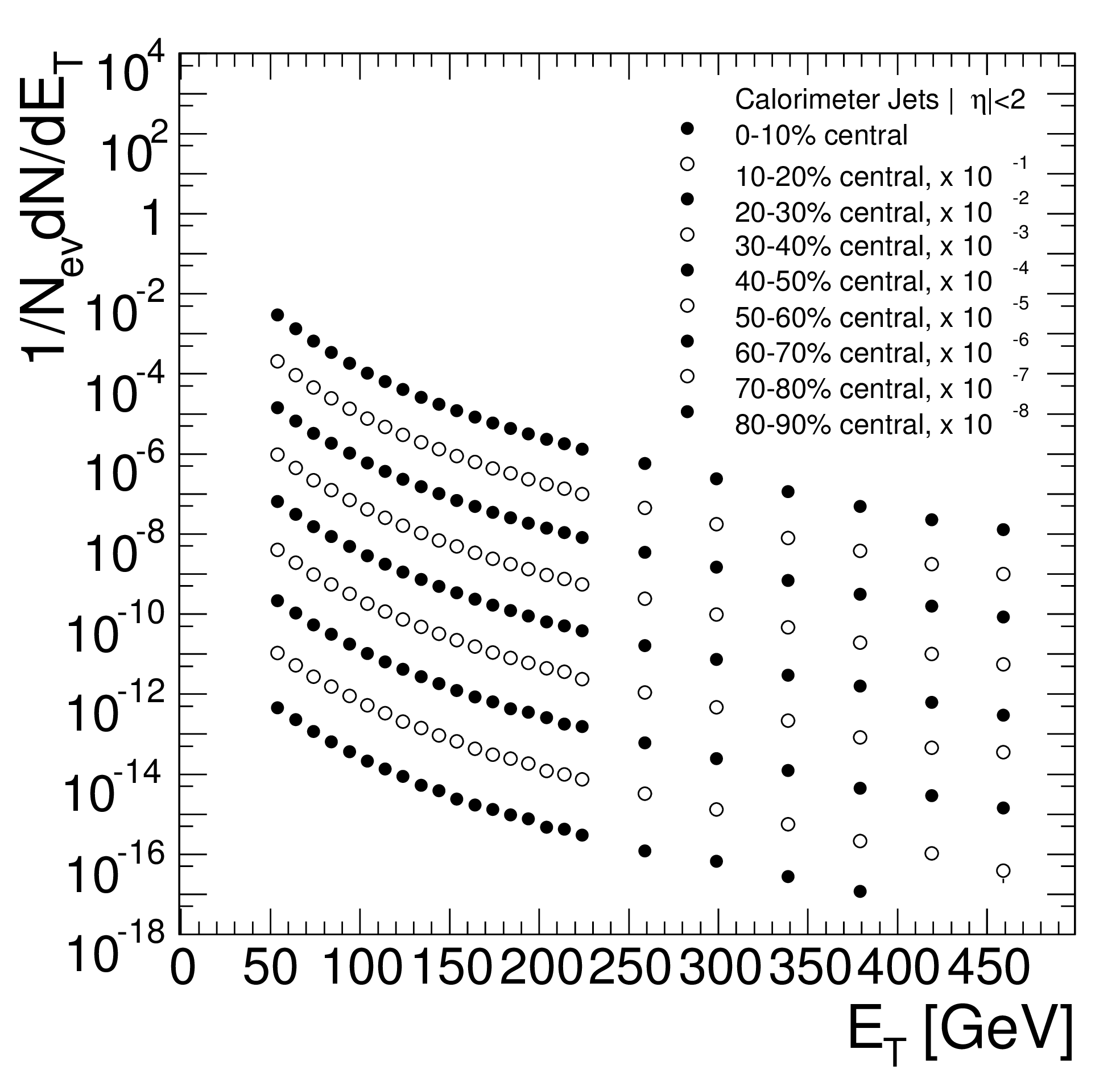}\hspace{0.05cm}
\includegraphics[width=6.cm,height=5.cm,clip]{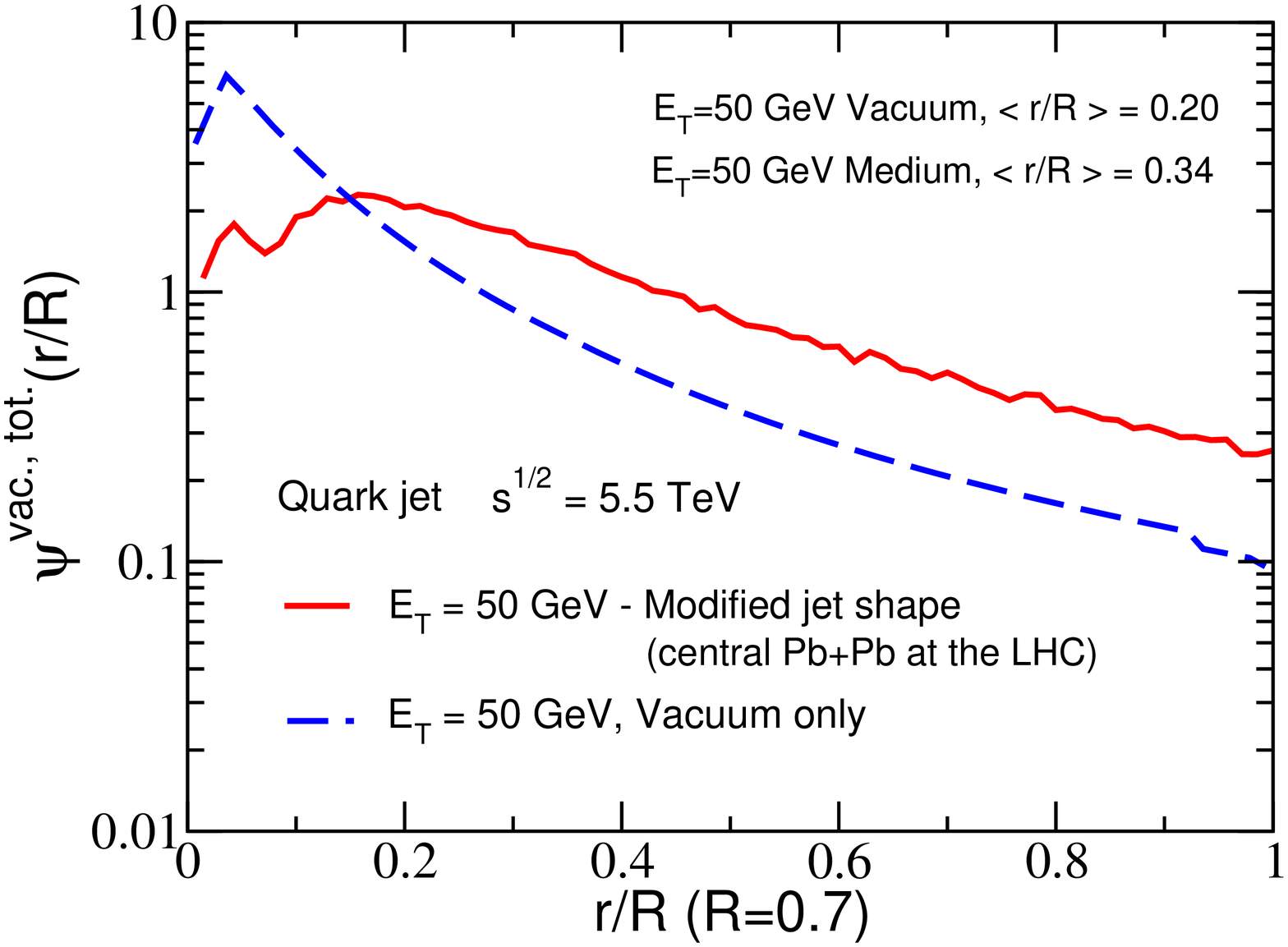}
\caption{Left: Jet spectra for various \PbPb\ centralities expected at 5.5~TeV in CMS 
($\int\mathcal{L}\,dt$~=~0.5~nb$^{-1}$)~\cite{D'Enterria:2007xr}. Right: Comparison of the vacuum and 
in-medium jet shapes for $E_T = 50$~GeV gluon jets in central \PbPb\ collisions at the LHC~\cite{Vitev:2008jh}.}
\label{fig:jetsPbPb}
\end{figure}

The direct comparison of the fully corrected $p_T$-differential jet spectra in \AaAa\ and \pp\ collisions 
will provide crucial first-hand information about the nature of the medium produced in heavy-ion 
collisions. The expected ALICE (ATLAS and CMS) jet $p_T$ range measured in \PbPb\ collisions for 
nominal integrated luminosities is $p_T\approx$~30~--~200~GeV/c (50~--~500~GeV/c, see Fig.~\ref{fig:jetsPbPb} left). 
A natural generalization of the nuclear modification factor, Eq.~(\ref{eq:R_AA}), for jets~\cite{Vitev:2008rz,Lokhtin:2007ga},
\begin{equation}
R_{AA}^{jet}(p_T; R^{\max},\omega^{\min}) = \frac{ dN^{AA}(p_T;R^{\max},\omega^{\min})/dy dp_T^2 }
{ \mean{T_{AA}} \; d\sigma_{pp}(p_T;R^{\max},\omega^{\min})/dy dp_T^2  } \; ,
\label{eq:RAAjet}
\end{equation}
is a sensitive measure of the nature of the medium-induced energy loss. The steepness of the 
spectra amplifies the observable effects and the varying values of the jet radius $R^{\max}$ and 
the minimum particle/tower energy $p_{T}^{\min} \approx \omega^{\min}$ will provide, 
through the evolution of $R_{AA}^{jet}(p_T; R^{\max},\omega^{\min})$ at any centrality, 
experimental access to the QGP response to  quark and gluon propagation~\cite{Vitev:2008rz}.
If the medium-induced energy loss of the parent parton is radiated {\it inside} the jet cone, 
one will observe $R_{AA}^{jet}\approx$~1 at variance with the leading-hadron spectra 
($R_{AA}\ll$~1). On the contrary, important large-angle radiation (as expected e.g. in 
some models of collisional energy loss~\cite{Lokhtin:2007ga}) will result in a {\it quenched}
jet spectrum in heavy-ion collisions ($R_{AA}^{jet} < 1$).

\vspace{-0.5cm}


\subsection{Jet shapes}

The study of the internal structure of jets -- via observables such as 
jet shapes and jet multiplicity distributions -- in \ppbar\ collisions at Tevatron has provided valuable 
tests of the models for parton branching and soft-gluon emission in the vacuum~\cite{Acosta:2005ix}.
The energy degradation of partons traversing a dense QCD plasma will be also directly reflected in the 
modification of such jet observables in heavy-ion collisions~\cite{Salgado:2003rv,Vitev:2008rz}. 
Three variables are useful in this context:
\begin{itemize}
\item the {\it differential jet shape}, 
$\rho(r)$, defined as the average fraction of the jet $p_T$ that lies inside an annulus 
of radius $r \pm \delta r/2$ (e.g. $\delta r$~=~0.1) around the jet axis: 
\begin{equation}
\rho(r) = \frac{1}{\delta r} \frac{1}{N_{\rm jet}}\sum_{\rm jets} \frac{p_T(r - \delta r/2,r + \delta r/2) }{p_T(0,R)}, 
\ \ \ \ 0 \leq  r = \sqrt{\Delta y^2 + \Delta \phi^2} \leq R \; 
\end{equation}
\item the {\it integrated jet shape}, $\Psi(r)$, i.e. 
the average fraction of the jet $p_T$ that lies inside a cone of radius $r$ concentric 
to the jet cone (by definition, $\Psi(r = R) = 1$):
\begin{equation}
\Psi(r) = \frac{1}{\rm N_{jet}} \sum_{\rm jets} \frac{p_T(0,r) }{p_T(0,R)}, \ \ \ \ 0 \leq  r \leq R
\end{equation}
\item the {\it thrust} variable, 
constructed from the 3-momenta ${\bf p}_i$ of all jet particles, characterises the global energy flow:
\begin{equation}
\mathcal{T} \equiv {\rm max}_{\,{\bf n}_\mathcal{T}} \frac{\sum_i \vert {\bf p}_i\cdot {\bf n}_\mathcal{T}\vert}{\sum_i \vert {\bf p}_i\vert}\, ,
\end{equation}
i.e. $\mathcal{T} =$ 1 (1/2) if the dijet event is pencil-like (spherical), i.e. if all particles are aligned (or not at all) 
along a thrust axis ${\bf n}_\mathcal{T}$. The projection of all particle momenta on the direction  ${\bf n}$ 
along which the momentum flow is maximal is the ``thrust major'' and on that orthogonal to the plane 
formed by ${\bf n}_\mathcal{T}$ and ${\bf n}$ is the ``thrust minor''~\cite{Heister:2003aj}. 
The \jewel\ MC predicts a broadening of these (perturbative and infrared-safe) $\mathcal{T}$, 
$\mathcal{T}_{\rm maj}$ and $\mathcal{T}_{\rm min}$ distributions inside a dense QGP~\cite{Zapp:2008gi}.
\end{itemize}
Interestingly, from an experimental point of view, all those observables are robust against jet energy scale/corrections/resolutions.
Medium-modified jet shapes in \PbPb\ collisions at LHC energies have been analytically investigated 
in~\cite{Salgado:2003rv,Vitev:2008rz} (see e.g. Fig.~\ref{fig:jetsPbPb}, right). More detailed studies using 
the recently available jet-quenching Monte Carlo's (Section~\ref{sec:Eloss_models}) are needed.


\subsection{Medium-modified fragmentation functions}

Due to the coherence and interference of gluon radiation {\it inside} a jet (resulting, 
on average, in {\it angular ordering}\footnote{Angular ordering (or coherence) implies
$\theta_{p_1p_2} \gg \theta_{k_1p_1} \gg \theta_{k_2k_1} \gg \theta_{k_3k_2} \gg ...$, where $\theta_{k_1p_1}$ 
is the emission angle of the primary soft gluon from the direction of the hard parton, 
$\theta_{k_2k_1}$ is that of the softer secondary gluon from the direction of the primary gluon, etc.}
of the sequential branching), not the softest partons but those with intermediate energies 
($E_h\propto E_{jet}^{0.3-0.4}$) multiply most effectively in QCD cascades~\cite{Dokshitzer:1991wu}. 
\begin{figure}[htbp]
\hspace{-0.2cm}
\includegraphics[width=6.0cm,height=4.3cm,clip]{figs/denterria_borghiniwiedemann_humpedback.eps}\hspace{0.25cm}
\includegraphics[width=5.8cm,height=4.3cm,clip]{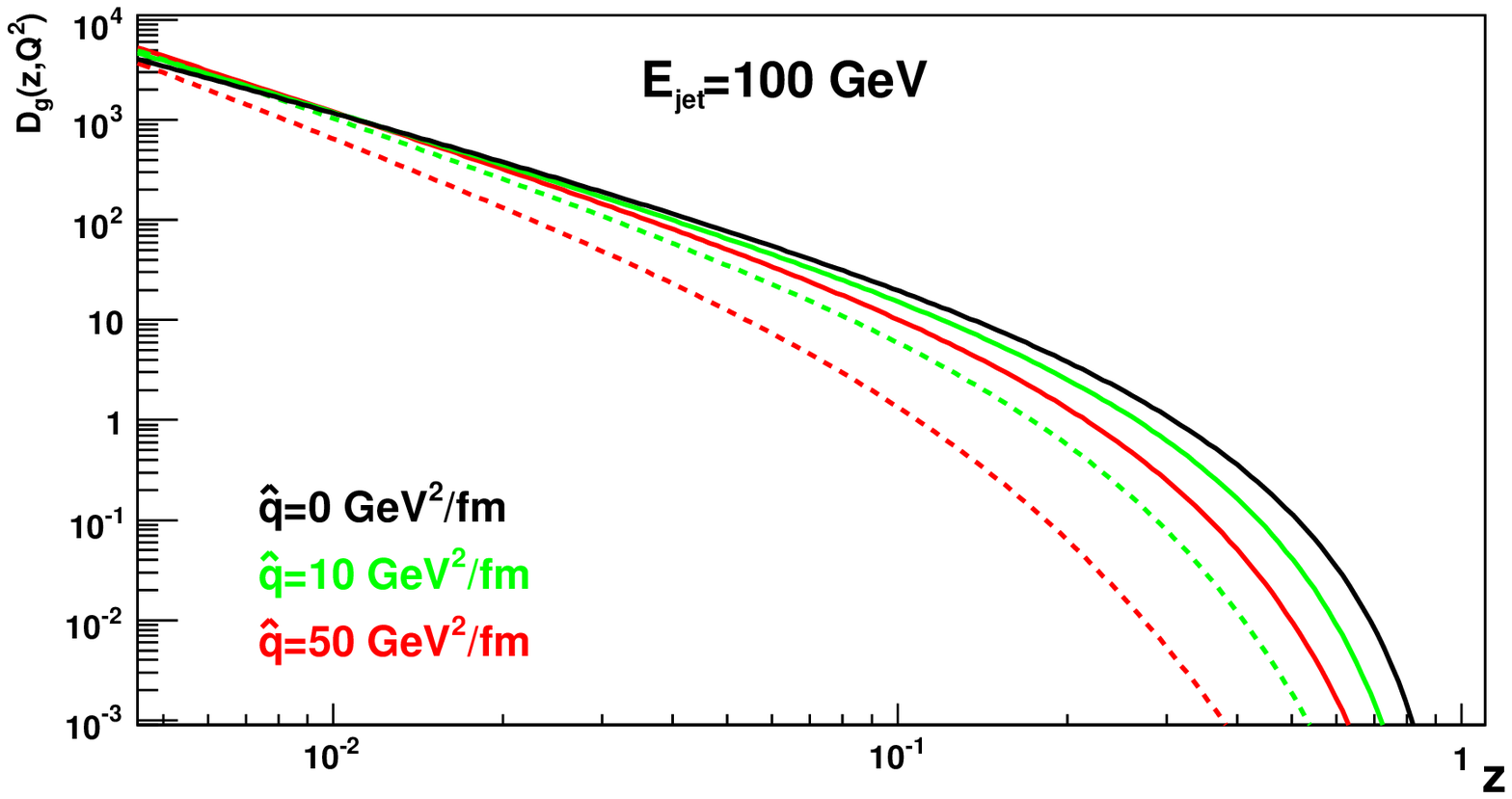}
\caption{Left: Single inclusive distribution of hadrons vs. $\xi=\ln\,(E_{jet}/p)$ for a 17.5-GeV jet in $e^+e^-$ 
collisions (TASSO data) compared to MLLA predictions in the vacuum (solid curve) and in-medium 
(dashed curve)~\cite{Borghini:2005em}. Right: Medium-modified pion FF for a 100-GeV gluon in 
a medium of length $L$~=~2~--~6~fm (solid--dashed lines) with increasing $\hat{q}$ parameter~\cite{Armesto:2007dt}.}
\label{fig:mFFs}
\end{figure}
This is best seen in the approximately Gaussian distribution in the variable $\xi=\log(E_{jet}/p)=\log(1/z)$ 
for particles with momentum $p$ in a jet of energy $E_{jet}$, which peaks at the 
so-called ``hump-back plateau'' at intermediate $\xi\approx$~3~--~4 values\footnote{More
generally, the peak is at $\xi = 0.5\,\ln(E_{jet}/\Lambda_{_{QCD}})$ at leading order (e.g.
$\xi$~=~3 for $E_{jet}$~=~100~GeV) but it is shifted by an amount $\propto\sqrt{\alpha_s}$
in MLLA~\cite{Arleo:2008dn}.} (Fig.~\ref{fig:mFFs}, left).
In a QCD medium, energy loss shifts parton energy from high-$z$ to low-$z$ hadrons and, as a result, 
leading hadrons are suppressed as seen in Fig.~\ref{fig:mFFs} (right) where, for increasing $\qhat$ 
coefficient, the fragmentation function  $\mathcal{D}_{i\to h}(z,Q^2)$ is increasingly depleted at high-$z$. 
Correspondingly, the number of low-$p_T$ hadrons rises, resulting in a {\it higher} humped-back 
in Fig.~\ref{fig:mFFs} (left).\\

Theoretically, the resummed (next-to) Modified Leading Logarithmic Approximation (N)MLLA
approach describes well, to (next-to)-next-to-leading order $\sqrt{\alpha_s}$ accuracy, the measured 
distributions of hadrons $\mathcal{D}_{i\to h}(z,Q^2)$ inside a jet (Fig.~\ref{fig:mFFs}, left) down to 
nonperturbative scales $\Qeff \approx \Lambda_{QCD}\approx$~0.2~GeV, provided that each 
parton is mapped locally onto a hadron (``Local Parton--Hadron Duality'', LPHD)~\cite{lphd} 
with a proportionality factor $\kappa\approx 1$. 
Various recent promising applications of the (N)MLLA 
approach~\cite{Borghini:2005em,Armesto:2008qe,Ramos:2008qb,Ramos:2008as,Borghini:2006fp} 
have investigated QCD radiation in the presence of a medium.

\subsubsection{Photon-jet correlations}

The $\gamma$-jet (and $Z$-jet) channel provides a very clean means to determine parton fragmentation
functions (FFs)~\cite{Wang:1996yh}. In the leading-order QCD processes of photon production 
(Compton: $qg\to q\,\gamma$, and annihilation: $q\bar{q}\to g\,\gamma$), because of momentum 
conservation the photon is produced back-to-back to the hard jet, with equal and opposite transverse momentum. 
Since the prompt $\gamma$ is not affected by final-state interactions, its transverse energy 
($\etg$) can be used as a proxy of the away-side parton energy ($\etj\approx\etg$) {\it before} 
any jet quenching has taken place in the medium. Once the quark fragments into a
given hadron $h$, the $\gampi$ momentum imbalance variable~\cite{Arleo:2004xj}, 
$\z \equiv - {\bf p}_{_{T,h}} . {\bf p}_{_{T,\gamma}}/|{\bf p}_{_{T,\gamma}}|^2$,
reduces at LO to the fragmentation variable, i.e. $\z \simeq z$. 
The FF, defined as the normalised distribution of hadron momenta $1/{N_{jets}}\,dN/dz$
relative to that of the parent parton $\etj$, can then be constructed using $\z$ 
or, similarly, $\xi=-\ln (\z)$, 
for all particles with momentum $p_T$ associated with a jet in heavy-ion collisions.

\begin{figure}[htbp]
\includegraphics[width=6.0cm,height=5.0cm,clip]{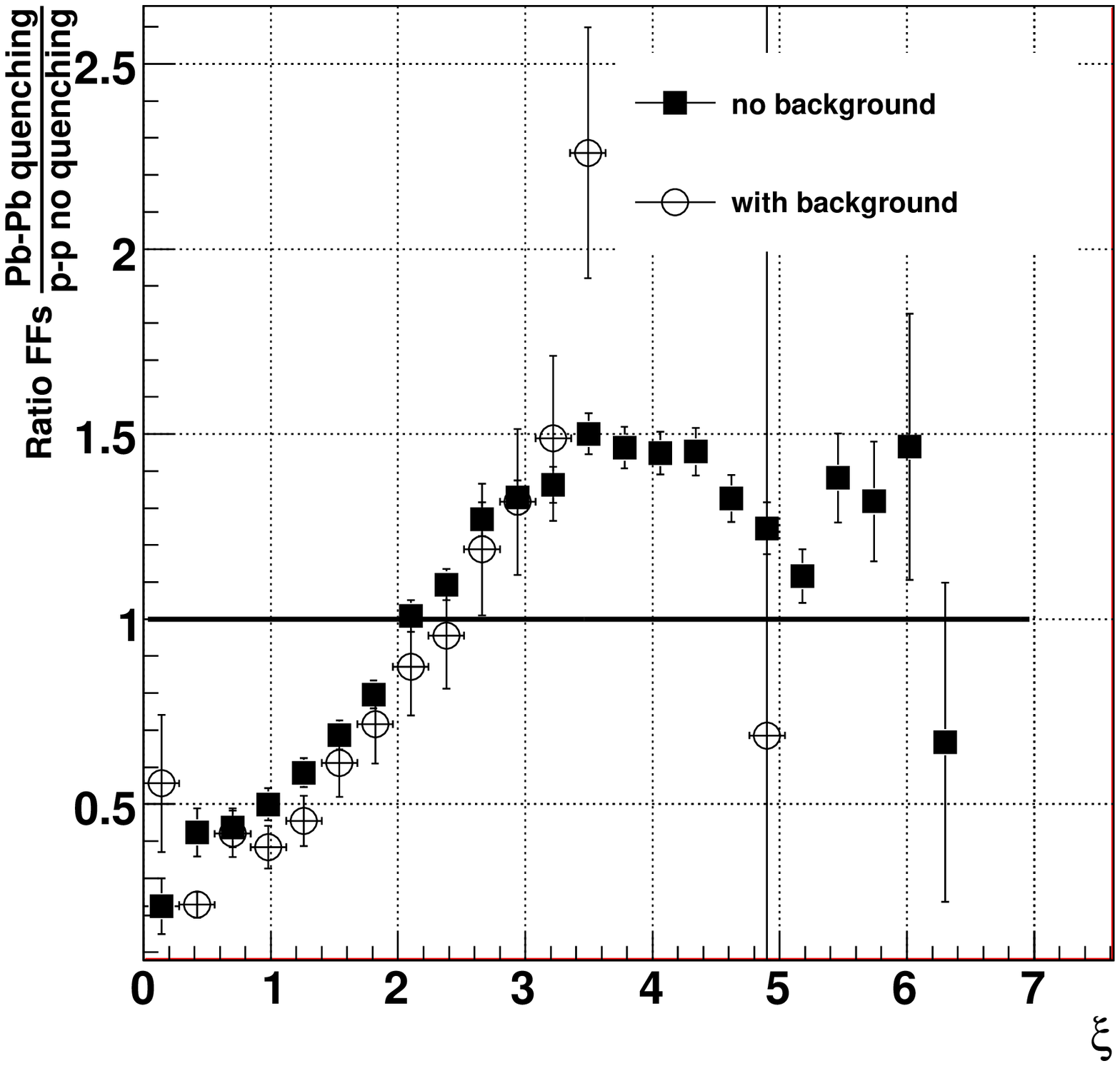} 
\hspace{0.1cm} 
\includegraphics[width=6.0cm,height=5.1cm,clip]{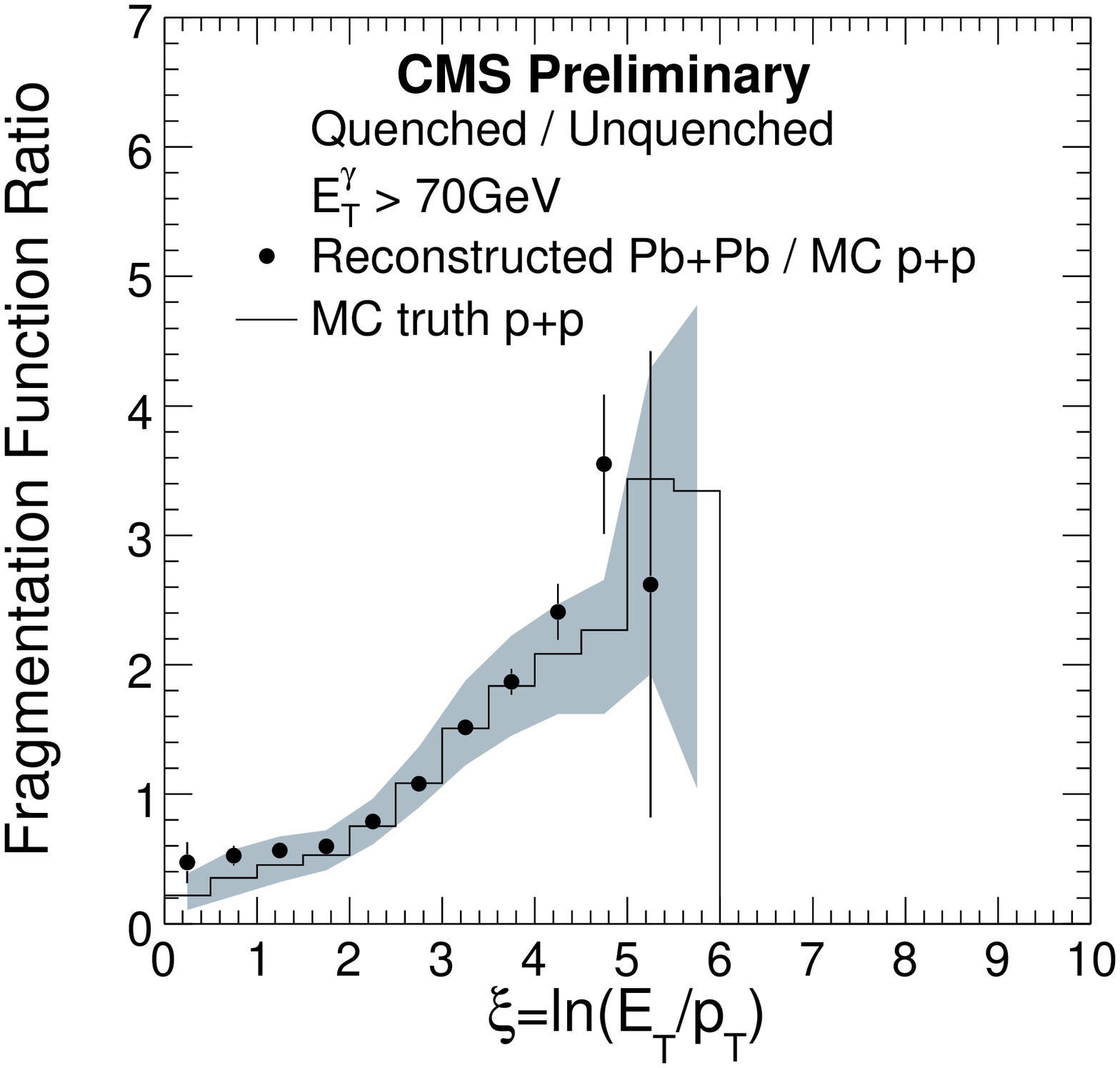}
\caption{
Ratio of medium over vacuum FFs as a function of  $\xi$ for quenched partons obtained in $\gamma-$jet simulations 
for central \PbPb\ at 5.5 TeV (0.5~nb$^{-1}$) in ALICE (left)~\protect\cite{Bourdaud:2008zz} and 
CMS (right)~\protect\cite{Loizides:2008pb}.}
\label{fig:mFFs_CMS_ALICE}
\end{figure}

ALICE~\cite{Bourdaud:2008zz,Conesa:2007nx} and CMS~\cite{Loizides:2008pb} have carried out simulation 
studies of the $\gamma$-jet channel, where the isolated $\gamma$  is identified in the EM calorimeter, 
the away-side jet {\it axis} ($\Delta\phi_{\gamma-jet}>$~3~rad) is reconstructed in the calorimeters, and the 
momenta of hadrons around the jet-axis ($R_{jet}<$~0.5) are measured in the tracker.
In the CMS acceptance and for $\etg>$~70~GeV, about 4500 $\gamma$-jet events are expected 
according to {\sc pythia} (scaled by the Glauber nuclear overlap) in one \PbPb\ year at the nominal 
luminosity. The obtained FFs for photon-jet events -- after subtraction of the underlying-event tracks --
are shown in Fig.~\ref{fig:mFFs_CMS_ALICE} for central \PbPb. Medium modified FFs are measurable with 
high significance in the ranges $z<$~0.7 and 0.2~$<\xi<$~6.


\section{Summary}

We have reviewed the main theoretical motivations behind the experimental
study of parton scattering and jet evolution and fragmentation in the hot and dense QCD matter
created in high-energy nucleus-nucleus collisions. The phenomenology of parton energy loss 
has been summarised as well as the main experimental results on single inclusive spectra and 
dihadron correlations measured at high transverse momentum, mainly in \AuAu\ reactions 
at RHIC collider energies, up to $\sqrtsnn$~=~200~GeV.
The analysis of jet structure modifications in heavy-ion collisions provides 
quantitative ``tomographic'' information on the thermodynamical and transport properties of the 
strongly interacting medium produced in the reactions.\\

Two chapters have discussed in detail two notable experimental results:
(i) the observed factor of $\sim$5 suppression of high-$p_T$ leading hadrons, 
and (ii) the strongly distorted azimuthal distributions of secondary hadrons 
emitted in the away-side hemisphere of a high-$p_T$ trigger hadron,
in central \AuAu\ relative to \pp\ collisions in free space.
Most of the properties of the observed high-$p_T$ single hadron and dihadron 
suppression (such as its magnitude, light flavour ``universality'', $p_T$, 
reaction centrality, path-length, and $\sqrtsnn$ dependences) 
are in quantitative agreement with the predictions of parton energy loss 
models in a very dense system of quarks and gluons.
The confrontation of these models to the data permits to derive the initial 
gluon density $dN^g/dy\approx$ 1400 and transport coefficient 
$\hat{q} =\mathscr{O}($10~GeV$^2$/fm$)$ of the produced medium at RHIC.\\

\noindent
In the last Chapter of this document, we have reviewed the details of jet reconstruction
in heavy-ion collisions: jet algorithms, underlying-event background subtraction, and
jet energy corrections. The analysis of jet spectra, jet shapes and the extraction of 
medium-modified parton-to-hadron fragmentation functions at low- and high- relative 
hadron momenta promise to fully unravel the mechanisms of parton energy loss 
in QCD matter, in particular at the upcoming LHC energies. The study of jet quenching 
phenomena proves to be an excellent tool to expand our knowledge of the dynamics 
of the strong interaction at extreme conditions of temperature and density.


\clearpage

\section*{Acknowledgments}

\noindent
Special thanks due to Fran\c{c}ois~Arleo, N\'estor Armesto, Constantin Loizides, St\'ephane Peign\'e, 
Carlos~Salgado and Marco van Leeuwen, for informative discussions and valuable comments on the manuscript. 
I am also indebted to Barbara Betz for her cooperation in various sections of a similar writeup of lectures 
given in the 2008 Jaipur QGP Winter School; and to Florian Bauer, Jorge Casalderrey-Solana, Rapha\"elle 
Ichou, Thorsten Renk, Bill Zajc and Slava Zakharov for useful feedback on the document. Last but not least, 
I would like to thank Reinhard Stock for his inspiring encouragement before and during the preparation of this work. 
Support by the 6th EU Framework Programme contract MEIF-CT-2005-025073 is acknowledged.



\printindex


\end{document}